\documentclass[11pt,twoside]{article}
\usepackage{graphicx} 
\usepackage{ulem}
\usepackage{amssymb}
\usepackage[mathscr]{eucal}
\usepackage{eufrak}
\usepackage{amsmath,amsthm}
\usepackage{physics}
\usepackage{mathrsfs}
\usepackage{lgreek}
\usepackage[shortlabels]{enumitem}
\usepackage[colorlinks=true,
   urlcolor=blue,           filecolor=green,      
   citecolor=green,      
   linkcolor=red,           bookmarks=true,
  unicode,
   plainpages=false,   ]{hyperref}
\usepackage{color}
\usepackage{comment}
\usepackage{amsfonts}
\font\bg=cmbx10 scaled 1300

\definecolor{royalblue}{rgb}{0,0,0.128}
\def\bp{\begin{proof}}
\def\ep{\end{proof}}
\def\n{\nabla}

\def\sfrac#1#2{\mbox{\Large$\frac{#1}{#2}$}}
\def\intl#1{\int\limits_{#1}}
\def\intll#1#2{\int\limits_{#1}^{#2}}
\def\dm{|\hskip-0.05cm|}

\def\OO{\Omega}
\def\displ{\displaystyle}
\def\VSE{\vspace{6pt}\\&\displ }
\def\VS{\vspace{6pt}\\\displ }
\def\rf#1{{\rm(\ref{#1})}}

\def\R{\Bbb R}
\def\N{\Bbb N}
\def\à{à}

\def\vep{\varepsilon}

\def\be{\begin{equation}}
\def\ba{\begin{array}}
\def\ea{\end{array}}
\def\ee{\end{equation}}
\def\vs1{\vspace{1ex}}

\def\ov{\overline}

\def\po{\partial\Omega}

\def\Ã©{\'{e}}
\def\Ãš{\`{e}}
\def\rot{{\rm curl}\hskip0.02cm}

\newtheorem{lemma}
{\bf Lemma}[section] 
\numberwithin{equation}{section}
\font\sc=cmcsc10
\date{\today}

\newtheorem{defi}
{\bf Definition}[section]
\newtheorem{tho}
{\bf Theorem}[section] 
\newtheorem{rem}
{\sc Remark}[section] 
\newtheorem{coro}
{\bf Corollary}[section] 

\title{\bg On the interaction between a rigid-body and a viscous-fluid:\\ existence of a weak solution and a suitable \textit{\bg Th\'{e}or\`{e}me de Structure}}
\author{\sc Paolo Maremonti and Filippo Palma
\thanks{Dipartimento di Matematica e Fisica,  
Universit\`{a} degli 
Studi della Campania
``L. Vanvitelli'', via Vivaldi 43, 81100 \null\hskip0.55cmCaserta,
 Italy.\newline\null\hskip0.55cm
paolo.maremonti@unicampania.it
\newline\null\hskip0.55cmfilippo.palma@unicampania.it
\newline\null\hskip0.55cm The  research activity  is performed under the
auspices of   GNFM-INdAM. }}
\begin{document}
\markboth{\footnotesize\rm    P. Maremonti and F. Palma} {\footnotesize\rm On...a rigid-body and a viscous-fluid: existence of a weak solution and a suitable\,\textit{Th\'{e}or\`{e}me de Structure}
}
\maketitle
\begin{abstract}
    In this paper, we prove the existence and a partial regularity of a weak solution to the system governing the interaction between a rigid body and a viscous incompressible Newtonian fluid. The evolution of the system body-fluid is studied in a frame attached to the body. The choice of this special frame becomes critical from an analytical point of view due to the presence of the term $\omega\times x\cdot \n u$ in the balance of momentum equation for the fluid. {As a consequence, we are forced to look for a technique that is different from the ones usually employed both for the existence and for the partial regularity of a weak solution to the Navier-Stokes problem. Hence, we prove the existence of a weak solution in an original way and give a new proof of the celebrated  }\textit{Th\'{e}or\`{e}me de Structure} due to Leray. {However, the regularity obtained for our weak solution is only for large times, hence our result is weaker compared to the one obtained by Leray.} 
\end{abstract}
\noindent
\section{Introduction}
In this article, we consider the initial boundary value problem for the motion of a rigid body in a viscous incompressible Newtonian fluid. We denote by $\mathcal B$ the rigid body and we also mean  $\mathcal{B}\subset\mathbb R^3$ as the region of space occupied by the body. We assume $\mathcal{B}$ closed, bounded, and with a sufficiently smooth boundary $\partial\mathcal{B}$. We assume that the fluid $\mathcal F$  occupies the region of space exterior to $\mathcal B$, that is the domain $\Omega:=\mathbb R^3\setminus \mathcal{B}$, so that $\partial\Omega=\partial\mathcal{B}$. \par The whole system is denoted by $\mathcal S:=\{\mathcal B,\mathcal F\}$.\par  Following Serre \cite{Serre} and Galdi \cite{Ga:Hand}, we study the dynamic interaction between the rigid body $\mathcal B$ and the fluid $\mathcal F$ in a principal inertial frame attached to the body, with the origin in its center of mass $\mathcal G$. In the absence of external forces and torques, the equations governing the dynamic interaction in the aforementioned frame are given by
\begin{equation} \label{eq:model}
    \begin{cases}
        & u_t-\Delta u= - [(u-V)\cdot \nabla u+\omega\times u]-\nabla\pi\,,\quad \forall (t,x) \in (0,T)\times \Omega\,, \\
        & \nabla \cdot u=0\,, \quad \forall (t,x) \in (0,T)\times \Omega\,, \\
        & u(t,x)=V(t,x)=\xi(t)+\omega(t)\times x\,, \quad \forall (t,x) \in (0,T)\times \partial \Omega\,, \\
        &\displ \lim_{\abs{x}\to \infty} u(t,x)=0\,, \quad \forall t\in (0,T)\,, \\
        &m\dot{\xi} + m\omega \times \xi +  \int_{\partial\Omega} \mathbb{T}(u, \pi)\cdot \nu=0\,, \quad \forall t\in (0,T)\,,\\
        & I \cdot \dot{\omega}+ \omega \times(I\cdot \omega)+ \int_{\partial\Omega} x \times \mathbb{T}(u,\pi) \cdot \nu=0\,, \quad \forall t\in (0,T)\,,\\
        & \xi(0)= \xi_0\,, \quad \omega(0)=\omega_{0}\,, \\
        & u(0,x)=u_0(x)\,, \quad \forall x \in \Omega\,.
    \end{cases}
\end{equation}
Here, $u:(0,T)\times \Omega \to \mathbb{R}^3$ is the velocity of the fluid, $\pi:(0,T)\times \Omega \to \mathbb{R}$ is the pressure field, $V:(0,T)\times \overline{\OO}\to \mathbb{R}^3$ is the velocity  of the rigid body expressed by $V(t,x)=\xi(t)+\omega(t)\times x$, where  $\xi:(0,T)\to\mathbb{R}^3$ is the translation velocity and $\omega:(0,T)\to\mathbb{R}^3$ is the angular velocity, $m$ is the total mass of the body, $I$ is the inertia tensor\footnote{Since we study the evolution of the system $\mathcal S$ in a principal inertial frame, with the origin in the center of mass of the body, the inertia tensor $I$ is diagonal and positive definite. Following the consolidated Euler's notation, we represent it as \mbox{\tiny $
I=\Bigg(\!\!\ba{lll}
    A\hskip-0.2cm&0 &\hskip-0.2cm0\\ 0\hskip-0.2cm& B&\hskip-0.2cm 0\\ 0\hskip-0.2cm& 0& \hskip-0.2cmC
\ea\!\!\Bigg)$}.}, $\nu$ is the outer unit normal vector of the surface $\partial\Omega$ and $\mathbb{T}$ is the Cauchy stress tensor of the fluid, defined by
\[
\mathbb{T}(u,\pi):= -\pi\mathbb{I}+2\mu\mathbb{D}(u),
\]
where $\mathbb{I}$ is the identity tensor, $\mu$ is the  dynamic viscosity and $\mathbb{D}(u)=\frac{1}{2}(\nabla u+ (\nabla u)^T)$ is the symmetric part of $\nabla u$. For the sake of the simplicity, we assume $\mu=1$.\footnote{ Actually, the topic of Fluid-Structure Interaction is widely studied. Although our focus is the interaction between a rigid body and a fluid, there are also studies concerning the case in which the body is elastic instead of rigid, see for instance \cite{Spe}.} \par In the formulation of problem \eqref{eq:model}, we did not consider the action of possible external forces and torques. Actually, under suitable assumptions, their introduction does not add analytical issues related to the proof of our results. \par
The model \eqref{eq:model}, related to $\mathcal S$, has been widely investigated in the last decades. The corresponding stationary problem has also been subject to several studies. \par
In the typical case in which the body is assumed to be in free fall, hence it is subject to the gravitational force, and the fluid motion is supposed to be steady, the first studies go back to Weinberger \cite{Wei}. \par
Some years later, Serre \cite{Serre} stated the existence of weak solutions, both in the stationary and in the nonstationary case. Another proof of the existence of weak solutions for the evolution problem is due to Silvestre \cite{Silv: tesi}. \par The stationary problem associated with system $\mathcal{S}$ was investigated by Galdi and Silvestre in \cite{GS-sta}. \par For the evolution problem, a first result on the existence of a unique local mild solution was obtained by Hishida in \cite{His}, in the case of prescribed rigid motion and of a purely rotating body, i.e. $\xi\equiv 0$ and $\omega$ is a constant vector. The result of Hishida was obtained in the $L^2$-setting and was extended to the $L^p$-setting by Geissert et al. in \cite{Gei-Hec-Hie}. \par In the entirety of the unknown functions, the evolution problem was investigated by Galdi and Silvestre in \cite{GS},  that showed the existence of a strong local solution in the $L^2$-setting. \par Recently, Galdi \cite{Ga:23} also proved the existence of a strong global solution under a suitable smallness assumption on the initial datum and he investigated the large-time behavior of the Dirichlet norm, proving an asymptotic property in time. \par The large-time behavior of the kinetic energy was discussed by Galdi and Maremonti in \cite{Gal-Mar}.  \par An intriguing problem that, to the best of our knowledge, is still open is the partial regularity of a weak solution. The aim of this article is to prove a result in this direction. \par However, our result is an original restatement of the Leray {\it  Th\'{e}or\`{e}me de Structure},  regarding both the existence and the partial regularity of a weak solution.\par
Let us recall the result by Leray, commonly known as the {\it Th\'{e}or\`{e}me de Structure} of a weak solution to the Navier-Stokes problem. By the symbol $\mathscr H^\frac12$ we refer to the $\frac12-$dimensional Hausdorff measure. \par {\it The ``Th\'{e}or\`{e}me de Structure" states that the Leray weak solution\footnote{\,We refer to the Leray weak solution since we are considering the IBVP in an unbounded domain.\par Actually, as far as we know, in unbounded domains for a Hopf weak solution no structure theorem is known. \par In our problem, although we propose a different proof of the partial regularity of a Leray weak solution, the distinction between a generic weak solution and  our weak solution is due.}, corresponding to the initial datum $u_0\in L^2(\mathbb R^3)$, divergence free in a weak sense, is regular in a time interval $(\theta,\infty)$, with $\theta\le c\dm u_0\dm_2^4$, and is also regular in a sequence of open time intervals $\{(\theta_\ell,T_\ell)\}_{\ell\in\mathbb N}$ such that\newline $\mathscr H^\frac12((0,\theta)\setminus \mbox{$\underset{\ell\in\N}\cup$}(\theta_\ell,T_\ell))$ is null.} (See \cite{Ler, Sch}) \par There exists a wide literature concerning the extension of the Leray Theorem to the more general case of the initial boundary value problem, {we do not cite it since we suppose that it is well-known to the reader}.\par  Let us mention the main tools that allowed Leray to prove his result.  First of all, for the Navier-Stokes Cauchy problem, it is required the validity of an energy inequality in a strong form, that is  
\be \label{eq: Energia-Leray}
\dm u(t)\dm_2^2+2\int_s^t\dm \n u(\tau)\dm_2^2\,d\tau\le \dm u(s)\dm_2^2\,, \quad \text{for all }t>0\,,\,\text{ a. a. } s\in(0,t)\, \,\text{and }s=0
\ee
and, setting $\mathcal T:=\{s\,:\,\eqref{eq: Energia-Leray}\, \text{holds}\}$, for all $s\in\mathcal T$
\be \label{IC}
\lim_{t\to s^+}\dm u(t)-u(s)\dm_2=0\,.
\ee
Moreover, the coincidence between two solutions, of which at least one is regular and another satisfies the energy relation \eqref{eq: Energia-Leray}, holds. For the last goal, the limit property \rf{IC} is crucial.\par
For the problem \eqref{eq:model}, the construction of a weak solution enjoying an energy inequality in the strong form \eqref{eq: Energia-Leray} seems still out of reach.\par Actually, the only available energy inequality seems to be the weaker energy inequality of Hopf type, which reads as
\[
\dm u(t)\dm_2^2+m|\xi(t)|^2+(I\cdot\omega(t))\cdot \omega(t)+4\int_0^t \dm \mathbb D(u)\dm_2^2\,d\tau\le \dm u_0\dm_2^2+|\xi_0|^2+(I\cdot\omega_0)\cdot\omega_0\,.
\]
{Furthermore, there are very challenging problems related to the uniqueness of a strong solution, in fact it is not achieved for any initial datum in $W^{1,2}(\Omega)$ that satisfies a suitable compatibility condition on the boundary.} \par Actually, in their last work \cite{Mar-Pal}, the authors prove a uniqueness result in the sense of Leray-Serrin. However, a further assumption on the strong solution is required; specifically, they need a weighted integrability property for $\n u$ that they prove possible for weighted initial data\footnote{A future goal is the investigation of the asymptotic behavior in the space variable of the regular solution corresponding to weighted initial data, as done in \cite{GM-peso, MV, AV} for the Stokes and for the Navier-Stokes problems.}. \par The foregoing considerations lead to the search of a suitable technique in order to obtain a result of partial regularity for a weak solution to equations \eqref{eq:model}. \par { We point out that our result ensures the regularity of a suitable weak solution on an interval of time $(0,\theta_0)\cup(\theta,\infty)$, with $\theta_0,\theta>0$. However, if $\theta_0<\theta$, the study of the regularity of our weak solution on the finite interval of time $(\theta_0,\theta)$ appears to us non trivial and still needs careful investigation. Therefore, there is an intriguing gap between the analytical theory for problem \,\eqref{eq:model} and the analytical theory for the Navier-Stokes problem. We stress that this is not due to the new techniques introduced in this work, as a matter of fact the same techniques are employed for the Navier-Stokes problem in the paper \cite{MP-NS}, leading to the Leray's {\it   Th\'{e}or\`{e}me de Structure} in its entirety. }\par {The main issue that does not allow us to get the desired regularity on $(\theta_0,\theta)$ consists in the fact that for  the approximating sequence $\{u^n\}$ of our weak solution it seems still out of reach to prove either a strong convergence in $L^2(0,T;L^2(\Omega))$ or a strong convergence of $\{\n u^n\}$ in $L^p(0,T;L^2(\Omega))$, for some $p\in (1,2)$. The aforementioned convergences are known for the Navier-Stokes Problem (see \cite{Ler,CGM,M-EE}).} \par
Before stating our main result, we briefly introduce the notation for the function spaces that will be used throughout the paper. 
\par {By $L^p(\Omega)$, $W^{m,p}(\Omega)$ and $\widehat W^{m,p}(\Omega)$ ($m\in\mathbb{N}$, $p\ge1$), we refer to the usual Lebesgue, Sobolev and homogeneous Sobolev spaces, whose norms are denoted by $\norm{\cdot}_p$\,, $\norm{\cdot}_{m,p}$ and $\dm D^m\cdot \dm_p$\,, respectively.} Moreover, we denote by $L^p_{w}(\Omega)$ the weak-$L^p$ Lebesgue space and we refer to its norm with the symbol $\dm \cdot\dm_{p_w}$. \par
Let $a\in\mathbb N_0$. We set
\[
\ba{rl}
L^p(\Omega,|x|^a)&\hskip-0.2cm:=\{u\,:|x|^au\in L^p(\Omega)\}\,,\VS
\widehat W^{1,p}(\Omega,|x|^a)&\hskip-0.2cm:=\{u : |x|^a\n u\in L^p(\Omega)\}\,.
\ea
\]
Setting
\[
\mathscr C_0(\Omega):=\{u\in C^\infty_0(\Omega)\,:\,\n\cdot u=0\}\,,
\]
we denote by $J^p(\Omega)$ the completion of $\mathscr C_0(\Omega)$ with respect to the norm $\dm\cdot\dm_p$.\par
{We denote by $H^\frac12(\partial\Omega)$ the trace space of $W^{1,2}(\Omega)$. We refer to the usual trace operator by the symbol $\gamma$. \par We denote by $E(\Omega)$ the completion of $C_0^1(\overline \Omega)$ with respect to the metric
\[
\dm u\dm_{E(\Omega)}=\dm u\dm_2+\dm \n\cdot u\dm_2\,.
\]
As it is well-known (see \autoref{sec: prel} for more details), we can define a trace in the sense of distributions for functions belonging to $E(\Omega)$. We refer to the trace space of $E(\Omega)$ by the symbol $H^{-\frac12}(\partial\Omega)$ and we denote the generalized trace operator by $\phi_\nu[u]$\,, for all $u\in E(\Omega)$. } \par
Let $X$ be a Banach space. The spaces $C([a,b];X)$ and $L^{p}(a,b;X)$ are defined as follows
\[
C([a,b];X):=\{u:[a,b]\to X \, : \, u\text{ is continuous in the norm} \, \, \norm{\cdot}_X\};
\]
\[
L^{p}(a,b;X):=\{ u:(a,b)\to X \, :\,  \biggl(\int_a^b \norm{u(t)}_{X}^p \, dt\biggr)^{\frac{1}{p}}<\infty \}\,. 
\]
We now introduce some function spaces that allow us to follow the evolution of both elements of the system $\mathcal{S}$.\par  We set
\[
\mathcal{R}:=\{\overline{u} \in C^{\infty}(\mathbb{R}^3)\, : \, \overline{u}(x)=\overline{u}_1+\overline{u}_2\times x, \,\, \overline{u}_1,\overline{u}_2\in \mathbb R^3\}.
\]
The vectors $\overline{u}_1$ and $\overline{u}_2$ are called \textit{characteristic vectors} of the rigid motion $\overline{u}$.\par
{Let us consider the set $\mathcal  K(\R^3)$ of functions $v$ such that:
\begin{itemize}
    \item $v\in C_0^\infty(B_R)$, for some $R>\text{diam}(\mathcal B)$;
    \item $v(x)=\overline v(x)=\overline v_1+\overline v_2\times x$, with $\overline v_1, \overline v_2\in \R^3$, in a neighborhood of $\mathcal B$.
\end{itemize}
We define the following scalar products on $\mathcal K(\R^3)$:\footnote{After straightforward computations, in view of the choice of a principal inertial frame, with the origin in $\mathcal G$, we get that the scalar products in \eqref{SPL2H1} are equivalent, respectively, to the usual $L^2(\R^3)$ and $W^{1,2}(\R^3)$ scalar products. }
\be \label{SPL2H1}
\ba{rl}
(v,w)_1 &\hskip-0.2cm\displ:= \int_\Omega v(x)\cdot w(x)\,dx +m\overline v_1\cdot \overline w_1+( I\cdot\overline v_2)\cdot\overline w_2\,,\VS
(v,w)_2 &\hskip-0.2cm\displ:=\int_\Omega v(x)\cdot w(x)\,dx+2\int_\Omega \mathbb D(v)\cdot\mathbb D(w)\,dx  +m\overline v_1\cdot \overline w_1+ ( I\cdot\overline v_2)\cdot\overline w_2\,.
\ea
\ee
We define the space $L(\R^3)$ as the completion of $\mathcal K(\R^3)$ with respect to the metric induced by the scalar product \eqref{SPL2H1}$_1$ and we define the space $W(\R^3)$ as the completion of $\mathcal K(\R^3)$ with respect to the metric induced by the scalar product \eqref{SPL2H1}$_2$. \par
 It is readily seen that the spaces $L(\R^3)$ and $W(\R^3)$ are characterized as follows:
    \[
    \ba{rl}
    L(\R^3)&\hskip-0.2cm=\{u\in L^2(\R^3)\,:\, u=\overline u\in\mathcal R \,\text{ in }\mathcal B\}\,,\VS
    W(\R^3)&\hskip-0.2cm=\{u\in W^{1,2}(\R^3)\,:\,u=\overline u\in\mathcal R \,\text{ in }\mathcal B\,, \,\,\gamma(u)=\overline u\}\,,
    \ea
    \]
    and they are separable Hilbert spaces. \par
We define the vector space $\mathcal{H}(\Omega)$ and its subspace $\mathcal{V}(\Omega)$ as 
\be \label{Hspace}
\mathcal H(\Omega):=\{ u \in L^2(\Omega)\,:\, \n\cdot u=0\,\,\text{in the weak sense, } \phi_\nu[u]=\phi_\nu[\overline u]\,,\,\,\text{for some }\overline u\in\mathcal R\}\,, 
\ee
\be \label{Vspace}
\mathcal V(\Omega):=\{ u \in W^{1,2}(\Omega)\,:\, \n\cdot u=0\,, \gamma(u)=\overline u\,,\,\,\text{for some }\overline u\in\mathcal R\}\,,
\ee
They are endowed, respectively, with the scalar products \eqref{SPL2H1}$_1$ and \eqref{SPL2H1}$_2$, and they are separable Hilbert spaces.\par It is clear that the presence of the total mass of the body does not add analytical issues to our purposes. Therefore, in order to avoid overloading the notation, we set $m=1$ in the following. \par }
Before giving the notion of weak solution to \eqref{eq:model}, we introduce the set of the space-time test functions $\mathscr C(\R^3_T)$. A function $\varphi\in\mathscr C(\mathbb R^3_T)$ if the following hold:
\begin{itemize}
    \item $\varphi\in C_0^\infty([0,T)\times B_R)$\,, for some $R>\text{diam}(\mathcal B)$;
    \item $\n\cdot \varphi=0$, for all $(t,x)\in[0,T)\times\Omega$;
    \item There exist $\overline\varphi_1,\overline\varphi_2\in C_0^\infty([0,T))$ such that  $\varphi(t,x)=\overline\varphi_1(t)+\overline\varphi_2(t)\times x$ in a neighborhood of $\partial \Omega$, for all $t\in [0,T)$;
\end{itemize}
We remark that in the following the symbol $c$ denotes a generic non-negative constant whose value may change from line to line, and it is inessential for our purposes.\par
Furthermore, following the notation adopted in the monograph \cite{Ga:book}, we denote by $B_R$ the open ball centered in the origin and of radius $R$, by $\OO_R$ the open subset $\OO \cap B_R$, and by $\OO^R$ the open subset $\OO \cap \{x \in \mathbb{R}^3 : |x| > R\}$.\par 
Denoting by $e_r$ the outer unit normal vector to $\partial B_R$,
we define the spaces $\mathcal H(\Omega_R)$ and $\mathcal V(\Omega_R)$, $R>\text{diam}(\mathcal B)$, as follows:{
\[
\mathcal H(\Omega_R):=\{u\in L^2(\Omega_R)\,:\, \n\cdot u=0\,\,\text{in the weak sense}\,,\, \phi_{e_r}[u]=0\,,\,  \phi_\nu[u]=\phi_\nu[\overline u]\,,\,\text{for a certain }\overline u\in\mathcal R\}\,,
\]
\[
\mathcal V(\Omega_R):=\{u\in W^{1,2}(\Omega_R)\,:\, \n\cdot u=0\,,\, \gamma_{ B_R}(u)=0\,,\, \gamma_{\Omega}(u)=\overline u\,,\,\text{for a certain }\overline u\in\mathcal R\}\,.
\]}
They are endowed, respectively, with the scalar products
\be\label{SPHR}
(u,v)_{\mathcal{H}(\Omega_R)}:= \overline{u}_1\cdot \overline{v}_1+(I\cdot\overline{u}_2) \cdot \overline{v}_2+ \int_{\Omega_R}u\cdot v \, dx
\ee
and
\be\label{SPVR}
(u,v)_{\mathcal{V}(\Omega_R)}:=2 \int_{\Omega_R}\mathbb D(u)\cdot \mathbb D( v) \, dx,
\ee
For more details on function spaces, we refer to \cite{Adams}, for the functional setting concerning the model, we quote \cite{Ga:Hand, GS, Silv: tesi}. \par
We now give the definition of a weak solution to system \eqref{eq:model}. 
\begin{defi} \label{def:WS}{\sl Let $(u_0,\xi_0,\omega_0)\in\mathcal H(\Omega)\times\mathbb R^3\times\mathbb R^3$, with $\phi_\nu[u_0-\xi_0-\omega_0\times x]=0$.
   For all $T>0$, the field  $u:(0,T)\times \OO\to\R^3$ and the fields $\xi:(0,T)\to\mathbb R^3$, $\omega:(0,T)\to\mathbb R^3$ 
    are said weak solution to \eqref{eq:model} if
    \begin{description}
        \item[(i)] $u\in L^2(0,T; \mathcal V(\Omega)) \cap L^{\infty}(0,T;\mathcal H(\Omega))$;
        \item[(ii)] $\gamma(u(t))=\xi(t)+\omega(t)\times x$ for a. a. $t\in (0,T)$, with $\xi,\omega\in L^2(0,T)$;
        \item[(iii)] $(u,\xi,\omega)$ satisfies the following integral equation:
        \begin{equation}
            \hskip-0.2cm\begin{array} {l} \displ
               \hskip-0.82cm \int_0^T\!\Bigg[\!\! \int_{\Omega} \!\Big[\!-u\!\cdot\! \varphi_t +2\mathbb D( u)\!\cdot \!\mathbb D( \varphi) - ((u-\!V)\cdot\! \nabla\! \varphi) \!\cdot\! u +\omega\!\times\! u\!\cdot \!\varphi\Big]dx-\!\frac{d\overline\varphi_1}{dt}\cdot \xi\! -\! \bigg(\!I\!\cdot\frac{d\overline\varphi_2}{dt}\!\bigg)\!\cdot \omega\!\Bigg]\!dt\VS \hskip0.48cm
                 = \!\int_0^T \!\Big[\overline\varphi_1\cdot\xi\times\omega + \overline\varphi_2\cdot(I\!\cdot\!\omega)\times\omega\Big]\, dt +\! \int_{\Omega}\!\varphi(0)\cdot u_0\, dx +\overline{\varphi}_1(0)\!\cdot\!\xi_0 + \! (I\!\cdot\overline{\varphi}_2(0)) \!\cdot\!\omega_0\,,
            \end{array}
        \end{equation}
        for all $\varphi\in\mathscr C(\mathbb R^3_T)$
        \item[(iv)] $\displ\lim_{t\to 0^{+}} (u(t)-u_0,\psi)_{1}=0$, for all $\psi\in\mathcal H(\Omega)$.
    \end{description} }
\end{defi}
 We introduce the notion of regular solutions. \begin{defi}\label{SOLR}{\sl Let $(u_0,\xi_0,\omega_0)\in \mathcal V(\OO)\times \mathbb R^3\times \mathbb R^3$, with $\gamma(u_0-\xi_0-\omega_0\times x)=0$. For some $T\in(0,\infty]$, the pair $(u(t),\pi(t))$, with $u:(0,T)\times\Omega\to \mathbb R^3$, $\pi\in L^2(0,T;\widehat W^{1,2}(\Omega))$, and the fields $\xi:(0,T)\to\mathbb R^3$, $\omega:(0,T)\to\mathbb R^3$, are  said to be a regular solution to the IBVP \rf{eq:model} if the four functions $(u,\pi,\xi,\omega)$ satisfy \eqref{eq:model} almost everywhere in $(t, x)\in(0,T)\times\OO$,  with
\be\label{SOLR-I}\ba{c}\xi(t)\,,\;\omega(t)\in C(0,T)\mbox{\; and\; }\dot\xi(t)\,,\;\dot\omega(t)\in L^2(0,T)\,,\VS u\!\in\! L^\infty\!(0,T;\mathcal V(\OO))\!\cap\! L^2(0,T;W^{2,2}(\OO)),u\in C(0,T; W^{1,2}(\Omega_R))\,\,\text{for all R}>\text{diam}(\mathcal B)\,,\VS\,u_t(1\!+\!|x|)^{-1}\!\!\in \!L^2(0,T;L^2(\OO)),\VS \lim_{t\to0}\xi(t)=\xi_0\,,\quad\lim_{t\to0}\omega(t)=\omega_0\,,\quad\lim_{t\to0}\dm u(t)-u_0\dm_{2}=0\,.\ea\ee }\end{defi}  The definition of  regular solution is  based on the results by Galdi-Silvestre \cite{GS} and Galdi \cite{Ga:23}. 
{\begin{tho}\label{GPD}{\sl For all $(u_0,\xi_0,\omega_0)\in\mathcal V(\Omega)\times\mathbb R^3\times\mathbb R^3$, with $\gamma(u_0-\xi_0-\omega_0\times x)=0$, there exists a regular solution $(u,\pi,\xi,\omega)$ to problem \eqref{eq:model}. Moreover, there exists $\delta>0$ such that, if the initial datum $(u_0,\xi_0,\omega_0)$ satisfies \be\label{GPD-I}\dm u_0\dm_{1,2}+|\xi_0|+\big[ (I\cdot\omega_0)\cdot\omega_0\big]^\frac12<\delta\,,\ee
there exists a regular solution defined for all $t>0$ and the following asymptotic property holds:
\be \label{ASPR}
\lim_{t\to\infty} \big[\dm \n u(t)\dm_2+|\xi(t)|+ |\omega(t)|\big]=0\,.
\ee
Finally, if $|x|\n u_0\in L^2(\OO)$, then the regular solution is unique.}\end{tho}} \bp The local existence is proved in \cite{GS}, the global existence and the asymptotic property \eqref{ASPR} are furnished by Galdi in \cite{Ga:23}. The uniqueness is given in \cite{Mar-Pal}.\ep 
 We denote by $A_0:=\dm u_0\dm_2^2+|\xi_0|^2+ (I\cdot\omega_0)\cdot\omega_0$, by $\chi$ the solution to the algebraic  equation $$\chi^\frac12(1+c)+\big[c(\dm \rot u_0\dm_1+A_0^\frac12+A_0)\big]^\frac16\chi^\frac16=\delta,$$ where  $\delta$ is given in Theorem\,\ref{GPD}, and $c$ is a numerical constant.\par{ We aim to establish the following  theorem.}
  \begin{tho} \label{thm:MR}{\sl
    Assume that \be \label{DI}
    (u_0,\xi_0,\omega_0)\!\in \!(\mathcal V (\Omega)\cap\widehat W^{1,2}(\Omega,|x|))\!\times\! \mathbb R^3\!\times\!\mathbb R^3,\text{ with }\gamma(u_0-\xi_0-\omega_0\times x)=0\text{ and }\rot u_0\!\in \!L^1(\Omega) \,.\ee 
    Then, there exists a weak solution $(u,\xi,\omega)$  to \eqref{eq:model}   such that, for all $t>0$,
    \be\label{REG} \norm{u(t)}_2^2 + \abs{\xi(t)}^2 + (I \cdot \omega (t))\cdot\omega(t)+ 4 \int_{0}^t \norm{\mathbb{D}(u)(\tau)}_2^2 \, d\tau \le \dm u_{0}\dm_2^2  + |\xi_{0}|^2+ (I \cdot \omega_{0})\cdot\omega_0 \,.
   \ee Moreover, 
   \be \label{CDI}
   \lim_{t\to 0^+}\dm u(t)-u_0\dm_2=\lim_{t\to 0^+}|\xi(t)-\xi_0|=\lim_{t\to 0^+}|\omega(t)-\omega_0|=0\,,
   \ee 
   and there exists a finite instant of time $\theta\le\exp [\frac{2A_0}{\chi}]-1$,
such that 
   $(u,\xi,\omega)$ becomes a regular solution for $t>\theta$ in the sense of Definition\,\ref{SOLR}. Hence, in particular, there exists a pressure field $\pi$ such that $(u,\pi,\xi,\omega)$ solves \rf{eq:model} almost everywhere in $(t, x)\in (\theta,\infty)\times\OO$.} 
\end{tho}
Since the asymptotic behavior in time of the Dirichlet norm of a regular solution has already been established in the statement of Theorem\,\ref{GPD}, in the following corollary we deal with the asymptotic behavior in time of the kinetic energy associated with the system $\mathcal S$.
\begin{coro}\label{ASBEH}
    {\sl Let $(u,\xi,\omega)$ be the solution to problem \eqref{eq:model} ensured by Theorem\,\ref{thm:MR}. Then, for all $q\in (\frac32,2)$, there exists $t_0\in [\theta,\theta+1]$ such that $(u,\xi,\omega)$ satisfies the following decay property:
    \[
    \dm u(t)\dm_2+|\xi(t)|+|\omega(t)|\le \frac{c(q, A_0,\dm \rot u_0\dm_1)}{(t+1)^{\frac{2-q}{4q}}}\,,\,\,\text{for all t}>t_0\,.
    \]}
\end{coro}

It is natural to inquire about the unusual assumptions imposed on the initial datum, namely 
\(\nabla u_{0}\in L^{2}(\Omega)\), \(|x|\nabla u_{0}\in L^{2}(\Omega)\), and \(\rot u_{0}\in L^{1}(\Omega)\), 
especially in a framework of weak solutions where, at least for the fluid part \(\mathcal F\), one typically 
requires only \(u_{0}\in L^{2}(\Omega)\). Our requirements on the initial datum arise from the specific construction of our weak 
solution and from the partial regularity result we want to establish. A preliminary explanation is already 
provided in the outline, in particular in item~ii. Here we only highlight the main ideas.\par { As far as the existence of a weak solution is concerned, we follow the approach proposed by Leray \cite{Ler} for the Navier-Stokes equations. Hence, we study a mollified version of system \eqref{eq:model} and we get a weak solution as the weak limit of $\{(u^n, \xi^n,\omega^n)\}_{n\in\mathbb{N}}$, where $(u^n,\pi^n,\xi^n,\omega^n)$ is the regular solution to the mollified system \eqref{eq:moll} that we will introduce in \autoref{sec: mollificatore}. However, as it will be clear in the following, the analytical theory for our mollified problem exhibits numerous differences with respect to the mollified Navier-Stokes problem. \par  We investigate the partial regularity of our weak solution by making use of {\it a priori} estimates for the sequence $\{(u^n, \xi^n,\omega^n)\}_{n\in\mathbb{N}}$. \par 
To sum up, our approach relies on the construction of a regular sequence \(\{u^{n}\}\) that converges to the weak solution. \par }
A suitable subsequence of the quoted sequence is required to be regular for all \(t>0\), regularity that is achieved through a suitable extension 
procedure. Such a requirement is unusual in the context of weak solutions, yet it is essential for our 
analysis. Moreover, this extension has to be uniquely defined. The regularity condition 
\(\nabla u_{0}\in L^{2}(\Omega)\) ensures the needed regularity, while the additional assumption 
\(|x|\nabla u_{0}\in L^{2}(\Omega)\) is required to ensure the uniqueness of the extension. These bounds 
are not uniform with respect to \(n\in\mathbb N\), but this lack of uniformity is inessential for our purposes.
Instead, concerning the assumption \(\rot u_{0}\in L^{1}(\Omega)\), it is used to ensure for all $n\ge n(\theta)$, the existence of 
an instant of time \(\theta_{n}\le \theta\) at which the quantity
\[
\|u^n(\theta_{n})\|_{1,2} + |\xi^{n}(\theta_{n})| + |\omega^{n}(\theta_{n})|
\]
is sufficiently small, smallness in the sense required by Theorem\,\ref{GPD}, which differs from the classical smallness conditions encountered in the  usual Navier-Stokes problem.
\subsection*{Outline of the proof.}\par  Our task is twofold. From one side, we look for an existence theorem for weak solutions. From another side, we look for the partial regularity of our weak solution in the sense of the Leray {\it   Th\'{e}or\`{e}me de Structure}. \par So that a first remark is that we do not prove the partial regularity for any weak solution.\par This points out a second remark. \par The fact is that we do not employ the classical approach developed in the analytical theory of the Navier-Stokes equations.\par In this connection we would like to say that we construct a suitable\footnote{\,Here, the term suitable has not to be meant in the sense of the celebrated Caffarelli-Kohn-Nirenberg paper \cite{CKN}.} weak solution.\par Our weak solution is constructed by means of an approximating  sequence which enjoys some a priori estimates that lead to the weak solution with partial regularity indicated in the statement of Theorem\,\ref{thm:MR}.\par  The particularity of the proof consists in the fact that the approximating solutions are  defined  for all $t>0$ only for the elements of the sequence with index $n$ up a suitable integer $n(\theta)$ that cannot be estimated a priori, in fact for each initial datum we state the existence of $n(\theta)$.   \par More precisely, we follow the following argument lines. \begin{itemize}
 \item[i. ] We consider the mollified equations \rf{eq:moll}. The mollification concerns the convective term $(u-V)\cdot\n u$ replaced by setting  $\mathbb J_n(u-V)\cdot \n u$, where $\mathbb J_n:=J_{\frac{1}{n}}$ is the  Friedrichs mollifier in the space variable. Following the arguments of Galdi and Silvestre \cite{GS}, suitably modified due to the mollification of the convective term, we are able to get, for all $n\in\mathbb N$, a solution $(u^n,\pi^n,\xi^n,\omega^n)$ to problem \rf{eq:moll} initially defined on a time interval $(0,T'_0)$, where $T'_0\ge2(cA_0^\frac12)^{-1}$, with $A_0:=\dm u_0\dm_2^2+|\xi_0|^2+(I\cdot\omega_0)\cdot\omega_0$. Since we choose an initial datum $(u_0,\xi_0,\omega_0)\in (\mathcal V(\Omega)\cap \widehat W^{1,2}(\Omega,|x|))\times\mathbb R^3\times\mathbb R^3$ then, in agreement with \cite{Mar-Pal}, we also get the uniqueness of the solution $(u^n,\pi^n,\xi^n,\omega^n)$. { Setting $T_0:=(cA_0^\frac12)^{-1}$, we get, for all $n\ge n_0\in\mathbb N$,
\be\label{I-B}\ba{l}\displ  \dm u^n(T_0)\dm_2^2+|\xi^n(T_0)|^2+(I\cdot\omega^n(T_0))\cdot \omega^n(T_0)+2\int_{0}^{T_0} \dm \n u^n(\tau)\dm_2^2\,d\tau\le \sfrac32A_0\,,\VS \dm \n u^n(T_0)\dm_2\leq c(n, A_0,\dm  u_0\dm_{1,2})\,,\VS \dm |x| \n u^n(T_0)\dm_2\leq c(n, A_0,\dm  |x|\n u_0\dm_{2})\,, \VS \mbox{with the boundary  compatibility condition \eqref{DI} in }T_0\,.\ea\ee The opportunity of estimate \rf{I-B} is due to the fact that, being the non linear term mollified,  we cannot obtain the classical energy relation. Actually, to the classical terms  there is summed, on the right-hand side, the term  
\be\label{II-B}-\intll0{T_0}\int_{\partial\Omega} \mathbb J_n(u^n-V^n)\cdot \nu |V^n|^2\,dS \,. \ee However, by virtue of the local existence on $(0,T_0]$, we can estimate \rf{II-B} by means of $\frac12A_0$ (see Lemma\,\ref{le:Stime per rel energia}) provided that $n\geq n_0$, $n_0$ suitable. As a consequence we get \rf{I-B}. \par In conclusion, problem \rf{eq:moll} has like first step the existence interval of $(u^n,\pi^n,\xi^n,\omega^n)$ restricted to $(0,T_0]$, with $T_0=(cA_0^{\frac12})^{-1}$ independent of $n\ge n_0$. Instead, depending on $n\ge n_0$ there are the bound of $\dm \n u^n(T_0)\dm_2$ and that of $\dm |x|\n u^n(T_0)\dm_2$.}
\item[ii.]Thanks to estimate \rf{I-B} the previous argument can be iterated by realizing a sequence of intervals $(T_{h-1},T_h]$, where   \be\label{III-B}\ba {rl}
T_h\hskip-0.2cm&:=T_{h-1}+(cA_h^{\frac{1}{2}})^{-1}\,,\VS
A_h\hskip-0.2cm&:=\big[\dm u^n(T_{h-1})\dm_2^2+|\xi^n(T_{h-1)}|^2+ (I\cdot\omega^n(T_{h-1}))\cdot\omega^n(T_{h-1})\big]\,,\,\,{ n\ge n_{h-1}}\,,
\ea\ee
such that
\be \label{IV-B}\left.\hskip-0.57cm\ba{l}\displ
A_{h+1}\!+\!2\!\int_0^{T_{h}} \!\!\dm \n u^n(\tau)\dm_2^2 \, d\tau \le A_0\mbox{$\underset{j=0}{\overset {h+1}\sum} 2^{-j},\,\, \dm \n u^n(T_h)\dm_2\!\leq c(n, A_0,\dm  \n u_0\dm_{2})$} \VS \mbox{and }\dm|x|\n u^n(T_h)\dm_2\le c(n,A_0, \dm |x|\n u_0\dm_2),\VS \mbox{with the boundary  compatibility condition \eqref{DI} in }T_h\,,\ea\!\!\!\!\right\}  \mbox{for all }n\geq n_h.
\ee
Here, a priori the sequence    $n_h$ is monotone increasing in $h\in\N$, being each $n_h$  chosen in relation to the estimate $$\Bigg|\intll{T_{h-1}}{T_h}\int_{\partial\Omega} \mathbb J_{n}(u^n-V^n)\cdot \nu |V^n|^2\,dS\Bigg|\leq \sfrac{A_0}{2^{\null^{h+1}}}\,\quad \forall n\ge n_h\, ,$$ in accord with Lemma\,\ref{le:convergenza integrale di superficie}. In other words, although our goal  is a weak solution, we are employing a procedure of extension   for the approximating sequence $(u^n,\pi^n,\xi^n,\omega^n)$  (that is in an univocal way, thanks to \rf{IV-B}$_2$, required for proving the Uniqueness Theorem\,\ref{thm: uniqueness}) by constructing a sequence of intervals of existence for which, uniformly in $n$, the unique metric preserved on the extension is the energy metric in \rf{IV-B}$_1$.
\item[iii.] In order to make consistent our construction, we prove that  the sequence $\{T_h\}_{h\in\mathbb N_0}$ is divergent for large $h$ in such a way that $T_h \ge T_0+ c\mbox{$\overset h{\underset{j=1}\sum}$} \Big[1-\frac{1}{2^{j+1}}\Big]^{-\frac{1}{2}}$ and contextually $A_0\mbox{$\underset{j=0}{\overset h\sum}$}2^{-j}$ in \eqref{IV-B}$_1$ is convergent.
\item[iv] Thanks to the assumption  $\dm \rot u_0\dm_1<\infty\,,$ we are able to prove the following estimate (see Theorem\,\ref{thm:rotoreL1}): \be\label{V-B}
\dm \rot u^n(t)\dm_{L^1(|x|\ge R_0+t^{\frac{1}{2}})}\le c(A_0,\dm\rot u_0\dm_1)(1+t^{\frac{1}{4}})\,,  \mbox{ for all }t\in [0,T_h]\,,
\ee with $c$ independent of $n\geq n_h$. \par Then, from estimate \eqref{V-B} and the energy bound, we get the bound
\be \label{eq: bound3/2}
\dm u^n(t)\dm_{\frac{3}{2}w}\le c(A_0,\dm \rot u_0\dm_1)(1+t^{\frac{1}{4}})\,, \quad \text{for all }n\ge n_h\,.
\ee
Via the interpolation of  the $L^2$-norm between $L^\frac32_w(\OO)$ and $L^6(\OO)$, recalling \eqref{eq: bound3/2} and Sobolev's inequality, via the \eqref{IV-B}$_1$ we arrive at
\be\label{vi-B}\hskip-0.27cm\begin{array}{l}\displ\hskip-0.5cm
\norm{u^n(t)}_2^2 \!+\!\abs{\xi^n(t)}^2\!+\! (I\! \cdot \!\omega^n(t))\!\cdot\!\omega^n(t)\! + \!\!\int_0^t\! \Big[(\tilde c(A_0,\dm\rot u_0\dm_1))^{-1}\frac{\norm{u^n(\tau)}_2^6}{(1+\tau^{\frac{1}{4}})^4}\!+\!\dm \n u^n(\tau)\dm_2^2\Big] d\tau\VS\hskip8cm \le \norm{u_0}_2^2\!+\!|\xi_0|^2\!+\!|\omega_0|^2\!+\!A_0\sum_{j=0}^h 2^{-j},
\end{array}
\ee for all $t\in (0,T_h)$ and $n\geq n_h$. 
\item[v.]Via a classical argument, the last estimate evaluated for $t\in (0,T_m)$, with $T_m\ge \exp[\frac{2A_0}\chi]-1$, furnishes the existence of a $\theta_n\in (0,\exp[\frac{2A_0}\chi]-1)$ such that\footnote{\,We would like to point out that the smallness expressed in \rf{vi-B} implies the one employed in the usual {\it Th\`{e}or\'{e}me de Structure} for solutions to the Navier-Stokes equations, that is $\dm \n u(\theta)\dm_2\dm u(\theta)\dm_2< \chi$.} \be\label{vi-B}\dm u^n(\theta_n)\dm_2^6+\dm\n u^n(\theta_n)\dm_2^2\leq \chi\,,\mbox{ for all } n\geq n_m\,.\ee
Subsequently, in turn  estimate \rf{vi-B} furnishes the smallness of \be\label{vii-B}|\xi^n(\theta_n)|+|\omega^n(\theta_n)|< c\chi^{\frac{1}{2}}\,,\ee with $c$ independent of $\theta_n$ and $n\geq n_m$.\item[vi.]By virtue of the results obtained in \cite{Ga:23} and the property $u^n\in  L^2(0,T,\widehat{W}^{1,2}(\Omega,|x|))$ (see Lemma\,\ref{le: peso-rot}), the smallness expressed by estimates \rf{vi-B} and \rf{vii-B} allows us to prove, for $t>\exp[\frac{2A_0}\chi]-1=:\theta$, the global existence and uniqueness of the sequence $\{(u^n,\pi^n,\xi^n,\omega^n)\}_{n\geq n_m}$ enjoying the regularity indicated in \eqref{SOLR-I}. Hence, we state $(u^n,\pi^n,\xi^n,\omega^n)$ enjoying the following estimates:
\be\label{viii-B}\ba{c}
    \xi^n,\omega^n\in C([0,T])\,,\mbox{ for all }T>0\,,\VS\hskip-0.2cm \norm{u^n(t)}_2^2 + \abs{\xi^n(t)}^2\! +\! (I \!\cdot \omega^n (t))\!\cdot\omega^n(t)\!+ \!2\!\! \int_{0}^t\! \norm{\n u^n(\tau)}_2^2  d\tau =\dm u_{0}\dm_2^2  + |\xi_{0}|^2\!+\! (I\! \cdot \omega_{0})\!\cdot\!\omega_0\VS\hskip7cm -\int_0^t\int_{\partial\Omega} \mathbb J_n(u^n-V^n)\cdot\nu |V^n|^2\,dS\,d\tau\,,
\ea    \ee for all $n\ge n_m$ and for all $t\in[0,T]$, and, in particular for $T>\theta$ and $n\ge n_m$, \be\label{ix-B}\hskip-0.5cm\ba{clc}\xi^n(t)\,,\;\omega^n(t)\in C(\theta,T)\mbox{\; and\; }\dot\xi^n(t)\,,\;\dot\omega^n(t)\in L^2(\theta,T)\,,\VS u^n\!\!\in\! L^\infty(\theta,T;\mathcal V(\OO))\!\cap\! L^2(\theta,T;W^{2,2}(\OO)),\!\,\n\pi^n\!\!\in\! L^2(\theta,T;\!L^2(\OO)),\VS u^n_t(1+|x|)^{-1}\!\in\! L^2(\theta,T;\!L^2(\OO)),\, u^n\in C(\theta,T; W^{1,2}(\Omega_R)),\,\text{for all }R>\text{diam}(\mathcal B),\VS \lim_{t\to\theta}\xi^n(t)=\xi^n(\theta)\,,\quad\lim_{t\to\theta}\omega^n(t)=\omega^n(\theta)\,,\quad\lim_{t\to\theta}\dm u^n(t)-u^n(\theta)\dm_{2}=0\,.\ea\ee  The right-hand side of \eqref{viii-B} admits the bound $2A_0$ for $t\in (0,\theta)$, uniformly in $n\ge n_m$. { Instead, for $t\geq\theta$ the uniform bound in $n\ge n_m$ is related to the metrics detected in \eqref{ix-B} (see Lemma\,\ref{le: globale mollificato}).\footnote{We point out that the regularity required to the initial datum $u_0$ ensures the existence of a unique regular solution to problem \eqref{eq:model} defined on a maximal interval of time $(0,\theta_0)$, $\theta_0>0$. }}\par In conclusion, we claim the existence of a first extract for $n\geq n_m$ which admits a weak limit  with respect to the metric of the energy and then of a second extract from the previous one that admits a weak limit with respect to the metrics detected in \rf{ix-B}, for $t\geq \theta$. By virtue of the uniqueness of the weak limit in the metric of the energy the first weak limit and the second weak limit coincide on $t>\theta$. This sentence formally completes the proof of {our} {\it Th\'{e}or\`{e}me de Structure}. \end{itemize}
\section{Preliminaries} \label{sec: prel}
In this section, we introduce some results that are preparatory to the analytical theory for the approximating system developed in \autoref{sec: mollificatore} and for the proof of our main result proposed in \autoref{sec: proof}.
\begin{lemma}\label{le:tracce-distribuzioni}
  {\sl  Let $D$ be a smooth domain and let $E(D)$ be the completion of $C^{1}_0(\overline{D})$ in the norm $\norm{u}_2+\norm{\nabla \cdot u}_2$. Then, for all $h\in E(D)$, it holds
    \[
    \abs{\int_{\partial D} h\cdot \nu \varphi\, d\sigma}\le c\norm{\varphi}_{1,2}(\norm{h}_2+\norm{\nabla\cdot h}_2),
    \]
   for all $\varphi\in W^{1,2}(D)$.}
\end{lemma}
\bp
The proof of the lemma can be found in \cite{Ga:book,Tem}.
\ep
The distribution $\phi_\nu[h]$ defined by
\[
\phi_\nu[h](\varphi):=\int_{\partial D} h\cdot \nu \varphi\, d\sigma\,,\quad \text{for all }\varphi\in C^1_0(\overline D)\,,
\]
is meant as a generalized trace of $h$. We denote the trace space of $E(D)$ by the symbol $H^{-\frac12}(\partial D)$. \par 
\begin{lemma} \label{le: proprieta spazi lebesgue}
 {\sl   Let $1\le q <\infty$ and let $h\in L^q(\mathbb{R}^n)$. Then
    \[
    \lim_{\abs{z}\to 0} \norm{h(x+z)-h(x)}_q =0.
    \]}
\end{lemma}
\bp
For the proof, the reader can consult \cite{Brezis, Mir}
\ep
\begin{lemma} \label{le:prel-BaseGal}
{\sl
    \begin{description}
        \item[(a)] Given $f\in \mathcal{H}(\Omega_R)$ and $\overline f(x)=\overline{f}_1+\overline{f}_2\times x$ such that $\phi_\nu [f]=\phi_\nu[\overline f]$, the problem
         \begin{equation}
             \begin{cases}
                 -\nabla\cdot \mathbb{T}(u,\pi)=f \quad \forall x\in \Omega_R\\
                 \nabla\cdot u=0 \quad \forall x\in \Omega_R\\
                 u=0 \quad \text{for all }x \,\text{such that }\abs{x}=R\\
                 \overline{f}_1=-\int_{\partial\Omega}\mathbb{T}(u,\pi)\cdot \nu \,d\sigma\\
                 I\cdot \overline{f}_2=-\int_{\partial\Omega}x\times \mathbb{T}(u,\pi)\cdot \nu \,d\sigma
             \end{cases}
         \end{equation}
         has a unique solution $(u,\pi)\in (\mathcal{V}(\Omega_R)\cap W^{2,2}(\Omega_R))\times W^{1,2}(\Omega_R)$.
         Moreover, the following estimates hold:
         \begin{equation} \label{eq:stimederseconde}
             \begin{aligned}
                 & \norm{\mathbb{D}(u)}_{L^2(\Omega_R)} \le c(\mathcal{B})R(\norm{f}_{L^2(\Omega_R)}^2+ \abs{\overline{f}_1}^2 + \overline{f}_2\cdot I \cdot \overline{f}_2)^{\frac{1}{2}}, \\
                 & \norm{\overline{u}}_{W^{\frac{3}{2},2}(\partial\Omega)}\le c(\mathcal{B})\norm{\mathbb{D}(u)}_{L^2(\Omega_R)},\\
                 & \norm{D^2u}_{L^2(\Omega_R)} +\norm{\nabla\pi}_{L^2(\Omega_R)} \le c(\mathcal{B})(\norm{f}_{L^2(\Omega_R)}+\norm{\mathbb{D}(u)}_{L^2(\Omega_R)}), \\
                 & \norm{\nabla u}_{L^3(\Omega_R)} \le c(\mathcal{B})(\norm{f}_{L^2(\Omega_R)}^{\frac{1}{2}}\norm{\mathbb{D}(u)}_{L^2(\Omega_R)}^{\frac{1}{2}} +\norm{\mathbb{D}(u)}_{L^2(\Omega_R)}),
             \end{aligned}
         \end{equation}
         where $\overline{u}=\gamma(u)\in \mathcal{R}$.
         \item[(b)] The problem
         \begin{equation}
             \begin{cases}
                 -\nabla\cdot \mathbb{T}(a,\pi)=\lambda a \quad \forall x\in \Omega_R \\
                 \nabla\cdot a=0 \quad \forall x\in \Omega_R\\
                 a=0 \quad \text{for all }x \,\text{such that }\abs{x}=R\\
                 \lambda \overline{a}_1=-\int_{\partial\Omega}\mathbb{T}(u,\pi)\cdot \nu \,d\sigma\\
                 \lambda I\cdot \overline{a}_2=-\int_{\partial\Omega}x\times \mathbb{T}(u,\pi)\cdot \nu \,d\sigma
             \end{cases}
         \end{equation}
         with $\lambda\in\mathbb{R}$, $a\in \mathcal{V}(\Omega_R)$ such that $\gamma(a)=\overline{a}_1+\overline{a}_2\times x$, admits a denumerable number of positive eigenvalues $\{\lambda_i\}_{i\in \mathbb{N}}$ clustering at infinity. The corresponding eigenfunctions $\{a_i\}_{i\in\mathbb{N}}$ are in $W^{2,2}(\Omega_R)$ and the pressure fields $\pi_i$ are in $W^{1,2}(\Omega_R)$. The vector fields $\{a_i\}_{i\in\mathbb{N}}$ are an orthonormal basis of $\mathcal{H}(\Omega_R)$.
         \item[(c)] Assume that every function in $\mathcal{H}(\Omega_R)$ is extended by zero in $\Omega^R$. Let $S=\{R_k, \, k\in\mathbb{N}\}$ be any increasing and unbounded sequence of positive numbers with $R_1>\text{diam}(\mathcal{B})$. Then $\displ{\underset{R_k\in S}\cup \mathcal H(\Omega_{R_k})}$ is dense in $\mathcal{H}(\Omega)$ and, denoting by $\{a_{i,k}\}_{i\in\mathbb N}$ the orthonormal basis of $\mathcal H(\Omega_{R_k})$ ensured by item\,{\bf (b)}, $\displ{\underset{R_k\in S}\cup \{a_{i,k}\}_{i\in\mathbb{N}}}$ is a basis of $\mathcal{H}(\Omega)$.
    \end{description}}
\end{lemma}
\bp
The reader can check the proof of items {\bf(a)}-{\bf(c)} in \cite{GS}.
\ep
{We denote by $\mathcal K_0(\R^3)$ the subset of divergence-free functions in $\mathcal K(\R^3)$.  
We define the space $H(\R^3)$ as the completion of $\mathcal K_0(\R^3)$ with respect to the metric induced by the scalar product \eqref{SPL2H1}$_1$.
We set
\be \label{DEF-G}
\ba{l}
G(\R^3)=\{u\in L(\R^3)\,: \, \text{there exists }p\in W^{1,2}_{loc}(\overline\Omega) \,\,\text{such that }u=\n p \,\,\text{in }\Omega \,\,\text{and  }\VS \hskip3cm \overline u_1=\int_{\partial\Omega}p\nu\,dS\,, \,\overline u_2=\bigg(\int_{\partial\Omega}px\times \nu\,dS\bigg)\cdot I^{-1}\,\}\,.
\ea
\ee
This space was introduced by Silvestre in \cite{Silv: tesi} in order to state a generalization of the well-known Helmholtz decomposition for the space $L(\R^3)$. In fact, the following result holds.
\begin{lemma} \label{le:HD}
    $L(\R^3)=H(\R^3)\oplus G(\R^3)$
\end{lemma}
\bp
We prove that $H(\R^3)^\perp=G(\R^3)$. We start by proving that, given $u\in H(\R^3)^\perp$, it belongs to $G(\R^3)$. For all $w\in H(\R^3)$ we have
\be \label{SP-HD}
0=(u,w)_1=\int_\Omega u(x)\cdot w(x)\,dx+\overline u_1\cdot \overline w_1+( I\cdot \overline u_2)\cdot\overline w_2
\ee
In particular, being $\mathscr C_0(\Omega)\subset \mathcal K_0(\R^3)$, we have, for all $w\in \mathscr C_0(\Omega)$,
\[
\int_\Omega u(x)\cdot w(x)\,dx=0\,.
\]
Hence, there exists $p\in W^{1,2}_{loc}(\overline\Omega)$ with $\n p\in L^2(\Omega)$ such that
\[
u=\n p\,\,\text{in }\Omega\,.
\]
Therefore, via \eqref{SP-HD}, we have
\be \label{PS-K0}
\int_\Omega \n p(x)\cdot w(x)\,dx+\overline u_1\cdot \overline w_1+(I\cdot \overline u_2)\cdot\overline w_2=0\,,
\ee
for all $w\in \mathcal K_0(\R^3)$.
Integrating by parts, we have
\be \label{PS-G}
\int_\Omega \n p(x)\cdot w(x)\,dx=\int_{\partial\Omega} p\nu\,dS\cdot \overline w_1+\int_{\partial\Omega} px\times\nu \,dS\cdot\overline w_2\,.
\ee
Replacing the first integral in \eqref{PS-K0} by virtue of \eqref{PS-G}, in view of the arbitrariness of $w\in \mathcal K_0(\R^3)$ and of the definition of the space $G(\R^3)$ given in \eqref{DEF-G}, we deduce that $u\in G(\R^3)$.\par
We now prove that given $u\in G(\R^3)$ we have $(u,w)_1=0$ for all $w\in H(\R^3)$. We prove this for all $w\in\mathcal K_0(\R^3)$ and get the claim using a density argument. Since $u\in G(\R^3)$, for all $w\in \mathcal K_0(\R^3)$ we have
\[
(u,w)_1=\int_\Omega \n p(x)\cdot w(x)\,dx-\int_{\partial\Omega} p\nu\,dS\cdot \overline w_1-\int_{\partial\Omega} px\times\nu \,dS\cdot I^{-1}\cdot I\cdot\overline w_2\,.
\]
Since $w\in \mathcal K_0(\R^3)$, it is divergence-free and \eqref{PS-G} holds. Replacing the first integral in the above identity in view of \eqref{PS-G}, we conclude the proof.
\ep
We can also deduce the following characterization of $H(\R^3)$.
\begin{lemma}{\sl
    $H(\R^3)=\{u\in L(\R^3)\,:\, \n\cdot u=0\,\,\text{in the weak sense, }\phi_\nu [u]=\phi_\nu [\overline u]\}.$}
\end{lemma}
\bp
Since $G(\R^3)=H(\R^3)^\perp$ and given $\psi \in C_0^\infty(\Omega)$ we have $\n\psi\in G(\R^3)$, we get, for all $u\in H(\R^3)$,
\[
\int_\Omega u\cdot\n\psi\,dx=0\,,
\]
that implies the weak divergence-free condition. We prove the generalized trace identity. Let $\psi \in H^\frac12(\partial\Omega)$ and $\varphi\in W^{1,2}(\Omega)$ such that $\gamma (\varphi)=\psi$. We set
\[
w=\n\varphi\,, \,\,\overline w_1=-\int_{\partial\Omega}\psi\nu \,dS\,,\,\,\overline w_2=-I^{-1}\cdot \int_{\partial\Omega}\psi x\times\nu\,dS\,.
\]
Clearly, $w\in G(\R^3)$ and therefore
\[
0=(u,w)_1=\int_\Omega u(x)\cdot \n\varphi(x)\,dx+\overline u_1\cdot \overline w_1+(I\cdot \overline u_2)\cdot\overline w_2=\int_{\partial\Omega}u\cdot \nu \psi\,dS-\int_{\partial\Omega}\overline u\cdot \nu \psi\,dS\,.
\]
The lemma is completely proved.
\ep
Let $R>\text{diam}(\mathcal B)$. We consider the spaces 
\[
L(B_R)=\{ u\in L^2(B_R)\,:\, u=\overline u\in\mathcal R \,\text{ in }\mathcal B\}
\]
and
\[
W(B_R)=\{u\in W^{1,2}(B_R)\,:\,u=\overline u\in\mathcal R \,\text{ in }\mathcal B\}\,.
\]
They are endowed with the scalar products
\be \label{SPL2H1-R}
\ba{rl}
(v,w)_{1,R} &\hskip-0.2cm\displ:= \int_{\Omega_R} v(x)\cdot w(x)\,dx +\overline v_1\cdot \overline w_1+  (I\cdot\overline v_2)\cdot\overline w_2\,,\VS
(v,w)_{2,R} &\hskip-0.2cm\displ:=\int_{\Omega_R} v(x)\cdot w(x)\,dx+2\int_{\Omega_R} \mathbb D(v)\cdot\mathbb D(w)\,dx  +\overline v_1\cdot \overline w_1+ (I\cdot\overline v_2)\cdot\overline w_2\,,
\ea
\ee
respectively.
We point out that the two spaces, endowed with the respective norms, are separable Hilbert spaces. \par
We state the following result, it is a trivial extension of the well-known Rellich-Kondrachov Theorem \cite{Brezis}.
\begin{lemma}{\sl
    $W(B_R)$ is compactly embedded into $L(B_R)$.}
\end{lemma}
\bp
The lemma is proved analogously to the Rellich-Kondrachov Theorem, slightly modified due to the presence of the rigid motion in $\mathcal B$ that can be handled easily. For the proof of the Rellich-Kondrachov Theorem we refer the reader to \cite{Brezis}.
\ep
As a consequence, we obtain the following lemma, it is an expected generalization of the Friedrichs Lemma. 
\begin{lemma} \label{FG}{\sl
    For all $\vep>0$ there exists $N\in\N$ such that
    \[
    \dm u\dm_{L(B_R)}\le (1+\vep)\big[\sum_{k=1}^N(u,a_k)_{L(B_R)}]^\frac12 +\vep \dm \n u\dm_{L^2(B_R)}\,,\quad \forall u\in W(B_R)\,,
    \]
    where $\{a_k\}$ is an orthonormal basis of $L(B_R)$.}
\end{lemma}
Following the arguments in \cite[p.176]{LADY}, suitably modified due to the presence of the rigid motion, we get the following corollary. We do not give the proof, actually the presence of the rigid motion does not add any issue to the extension of the arguments in \cite{LADY}.
\begin{coro} \label{GEN-CF}{\sl
    Assume that $\{v^k(t,x)\}$ is a sequence such that, for all $k\in\N$,
    \[
    \text{ess}\sup_{(0,T)}\dm v^k(t)\dm_{L(B_R)}^2+\int_0^T\big[\dm v^k(t)\dm_{L(B_R)}^2+\dm \n v^k(t)\dm_{L^2(B_R)}^2]\,dt \le M<\infty\,.
    \]
    Assume also that, for all $\varphi\in L(B_R)$ $$(v^k(t),\varphi)_1\to (v(t),\varphi)_1\,,\,\,\text{for almost all }t\in (0,T)\,. $$ 
    Then, $\{v^k\}$ converges to $v$ strongly in $L^2(0,T;L(B_R))$. Furthermore, if $\{v^k\}\subset L^2(0,T;W^{2,2}(\Omega_R))$ and 
    \[
    \int_0^T \dm v^k(t)\dm_{2,2}^2\,dt \le C,
    \]
    with $C>0$ independent of $v_k$, then 
    \[
    v^k\to v \,\,\text{strongly in }L^2(0,T;W^{1,2}(\Omega_R))\,.
    \]}
\end{coro}}
\begin{lemma} \label{le:prel-symgrad}
 {\sl    Let $u\in\mathcal{V}(\Omega)$, with $u|_{\partial\Omega}=\overline u\in\mathcal R$. Then,
    \begin{equation} \label{eq:gradsimmetrico}
        \norm{\n u}_2=\sqrt{2}\norm{\mathbb{D}(u)}_2.
    \end{equation}
    Moreover, 
    \be \label{eq: moto rigido-gradiente}
    |\overline u_1|+|\overline u_2|\le c(\text{diam}(\mathcal{B}))\dm \n u\dm_2\,.
    \ee}
\end{lemma}
\bp
For the proof, we refer to \cite{Ga:Hand}
\ep
\begin{lemma} \label{le: Stokes Exterior Domains}
 {\sl  Let $f\in L^2(\Omega)$, $u\in  W^{2,2}(\Omega)\cap \mathcal V(\Omega)$ and $\n \pi\in L^2(\Omega)$ such that  
    \be\label{SED}
    \begin{cases}
    \ba{rl}
        \Delta u-\n\pi&\hskip-0.2cm=f\,,\quad \text{in }\Omega\,,\VS
        \n\cdot u&\hskip-0.2cm=0\,, \quad \text{in }\Omega\,,\VS
        u&\hskip-0.2cm=V\,,\quad \text{in }\partial\Omega\,.
        \ea
    \end{cases}
    \ee
    Then, 
    \be \label{eq: Stokes-stima}
    \dm D^2u\dm_2+\dm \n \pi\dm_2\le c(\dm f\dm_2+\dm V\dm_{H^{\frac{3}{2}}(\partial\Omega)}+\dm \n u\dm_{L^2(\OO_\eta)}+\dm u\dm_{L^2(\Omega_\eta)})\,,
    \ee
    for some positive constant $c$, independent of $u$, and a suitable neighborhood $\Omega_\eta$ of $\partial\Omega$.}
\end{lemma}
\bp{Actually, the result is  contained { in }\cite{Ga:book}, Chapter V. However, we sketch the argument of the proof. In Lemma\,{V.4.3} of \cite{Ga:book} is proved the following estimate:
\be\label{GPG-I} \dm D^2 u\dm_2+\dm\n \pi\dm_2\leq c\big[\dm f\dm_2+\dm V\dm_{\frac32,2}+\dm u\dm_{L^2(\OO_\eta)}+\dm \pi\dm_{L^2(\OO_\eta)}\big].\ee  Now, our goal is to estimate the $\dm \pi\dm_{L^2(\OO_\eta)}$. Without losing the generality of the estimate, we assume that the pressure field $\pi$ is defined with zero mean on $\OO_\eta$, that is $ \pi=\ov\pi-\sfrac1{|\OO_\eta|}\mbox{\Large$\intl{\Omega_\eta}$}\ov\pi dx$. Let us consider a solution of the Bogovski problem: 
$$\n\cdot B(x)= \ov h(x)\,,\mbox{ in }\OO_\eta\,,\mbox { with }B=0\mbox{ on }\po_\eta\,,$$ where,{ in order to satisfy the compatibility  condition of the problem,}  by  the symbol $\ov h$ we mean $$h-\sfrac1{|\OO_\eta|}\intl{\Omega_\eta}h(x)dx\,,$$ with $h(x)\in C_0(\OO_\eta)$. It is well known (see \cite{Ga:book}) that there exists a solution such that, for some constant  independent of $h$, \be\label{GPG-II}\dm\n B\dm_{L^{2}(\OO_\eta)}\leq c\dm h\dm_2\,.\ee Multiplying by $B$ equation  \rf{SED}$_1$, and integrating on $\OO_\eta$, we get  
$$ (\pi, h)=(\n u,\n B)-(f,B)\,.$$
Applying Holder's inequality, Poincar\'{e} inequality, and employing \rf{GPG-II}, we arrive at
$$|(\pi,h)|\leq c(\OO_\eta)\dm h\dm_2\big[\dm\n u\dm_{L^2(\OO_\eta)}+\dm f\dm_2\big]\,.$$ Since $h$ is arbitrary we have proved the estimate related to $\pi$. Considering again \rf{GPG-I} and increasing the right-hand side, we conclude the proof.}\ep
{\begin{coro}\label{COR-I}{\sl Under the same assumption of Lemma\,\ref{le: Stokes Exterior Domains} we get  \be \label{eq: stima gradiente pressione}
\dm D^2u\dm_2+\dm \n \pi\dm_2\le C(\dm f\dm_2+\dm \n u\dm_2)\,.\ee 
}\end{coro} \bp Employing H\"{o}lder inequality for the $L^2$-norm of $u$ on the right hand side of \rf{eq: Stokes-stima} and the Sobolev inequality, estimate \rf{eq: moto rigido-gradiente} of Lemma\,\ref{le:prel-symgrad} for the rigid motion $V$,   we arrive at \rf{eq: stima gradiente pressione}.\ep} Since we are interested in  fields $u\in\mathcal V(\OO)$  that become rigid motions $V$ on the boundary $\po$, we assume that the fields $u-V$ is extended by zero inside  $\mathcal B$. \par We recall that the notation $J_\frac1n$ is adopted for a Friedrich mollifier in the space variable.
\begin{lemma} \label{le:Stime per rel energia}
{\sl    Let $u\in \mathcal V(\Omega)$ such that $u|_{\mathcal B}=V$, with $V(x)=\xi+\omega\times x$, $\xi,\omega\in\mathbb{R}^3$. Let $\varphi_\rho$ be a smooth cut-off function such that $\varphi_\rho(x)=1$ if $|x|\le \rho$ and $\varphi_\rho(x)=0$ if $|x|\ge 2\rho$, with $\rho>\text{diam}(\mathcal{B})$. Then
    \begin{description} 
        \item[i)] 
        there exists $C=C(\mathcal{B})$ such that
        $$\Bigg|\int_{\partial\Omega} J_{\frac{1}{n}}(u-V)\cdot \nu |V|^2\,dS\Bigg|\le C(\mathcal{B})(\dm u\dm_{L^2(\Omega_\eta)}+\dm V\dm_{L^2(\Omega_\eta)})(|\xi|^2+|\omega|^2)\,,$$
        where $\Omega_\eta$ is a smooth neighborhood of $\partial\Omega$;
        \item[ii)] there exists $c>0$ such that $$
       \Bigg| \int_\Omega \n \cdot J_{\frac{1}{n}}(u-\varphi_\rho V)|u|^2\,dx\Bigg|\le c|\xi|\dm u\dm_2^2\,.
        $$
    \end{description}}
\end{lemma}
\bp
We prove the first item. We have
\[
\Bigg|\int_{\partial\Omega} J_{\frac{1}{n}}(u-V)\cdot \nu |V|^2\,dS\Bigg|\le \dm J_{\frac{1}{n}}(u-V)\cdot\nu \dm_{H^{-\frac{1}{2}}(\partial\Omega)}\dm |V|^2\dm_{H^{\frac{1}{2}}(\partial\Omega)}\,.
\]
Lemma\,\ref{le:tracce-distribuzioni} and the divergence-free condition allow us to say that
\[
\dm J_{\frac{1}{n}}(u-V)\cdot\nu \dm_{H^{-\frac{1}{2}}(\partial\Omega)}\le c\dm u-V\dm_{L^2(\Omega_\eta)}\le c(\dm u\dm_{L^2(\Omega_\eta)}+\dm V\dm_{L^2(\Omega_\eta)})\,,
\]
with $\Omega_\eta$ smooth neighborhood of $\partial\Omega$.\par
On the other hand, it is easy to check that
\[
\dm |V|^2\dm_{H^{\frac{1}{2}}(\partial\Omega)}\le C(\text{diam}(\mathcal{B}))(|\xi|^2+|\omega|^2)\,.
\]
Hence, the proof of item {\bf i)} is complete. \par We now prove {\bf ii)}.  Integrating by parts and using the divergence-free condition, we get
\[
\ba{l} \displ
\n\cdot J_{\frac{1}{n}}(u-\varphi_\rho V)=\partial_{x_i}\int_{\mathbb{R}^3}J_{\frac{1}{n}}(x-y)(u-\varphi_\rho V)^i(y)\,dy=\int_{\mathbb{R}^3}\partial_{x_i}J_{\frac{1}{n}}(x-y)(u-\varphi_\rho V)^i(y)\,dy\VS\hskip 3cm =-\int_{\mathbb{R}^3}\partial_{y_i}J_{\frac{1}{n}}(x-y)(u-\varphi_\rho V)^i(y)\,dy=\int_{\mathbb{R}^3}J_{\frac{1}{n}}(x-y)\partial_{y_i}(u-\varphi_\rho V)^i(y)\,dy\VS \hskip 10cm=\int_{\mathbb{R}^3}J_{\frac{1}{n}}(x-y)(V\cdot \n \varphi_\rho)(y) \,dy\,.
\ea
\]
Considering that $|\n \varphi_\rho(x)|\le c|\rho|^{-1}$, for all $x\in\mathbb{R}^3$, by an elementary property of the scalar triple product, we find that
\[
|\n \cdot J_{\frac{1}{n}}(u-V)|=\Bigg|\int_{\mathbb{R}^3}J_{\frac{1}{n}}(x-y)(V\cdot \n \varphi_\rho)(y) \,dy\Bigg|=\Bigg|\int_{\mathbb{R}^3}J_{\frac{1}{n}}(x-y)(\xi\cdot \n \varphi_\rho)(y) \,dy\Bigg|\le c|\xi|\,.
\]
Hence, we can conclude the proof of {\bf ii)}.
\ep
We recall that given $g\in \widehat W^{1,2}(\Omega)$, there exists $g_0\in\mathbb R$ such that
\be \label{Hardy}
\dm |x|^{-1}(g-g_0)\dm_2 \le c\dm \n g\dm_2\,,
\ee
with $c>0$ independent of $g$. See, \textit{e. g.}, \cite{Ga:book}.\par{ Given a rigid motion $V(x)=\xi+\omega\times x$, $\xi,\omega\in\mathbb R^3$, we set
\[
V_2(x):=\omega\times x\,.
\]}
\begin{lemma} \label{le: I1}
{\sl Let $(u,\pi)\in (\mathcal V(\Omega)\cap W^{2,2}(\Omega))\times\widehat W^{1,2}(\Omega)$, with $u|_{\mathcal B}=V$ and $V(x)=\xi+\omega\times x$, $\xi,\omega\in\mathbb R^3$. Then  \[
\Bigg|\int_{\partial\Omega} \big(J_\frac1n (V_2)\cdot \n u-\omega\times V\big)\cdot\mathbb T(u,\pi)\cdot\nu \,dx\Bigg|\!\le \!\vep \dm \n\cdot\mathbb T(u,\pi)\dm_2^2+c(|\omega|+|\omega|^\frac43+|\omega|^2+|\omega|^4)\dm \n u\dm_2^2\,,
\]
for all $\vep>0$ and some $ c>0$ independent of $u$ and $\pi$.}
\end{lemma}
\bp 
We set 
\[
\ba {rl}
I&\hskip-0.2cm:=\displ \int_{\partial\Omega} \big(J_\frac1n (V_2)\cdot \n u-\omega\times V\big)\cdot\mathbb T(u,\pi)\cdot\nu \,dx \VS 
I_{1}&\hskip-0.2cm:=\displ\int_{\partial\Omega} (J_\frac1n(V_2)\cdot \n u- \omega\times V)\cdot \mathbb D(u)\cdot \nu \,dS\,,\VS
I_{2}&\hskip-0.2cm:=\displ \int_{\partial\Omega} (J_\frac1n(V_2)\cdot \n u-\omega\times V)\cdot ( \pi-\pi_0)\mathbb I\cdot \nu \,dS\,.
\ea
\]
We have that
\[
|I|\le 2|I_{1}|+|I_{2}|\,.
\]
Then, we see that
\be \label{eq: estimates surface integral}
\ba{rl}
|I_{1}|&\hskip-0.2cm\le c(\text{diam}(\mathcal B))[|\omega|\dm \n u\dm_{L^2(\partial\Omega)}^2+(|\omega||\xi|+|\omega|^2)\dm \n u\dm_{L^2(\partial\Omega)}]\,,\VS
|I_{2}|&\hskip-0.2cm\le c(\text{diam}(\mathcal B))[|\omega|\dm \n u\dm_{L^2(\partial\Omega)}\dm (\pi-\pi_0)\dm_{L^2(\partial\Omega)}+(|\omega||\xi|+|\omega|^2)\dm (\pi^n-\pi^n_0)\dm_{L^2(\partial\Omega)}]\,.
\ea
\ee
By Gagliardo trace inequalities, we have
\be \label{eq:traccia-pressione}
\ba {l}
\dm \pi-\pi_0\dm_{L^2(\partial\Omega)}\le c(\dm \pi-\pi_0\dm_{L^2(\Omega_\eta)}^{\frac{1}{2}}\dm \n\pi\dm_{L^2(\Omega_\eta)}^{\frac{1}{2}}+\dm \pi-\pi_0\dm_{L^2(\Omega_\eta)})\,.
\ea
\ee
Since $\Delta u-\n \pi=\n \cdot\mathbb T$, by virtue of estimate \eqref{eq: stima gradiente pressione}, we get
\be\label{eq:traccia-gradiente}
\ba {l}
\dm \n u\dm_{L^2(\partial\Omega)}\le c(\dm \n u\dm_{L^2(\Omega_\eta)}^{\frac{1}{2}}\dm D^2 u\dm_{L^2(\Omega_\eta)}^{\frac{1}{2}}+\dm \n u\dm_{L^2(\Omega_\eta)})\VS\hskip5.5cm\le c(\dm \n u\dm_{L^2(\Omega_\eta)}^{\frac{1}{2}}\dm \n\cdot\mathbb T(u,\pi)\dm_{L^2(\Omega_\eta)}^{\frac{1}{2}}+\dm \n u\dm_{L^2(\Omega_\eta)})\,.
\ea
\ee
Employing \eqref{Hardy}, we find 
\[
\dm \pi-\pi_0\dm_{L^2(\Omega_\eta)}\le c(\eta)\dm \frac{\pi-\pi_0}{|x|}\dm_{L^2(\Omega_\eta)}\le c(\eta)\dm\n\pi\dm_2\,.
\]
Hence, employing again \eqref{eq: stima gradiente pressione}, we arrive at
\be \label{eq: stima pressione post Hardy}
\dm \pi-\pi_0\dm_{L^2(\partial\Omega)}\le c \dm \n \pi\dm_2\le c(\dm \n u\dm_2+\dm \n\cdot\mathbb T(u,\pi)\dm_2)\,.
\ee
Going back to \eqref{eq: estimates surface integral}, employing Young's inequality and recalling Lemma\,\ref{le:prel-symgrad}, we finally find\footnote{We see that
\[
\ba{c}
|\omega|\dm \n u\dm_{L^2(\partial\Omega)}^2\le c |\omega|(\dm \n u\dm_2\dm \n\cdot\mathbb T\dm_2+\dm \n u\dm_2^2)\le \vep \dm \n\cdot\mathbb T\dm_2^2+c(|\omega|+|\omega|^2)\dm \n u\dm_2^2\,,\VS
(|\omega||\xi|+|\omega|^2)\dm \n u \dm_{L^2(\partial\Omega)}\le 
c(|\omega||\xi|+|\omega|^2)(\dm \n u \dm_2^\frac12\dm \n\cdot\mathbb T\dm_2^\frac12+\dm\n u\dm_2)\le\vep \dm \n\cdot\mathbb T\dm_2^2+c(|\omega|+|\omega|^\frac43)\dm \n u\dm_2^2\,,\VS
|\omega|\dm \n u\dm_{L^2(\partial\Omega)}\dm (\pi-\pi_0)\dm_{L^2(\partial\Omega)}\le c|\omega|(\dm \n u \dm_2^\frac12\dm \n\cdot\mathbb T\dm_2^\frac12+\dm\n u\dm_2)(\dm \n u\dm_2+\dm \n\cdot\mathbb T\dm_2)\VS\hskip5cm\le \vep\dm \n \cdot\mathbb T\dm_2^2+c(|\omega|^\frac43+|\omega|^4+|\omega|+|\omega|^2)\dm\n u\dm_2^2\,,\VS
(|\omega||\xi|+|\omega|^2)\dm (\pi^n-\pi^n_0)\dm_{L^2(\partial\Omega)}\le (|\omega||\xi|+|\omega|^2)(\dm\n\cdot\mathbb T\dm_2+\dm \n u\dm_2)\le \vep\dm\n\cdot\mathbb T\dm_2^2+c(|\omega|^2+|\omega|)\dm\n u\dm_2^2\,.
\ea
\]}
\[
|I|\le \vep\dm \n\cdot \mathbb T\dm_2^2+c(|\omega|+|\omega|^\frac43+|\omega|^2+|\omega|^4)\dm \n u\dm_2^2\,.
\]
\ep
\begin{lemma} \label{le:J1}
 {\sl   Let $(u,\pi)\in (\mathcal V(\Omega)\cap W^{2,2}(\Omega))\times\widehat W^{1,2}(\Omega)$, with $u|_{\mathcal B}=V$ and $V(x)=\xi+\omega\times x$, $\xi,\omega\in\mathbb R^3$. Let $\varphi_\rho$ be a smooth cut-off function such that $\varphi_\rho(x)=1$ if $|x|\le \rho$ and $\varphi_\rho(x)=0$ if $|x|\ge 2\rho$. Assume that $\text{dist}(x,B_{2\rho})>1$ for all $x\in\partial\Omega$. Then
    \[
    \ba{l}\displ
    \Bigg|\int_\Omega\n \big[ J_\frac1n(\varphi_\rho V_2)\cdot \n u- J_\frac1n(\varphi_\rho)\omega\times u\big]\cdot\n u\,dx \Bigg|\VS\hskip2cm\le \vep\dm\n\cdot\mathbb T(u,\pi)\dm_2^2+c|\omega|\dm u\dm_{W^{1,2}(\Omega^\rho)}^2+c(|\omega|+|\omega|^2)\dm \n u\dm_2^2\,,
    \ea
    \] for all $\vep>0$ and some $c>0$ independent of $u$ and $\pi$.}
\end{lemma}
\bp
We set
\[
J:=\int_\Omega\n \big( J_\frac1n(\varphi_\rho V_2)\cdot \n u- J_\frac1n(\varphi_\rho)\omega\times u\big)\cdot\n u\,dx 
\]
{Firstly, let us point out that the following identity holds:\footnote{By the symbol $<\cdot,\cdot>$ we refer to the usual matrix product.\par We spend some words on the identity \eqref{TPROD}. We have
\[
\n \big[ J_\frac1n(\varphi_\rho V_2)\cdot \n u- J_\frac1n(\varphi_\rho)\omega\times u\big]_{i,j}=\partial_{x_j}\big[\partial_{x_h}u_i(J_\frac1n(\varphi_\rho V_2))_h-J_\frac1n(\varphi_\rho)(\omega\times u)_i\big]\,.
\]
By employing elementary derivation properties, we arrive at \eqref{TPROD}.}
\be \label{TPROD}
\ba{l}
\n \big[ J_\frac1n(\varphi_\rho V_2)\cdot \n u- J_\frac1n(\varphi_\rho)\omega\times u\big]= <\n u,J_\frac1n(V_2\otimes \n\varphi_\rho)>+<\n u,J_\frac1n(\varphi_\rho \n V_2)>\VS +J_\frac1n(\varphi_\rho V_2)\cdot\n\n u- J_\frac1n(\varphi_\rho)\n (\omega\times u)- (\omega\times u)\otimes J_\frac1n(\n \varphi_\rho)\,.
\ea
\ee
We also remark that
\be \label {PM}
<\n u,J_\frac1n(\varphi_\rho \n V_2)>\cdot \n u=\sum_{i=1}^3\omega\times \n u_i\cdot \n u_i=0\,.\footnote{We denote by $u_i$ the $i-$th component of the vector field $u$.}
\ee
Hence, we have
\[
\ba{l}\displ
J=\int_{\Omega} <\n u,J_\frac1n(V_2\otimes \n\varphi_\rho)>\cdot\n u\,dx+\int_{\Omega}J_\frac1n(\varphi_\rho V_2)\cdot\n\n u\cdot\n u\,dx\VS\hskip4cm-\int_\Omega J_\frac1n(\varphi_\rho)\n(\omega\times u)\cdot\n u\,dx-\int_{\Omega} (\omega\times u)\otimes J_\frac1n(\n \varphi_\rho)\cdot \n u\,dx\,.
\ea
\]}
By virtue of H\"{o}lder's inequality, we get
\[
\Bigg|\int_{\Omega}<\n u,J_\frac1n(V_2\otimes \n\varphi_\rho)>\cdot\n u\,dx\Bigg|\le c|\omega|\dm \n u\dm_{L^2(\Omega^\rho)}^2\,,
\]
\[ 
\Bigg| \int_\Omega J_\frac1n(\varphi_\rho)\n(\omega\times u)\cdot\n u\,dx\Bigg|\le c|\omega|\dm \n u\dm_2^2
\]
and
\[
\Bigg|\int_{\Omega} (\omega\times u)\otimes J_\frac1n(\n \varphi_\rho)\cdot \n u\,dx\Bigg|\le c|\omega|\dm u\dm_{L^2(\Omega^\rho)}\dm \n u\dm_{L^2(\Omega^\rho)}\,.
\]
Moreover, we see that
\[
\int_{\Omega} J_\frac1n(\varphi_\rho V_2)\cdot\n\n u\cdot\n u\,dx=\frac12\int_{\Omega} J_\frac1n(\varphi_\rho V_2)\cdot\n|\n u|^2\,dx\,.
\]
Remarking that, by the elementary properties of the scalar triple product, $\n\cdot (\varphi_\rho\omega\times x)=0$, integration by parts yields to
\[
\int_{\Omega}J_\frac1n(\varphi_\rho V_2)\cdot\n|\n u|^2\,dx=\int_{\partial\Omega} J_\frac1n(V_2)\cdot\nu|\n u|^2\,dS\,.
\]
{By Gagliardo trace inequalities and estimate \eqref{eq: stima gradiente pressione} with $f=\n\cdot\mathbb T(u,\pi)$, we find}
\[
\Bigg|\int_{\partial\Omega} J_\frac1n(V_2)\cdot\nu|\n u|^2\,dS\Bigg|\le c|\omega|\dm \n u\dm_{L^2(\partial\Omega)}^2\le \vep\dm\n\cdot\mathbb T(u,\pi)\dm_2^2+c(|\omega|+|\omega|^2)\dm \n u\dm_2^2\,.
\]
We conclude that
\[
|J|\le \vep\dm\n\cdot\mathbb T(u,\pi)\dm_2^2+c|\omega|\dm u\dm_{W^{1,2}(\Omega^\rho)}^2+c(|\omega|+|\omega|^2)\dm \n u\dm_2^2\,.
\]
\ep
\begin{lemma}\label{le:J3}
    {\sl Under the same assumptions of Lemma\,\ref{le:J1}, we get
    \[
    \ba{l}\displ
    \Bigg| \int_\Omega \n \big[ J_\frac1n(\varphi_\rho V_2)\cdot \n u-J_\frac1n(\varphi_\rho)\omega\times u\big]\cdot (\pi-\pi_0)\mathbb I\,dx \Bigg|\VS\hskip3cm\le \vep\dm\n\cdot\mathbb T(u,\pi)\dm_2^2+c|\omega|^2\dm u\dm_{L^2(\Omega^\rho)}^2+c|\omega|\dm u\dm_{L^2(\Omega^\rho)}\dm \n u\dm_2\,,
    \ea
    \]
    for all $\vep>0$ and some $c>0$ independent of $u$ and $\pi$.}
\end{lemma}
\bp
We set 
\[
H:=\int_\Omega \n \big[ J_\frac1n(\varphi_\rho V_2)\cdot \n u-J_\frac1n(\varphi_\rho)\omega\times u\big]\cdot (\pi-\pi_0)\mathbb I\,dx\,.
\]
Recalling the validity of the identity \eqref{TPROD} and remarking that
\[ 
\ba{rl}
\big[<\n u,J_\frac1n(\varphi_\rho \n V_2)>-J_\frac1n(\varphi_\rho)\n(\omega\times u)\big]\cdot \mathbb I&\hskip-0.2cm=0\,,\VS
J_\frac1n(\varphi_\rho \n V_2)\cdot\n \n u\cdot (\pi-\pi_0)\mathbb I&\hskip-0.2cm=0\,,
\ea
\] we have
\[
H=\int_{\Omega} [<\n u, J_\frac1n( V_2\otimes \n\varphi_\rho)>-(\omega\times u)\otimes J_\frac1n(\n\varphi_\rho)]\cdot (\pi-\pi_0)\mathbb I\,dx\,.
\]
Integrating by parts, we have
\[
\ba{l}\displ
\int_{\Omega} <\n u, J_\frac1n( V_2\otimes \n\varphi_\rho)>\cdot(\pi-\pi_0)\mathbb I\,dx=
-\int_{\Omega} J_\frac1n(\Delta\varphi_\rho V_2)\cdot u(\pi-\pi_0)\,dx\VS \hskip 8cm-\int_{\Omega} J_\frac1n( V_2\otimes \n\varphi_\rho) \cdot u\cdot \n\pi\,dx\,.
\ea
\]
Using \eqref{Hardy} and \eqref{eq: stima gradiente pressione}, we arrive at
\[
\ba {l}\displ
\Bigg|\int_{\Omega} J_\frac1n(\Delta\varphi_\rho V_2)\cdot u(\pi-\pi_0)\,dx\Bigg|\le c|\omega|\dm \frac{\pi-\pi_0}{|x|}\dm_{L^2(\Omega^\rho)}\dm u\dm_{L^2(\Omega^\rho)}\le c|\omega|\dm \n \pi\dm_2\dm u\dm_{L^2(\Omega^\rho)}\VS\hskip9cm\le c|\omega|(\dm \n\cdot\mathbb T\dm_2+\dm \n u\dm_2)\dm u\dm_{L^2(\Omega^\rho)}\,,
\ea
\]
while
\[
\ba{l}\displ
\Bigg|\int_{\Omega} J_\frac1n( V_2\otimes \n\varphi_\rho) \cdot u\cdot \n\pi\,dx\Bigg|\le c|\omega|\dm u\dm_{L^2(\Omega^\rho)}\dm \n \pi\dm_2\VS\hskip8cm\le c|\omega|\dm u\dm_{L^2(\Omega^\rho)}(\dm \n u\dm_2+\dm \n\cdot \mathbb T(u,\pi)\dm_2)\,.
\ea
\]
Finally, using again \eqref{Hardy} and \eqref{eq: stima gradiente pressione}, we have
\[
\ba{l} \displ
\Bigg|\int_{\Omega}(\omega\times u)\otimes J_\frac1n(\n\varphi_\rho)\cdot (\pi-\pi_0)\mathbb I\,dx\Bigg|\le c|\omega|\dm u\dm_{L^2(\Omega^\rho)}\dm \frac{\pi-\pi_0}{|x|}\dm_2\VS\hskip6cm\le c|\omega|\dm u\dm_{L^2(\Omega^\rho)}(\dm \n u\dm_2+\dm \n\cdot \mathbb T(u,\pi)\dm_2)\,.
\ea
\]
Collecting all the estimates and employing Young'sinequality, we conclude that
\[
|H|\le \vep\dm \n \cdot\mathbb T(u,\pi)\dm_2^2+c|\omega|^2\dm u\dm_{L^2(\Omega^\rho)}^2+c|\omega|\dm u\dm_{L^2(\Omega^\rho)}\dm \n u\dm_2\,.
\]
\ep
\begin{lemma} \label{le: Rotore termine convettivo-u}
 {\sl   Let $u\in W^{2,2}(\Omega)$. Then
    \[
          \rot ( J_{\frac1n}(u)\cdot \n u)=J_{\frac1n}(u)\cdot \n \rot u+v   \,,
    \]
    where the components of the vector field $v$ are given by
    \[
    v^i= \vep_{ijk}<(\n u)_k,(\n J_{\frac1n}(u))^j>\footnote{By the symbols $(\n u)_k$ and $(\n J_{\frac1n}(u))^j$ we refer to the $k-$th line of the matrix $\n u$ and the $j-th$ column of the matrix $\n J_{\frac1n}(u)$, respectively. We recall that we make use of the symbol $<\cdot ,\cdot>$ to denote the usual matrix multiplication\,.}
    \]}
\end{lemma}
\bp
Consider a scalar function $\psi\in C^\infty_0(\Omega)$. Then
\[
\int_\Omega \psi\rot (J_{\frac1n}(u)\cdot \n u)\,dx=\int_\Omega \rot (\psi J_{\frac1n}(u)\cdot \n u)\,dx-\int_\Omega \n\psi\times J_{\frac1n}(u)\cdot \n u\,dx\,.
\]
By the Stokes Theorem, the first integral is clearly equal to $0$, while integration by parts yields to 
\[
\ba {l}\displ
-\bigg(\int_\Omega \n\psi\times J_{\frac1n}(u)\cdot \n u\,dx\,\bigg)^i=-\int_\Omega \vep_{ijk}\partial_{x_j}\psi \partial_{x_h}u_kJ_{\frac1n}(u_h)\,dx=\int_{\Omega}\vep_{ijk} \psi \partial_{x_j}\partial_{x_h}u_kJ_{\frac1n}(u_h)\,dx\VS\hskip10.5cm+\int_{\Omega} \vep_{ijk} \psi \partial_{x_h}u_k\partial_{x_j}J_{\frac1n}(u_h)\,dx\,.
\ea
\]
Then, we remark that
\[
\int_{\Omega}\vep_{ijk} \psi \partial_{x_j}\partial_{x_h}u_kJ_{\frac1n}(u_h)\,dx=\int_{\Omega}\vep_{ijk} \psi \partial_{x_h}\partial_{x_j}u_kJ_{\frac1n}(u_h)\,dx=\int_\Omega \psi (J_{\frac1n}(u)\cdot \n \rot u)^i\,dx\,,
\]
and
\[
\int_{\Omega} \vep_{ijk} \psi \partial_{x_h}u_k\partial_{x_j}J_{\frac1n}(u_h)\,dx=\int_{\Omega} \vep_{ijk} \psi<(\n u)_k,(\n J_{\frac1n}(u))^j>\,dx\,.
\]
Hence, we find that
\[
\bigg(\int_\Omega \psi\rot (J_{\frac1n}(u)\cdot \n u)\,dx\bigg)^i=\int_\Omega \psi(J_{\frac1n}(u)\cdot \n \rot u)^i\,dx+\int_{\Omega} \vep_{ijk} \psi<(\n u)_k,(\n J_{\frac1n}(u))^j>\,dx\,.
\]
Since $\psi$ was chosen arbitrarily, the lemma is proved.
\ep
\begin{lemma}\label{le: rotore termine convettivo-V}
{\sl    Let $V(x)=\xi+\omega\times x$, $\xi,\omega\in\mathbb R^3$ and $u\in W^{2,2}(\Omega)$. Then
    \[
    \rot (J_{\frac1n}(V)\cdot\n u)=J_{\frac1n}(V)\cdot\n\rot u-\omega\times\rot u-\omega\cdot \n u\,.
    \]}
\end{lemma}
\bp
The lemma can be proved by a similar argument as in Lemma\,\ref{le: Rotore termine convettivo-u}. In fact, given a scalar function $\psi\in C_0^\infty(\Omega)$, we have
\[
\int_\Omega \rot(J_{\frac1n}(V)\cdot \n u)\psi\,dx=\int_\Omega \rot(\psi J_{\frac1n}( V)\cdot\n u)\,dx-\int_\Omega \n \psi\times J_{\frac1n}(V)\cdot \n u\,dx\,.
\]
The Stokes Theorem ensures that the first integral is equal to $0$, while, integrating by parts, we find
\[
\ba {l} \displ
-\bigg(\int_\Omega \n \psi\times J_{\frac1n}(V)\cdot \n u\,dx\bigg)^i=-\int_\Omega \vep_{ijk}\partial_{x_j}\psi \partial_{x_h}u_kJ_{\frac1n}(V)_h\,dx\VS \hskip 6cm =\int_\Omega \vep_{ijk}\psi (\partial_{x_h}\partial_{x_j}u_kJ_{\frac1n}(V)_h+\partial_{x_h}u_k\partial_{x_j}J_{\frac1n}(V)_h)\,dx\,.
\ea
\]
Moreover, we can say that 
\[
\partial_{x_j}J_{\frac1n}(V)_h=J_{\frac1n}(\partial_{x_j}V_h)=J_{\frac1n}(\partial_{x_j}(V_2)_h)\,.
\]
Hence, similarly to \cite[Proposition 5]{Mar-Pal}, we arrive at
\[
\ba {l}\displ
\int_\Omega \vep_{ijk}\psi (\partial_{x_h}\partial_{x_j}u_kJ_{\frac1n}(V)_h+\partial_{x_h}u_k\partial_{x_j}J_{\frac1n}(V)_h)\,dx=\int_\Omega \psi ((J_{\frac1n} (V)\cdot \n\rot u)^i-(\n(\omega\cdot u))^i)\,dx\VS \hskip5.9cm =\int_\Omega \psi ((J_{\frac1n} (V)\cdot \n\rot u)^i-(\omega\times\rot u)^i-(\omega\cdot \n u)^i)\,dx\,,
\ea
\]
which leads to the claimed property.
\ep
\begin{lemma} \label{CM-Lemma}
 {\sl   Let $u\in W^{2,2}(\Omega)$. Then
    \[
    \dm u\dm_\infty\le c\dm D^2 u\dm_2^{\frac12}\dm \n u\dm_2^{\frac12}\,,
    \]
    for some constant $c$ independent of $u$.}
\end{lemma}
\bp
The proof is a consequence of the Gagliardo-Nirenberg interpolation inequality    $\dm u\dm_\infty\leq c\dm D^2 u\dm_2^\frac12\dm u\dm_6^\frac12$, and of Sobolev's embeddings, both properties with non homogeneous boundary value, see \cite{CM} and \cite{Ga:book}, respectively.
\ep
\begin{lemma} \label{le: peso-prel}
{\sl Let $g$ be a function that lies in the closure of $C_0^\infty(\Omega)$ with respect to the metric
\[
\dm |x|\rot g\dm_2+\dm |x|\n\cdot g\dm_2\,.
\]
Then
    \be\label{RGW}\dm |x|\n g\dm_2\leq c\Big[\dm |x|\rot g\dm_2+\dm |x|\n\cdot g\dm_2\Big]\,,\ee with $c$ independent of $g$.}
\end{lemma}
\bp
See \cite[Section 2]{Mar-Pal}
\ep
\begin{lemma} \label{le: interpolazione weak}
  {\sl  Let $D$ be a Lebesgue measurable set of $\mathbb R^n$ and let $g\in L^{p_0}_w(D)\cap L^{p_1}_w(D)$, with $1\le p_0<p_1\le +\infty$. Then, for all $p\in (p_0,p_1)$, $g\in L^p(D)$ and there exists a constant $K$ independent of $p$, $p_0$ and $p_1$ such that
    \[
    \dm g\dm_p\le K\dm g\dm_{p_{0_w}}^{1-\theta}\dm g\dm_{p_{1_w}}^{\theta}\,,
    \]
    where $\theta$ is given by the relation
    \[
    \frac{1}{p}=\frac{1-\theta}{p_0}+\frac{\theta}{p_1}\,.
    \]}
\end{lemma}
\bp
For the proof we quote \cite{Mir, Stein: book}
\ep
\begin{lemma} \label{le: Gronwall generalizzato}
 {\sl   Let $g:[0,T)\to [0,\infty)$, $T>0$, be absolutely continuous, such that
    \[
    g'(t)\le c_1g(t)+c_2g^{\alpha}(t)\,,\quad \alpha>1\,,\quad t\in (0,T)\,,
    \]
    with $c_1,c_2\geq0$. Let $M:=2\max\{1,c_1,c_2\}$, and assume 
    \[
    g(0)+\int_0^Tg(\tau)\,d\tau \leq \eta\,,
    \]
    for some $\eta\in (0, M^{-\frac{1}{\alpha-1}})$.
    Then, the following holds:
    \[
    g(t)<M\eta\,,\quad t\in [0,T)\,.
    \]}
\end{lemma}
\bp
The proof can be found in \cite{Ga:23}, where it is presented in a more general setting.
\ep
\section{The IBVP for the mollified equations} \label{sec: mollificatore}
In this section, we focus on the following system
\begin{equation} \label{eq:moll}
    \begin{cases}
         u^{n}_t-\Delta u^n= - \mathbb{J}_n(u^n- V^n)\cdot \nabla u^n-\omega^n\times u^n-\nabla\pi^n\,, \quad \forall (t,x) \in (0,T)\times \Omega\,, \\
         \nabla \cdot u^n=0\,, \quad \forall (t,x) \in (0,T)\times \Omega\,, \\
         u^n(t,x)=V^n(t,x)=\xi^n(t)+\omega^n(t)\times x\,, \quad \forall (t,x) \in (0,T)\times \partial \Omega\,, \\
        \displ \lim_{\abs{x}\to \infty} u^n(t,x)=0\,, \quad \forall t\in (0,T)\,, \\
        \dot{\xi}^n + \omega^n \times \xi^n +  \int_{\partial\Omega} \mathbb{T}(u^n, \pi^n)\cdot \nu=0\,,\quad \forall t\in (0,T)\,, \\
         I \cdot\dot{\omega}^n+ \omega^n \times(I\cdot \omega^n)+ \int_{\partial\Omega} x \times \mathbb{T}(u^n,\pi^n) \cdot \nu=0\, \quad \forall t\in(0,T)\,, \\
         \xi^n(0)= \xi_0\,, \quad \omega^n(0)=\omega_{0}\,, \\
         u^n(0,x)=u_0(x)\,, \quad \forall x \in \Omega\,,
    \end{cases} 
\end{equation}
for some $T\in(0, +\infty]$, where $\mathbb{J}_n:=J_{\frac{1}{n}}$ is a Friedrichs mollifier in the space variable. \par{ Moreover, we decompose the rigid motion $V^n$ as $V^n(t,x)=V^n_1(t)+V^n_2(t,x)$, where
\[
\ba{rl}
V^n_1(t)&\hskip-0.2cm:=\xi^n(t)\,,\VS V^n_2(t,x)&\hskip-0.2cm:=\omega^n(t)\times x\,.
\ea
\]}
For system \eqref{eq:moll} we will first prove a result of { local existence and uniqueness of a regular solution.} 
In order to reach our goal, we first prove the local { existence and uniqueness of a regular solution} to the following approximating system
\begin{equation} \label{eq:moll1}
    \begin{cases}
         u^{n}_t-\Delta u^n= - \mathbb{J}_n(u^n- \varphi_\rho V^n)\cdot \nabla u^n-\mathbb J_n(\varphi_\rho)\omega^n\times u^n-\nabla\pi^n\,, \quad \forall (t,x) \in (0,T)\times \Omega\,, \\
         \nabla \cdot u^n=0\,, \quad \forall (t,x) \in (0,T)\times \Omega\,, \\
         u^n(t,x)=V^n(t,x)=\xi^n(t)+\omega^n(t)\times x\,, \quad \forall (t,x) \in (0,T)\times \partial \Omega\,, \\
         \displ\lim_{\abs{x}\to \infty} u^n(t,x)=0\,, \quad \forall t\in (0,T)\,, \\
        \dot{\xi}^n + \omega^n \times \xi^n +  \int_{\partial\Omega} \mathbb{T}(u^n, \pi^n)\cdot \nu=0\,,\quad \forall t\in (0,T)\,, \\
         I \cdot \dot{\omega}^n+ \omega^n \times(I\cdot \omega^n)+ \int_{\partial\Omega} x \times \mathbb{T}(u^n,\pi^n) \cdot \nu=0\, \quad \forall t\in(0,T)\,, \\
         \xi^n(0)= \xi_0\,, \quad \omega^n(0)=\omega_{0}\,, \\
         u^n(0,x)=u_0(x)\,, \quad \forall x \in \Omega\,,
    \end{cases}
\end{equation}
where $\varphi_\rho$ is a sufficiently smooth cut-off function of radius $\rho>\text{diam}(\mathcal B)$, specifically $\varphi_\rho(x)=1$ if $|x|\le \rho$ and $\varphi_\rho=0$ if $|x|\ge 2\rho$ and, denoting by $B_\rho$ the ball of radius $\rho$ centered in the origin, we assume that
\be \label{eq:condizione su rho}
\text{dist}(x,B_{2\rho})>1\,,\quad \text{for all }x\in\partial\Omega\,.
\ee
{It is natural to inquire about the opportunity of investigating the system \eqref{eq:moll1}. Actually, as it is shown in \cite{GS}, in the energy estimates for the Galerkin approximating sequence associated to problem \eqref{eq:model} on an invading domain $\Omega_R$, there is no contribution arising from the spherical surface $\partial B_R$. This last sentence is false as soon as we consider a mollification of the convective term. We get over this difficulty by adding as intermediate step to the analytical theory of problem \eqref{eq:moll} the study of the approximating system \eqref{eq:moll1}. }
\subsection{Local existence and uniqueness of a solution to \eqref{eq:moll1}}
In this subsection we prove the following result.
\begin{tho} \label{thm: Esistenza-moll-1step}
{\sl    Let $\mathscr A:=(u_0,\xi_0,\omega_0) \in \mathcal{V}(\Omega)\times\mathbb R^3\times \mathbb R^3$, with $\gamma(u_0- \xi_0- \omega_0\times x)=0$ and set $A_0:=\dm u_0\dm_2^2+|\xi_0|^2+\omega_0\cdot (I\cdot\omega_0)$. There exist $c>0$, independent of $\mathscr A$, and a unique solution $(u^n_\rho,\pi^n_\rho, \xi^n_\rho,\omega^n_\rho)$ to problem \eqref{eq:moll1}, a. e. in $(0,T_0)\times\Omega$, with $T_0:=(cA_0^{\frac{1}{2}})^{-1}$, such that
\begin{equation} \label{eq:SM-i}
\begin{aligned}
&u^n_\rho \in C([0,T_0]; \mathcal{V}(\Omega)) \cap L^2(0,T_0; W^{2,2}(\Omega))\,,\\
& \xi^n_\rho,\omega^n_\rho \in W^{1,2}(0,T_0)\cap C[0,T_0]\,, \\
& \nabla\pi^n_\rho \in L^2(0,T_0; L^2(\Omega))\,, \\
& \partial_tu^{n}_\rho \in L^2(0,T_0; L^2(\Omega))\,.
\end{aligned}
\end{equation}}
\end{tho}

\bp
In order to prove the theorem, we just adapt to the mollified problem \eqref{eq:moll1} the proof given by Galdi and Silvestre in \cite{GS} for the existence of a local solution to \eqref{eq:model}. Hence, we employ the Galerkin method and the invading domains technique.\par 
We consider a positive, increasing and divergent sequence $\{R_m\}_{m\in\mathbb{N}}$, with $\text{diam}(\mathcal{B})<\rho<2\rho<R_1$.
 Next, we consider a sequence $\{u_{0_{R_{m}}}\}_{m\in\mathbb{N}}\subset \mathcal{V}(\Omega)$ such that $u_{0_{R_{m}}}(x)=0$ for $\abs{x}\ge R_m$, $\norm{u_{0_{R_m}}}_{\mathcal{V}(\Omega)}\le \norm{u_0}_{\mathcal{V}(\Omega)}$, for all $m\in\mathbb{N}$, and $u_{0_{R_{m}}}\to u_0$ in $\mathcal{V}(\Omega)$ as $m\to\infty$. For each $R\in\{R_m\}$, we consider $\{a_{i}\}_{i\in\mathbb{N}}\subset \mathcal{V}(\Omega_R)$ as the basis of $\mathcal{H}(\Omega_R)$ ensured by Lemma\,\ref{le:prel-BaseGal}, and we construct the sequences of approximating solutions as follows: 
\[
u^n_{k}(t,x)=\sum_{j=1}^k c_{k}^j(t)a_{j}(x)\,, \quad \xi^n_{k}(t)=\sum_{j=1}^k c_{k}^j(t)\overline{a}_{j_{1}}\,, \quad \omega^n_{k}(t)=\sum_{j=1}^k c_{k}^j(t)\overline{a}_{j_2}\,,
\]
{where $\overline a_{j_1}$ and $\overline a_{j_2}$ are the characteristic vectors associated to $a_j$. }\par
We require that $c_{k}^j(t)$, $j=1,\dots,k$, are solution to 
\begin{equation} \label{eq:Gal}\hskip-0.1cm
\begin{array}{l}
(\partial_tu^n_{k},a_{j})_{L^2(\Omega_R)}+2 (\mathbb D( u^n_{k}),\mathbb D( a_{j}))_{L^2(\Omega_R)}+ \dot{\xi}^n_{k} \cdot \overline{a}_{j_1} + \dot{\omega}^n_{k} \cdot (I\cdot\overline{a}_{j_2})\VS\hskip0.07cm =-(\mathbb{J}_n(u^n_{k}\!-\!\varphi_{\rho}V^n_{k})\!\cdot \!\nabla u^n_{k} \!+\!\mathbb J_n(\varphi_\rho)\omega^n_{k}\times u^n_{k}, a_{j})_{L^2(\Omega_R)}+ [(\overline{a}_{j_1}\!\cdot \!\xi^n_{k}\times\omega^n_k\!+\! \overline{a}_{j_2}\cdot(I\cdot \omega^n_{k}))\times \omega^n_{k}]\,.
\end{array}
\end{equation}
We set $V^n_{k}(t,x):=\xi^n_{k}(t)+\omega^n_{k}(t)\times x$. \par
System \eqref{eq:Gal} is endowed with the initial condition
\begin{equation} \label{eq:CI}
c^{j}_{k}(0)=(u_{0_R}, a_{j})_{L^2(\Omega_R)}+ \overline{a}_{j_1}\cdot \xi_{0_R}+ \overline{a}_{j_2}\cdot (I\cdot \omega_{0_R})= (u_{0_R}, a_{j})_{1,R}\, \quad j=1,\dots,k\,.
\end{equation}
Then, the Cauchy theorem of local existence and uniqueness ensures that the system of ordinary differential equations \eqref{eq:Gal} admits a unique solution corresponding to the initial datum \eqref{eq:CI} in a time interval $(0,T_{k})$. \par Moreover, we set
\[
u_{0_{k}}(x):=u^n_{k}(0,x)\,,\quad \xi_{0_{k}}:=\xi^n_{k}(0)\,, \quad \omega_{0_{k}}:=\omega^n_{k}(0)\,.
\]
Since $u_{0_k}$ is the projection of $u_{0_R}$ in the subspace $\text{span}\{ a_{1},\dots,a_{k}\}\subset \mathcal{V}(\Omega)$ and
\[
\dm u_{0_R}\dm_{\mathcal{V}(\Omega)}\le \dm u_0\dm_{\mathcal{V}(\Omega)}\,,
\]
recalling the definitions of the scalar products \eqref{SPL2H1-R}$_1$ and \eqref{SPL2H1-R}$_2$, we get
\begin{equation}
    \dm u_{0_{k}}\dm_{L^2(\Omega_R)}^2+|\xi_{0_{k}}|^2+\omega_{0_{k}}\cdot (I \cdot \omega_{0_{k}})\le\dm u_0\dm_{\mathcal{V}(\Omega)}^2\,,
\end{equation}
and
\begin{equation}
    \dm \mathbb{D}(u_{0_{k}})\dm_{L^2(\Omega_R)}^2\le \dm u_0\dm_{\mathcal{V}(\Omega)}^2\,.
\end{equation}
We now provide energy estimates for $u^n_{k}$.
Multiplying in \eqref{eq:Gal} by $c_{k}^j(t)$ and summing on $j$, recalling that, by construction, $u^n_{k}(t,x)=0$ if $\abs{x}\ge R$, and using assumption \eqref{eq:condizione su rho},  we get
\begin{equation} \label{eq:en1}\hskip-0.1cm
\ba {l} \displ
\sfrac{1}{2} \sfrac{d}{dt}(\norm{u^n_{k}(\tau)}_{L^2(\Omega_R)}^2     \!+ \abs{\xi^n_{k}(\tau)}^2 \!+ \omega^n_{k}(\tau) \cdot( I \cdot \omega^n_{k} (\tau)))+ 2\norm{\mathbb{D} (u^n_{k})(\tau)}_{L^2(\Omega_R)}^2\VS\hskip 2.2cm=\!-\sfrac12\!\int_{\partial\Omega} \!\!\!\nu\cdot \mathbb{J}_n(u^n_{k}-V^n_{k})(\tau) \abs{V^n_{k}}^2 \,d\sigma+\sfrac12\int_{\Omega_R}\!\!\!\n \cdot \mathbb{J}_n(u^n_k-\varphi_\rho V^n_k)(\tau)|u^n_k(\tau)|^2\,dx\,.
\ea
\end{equation}
The last two integrals in \eqref{eq:en1} can be estimated using Lemma\,\ref{le:Stime per rel energia}. Denoting by $\Omega_{\eta}$ a suitable smooth neighborhood of $\partial\Omega$, we have
\[
\ba {l} \displ
\Bigg|\int_{\partial\Omega} \nu\cdot \mathbb{J}_n(u^n_{k}-V^n_{k})(t) |V^n_{k}(t)|^2 \, d\sigma \Bigg|\le c\biggl(\norm{u^n_{k}(t)}_{L^2(\Omega_{\eta})} + \norm{V^n_{k}(t)}_{L^2(\Omega_\eta)}\biggr)(|\xi^n_{k}(t)|^2+ |\omega^n_{k}(t)|^2)\VS \hskip 8cm\le c(\norm{u^n_{k}(t)}_{L^2(\Omega_R)}^2 +\abs{\xi^n_{k}(t)}^2+ \abs{\omega^n_{k}(t)}^2)^{\frac{3}{2}}\,,
\ea
\]
and
\[
\Bigg|\int_{\Omega_R}\n \cdot \mathbb{J}_n(u^n_k-\varphi_\rho V^n_k)|u^n_k|^2\,dx\Bigg|\le c|\xi^n_k|\dm u^n_k\dm_{L^2(\Omega_R)}^2\le c(\dm u^n_k\dm_{L^2(\Omega_R)}^2+|\xi^n_k|^2+|\omega^n_k|^2)^{\frac{3}{2}}\,.
\]
Hence, we get
\begin{equation} \label{eq:diff in}
    \sfrac{d}{dt}\bigl(\dm u^n_k\dm_2^2+|\xi^n_k|^2+\omega^n_k\cdot (I\cdot\omega^n_k))+2\dm \n u^n_k\dm_2^2\le c(\dm u^n_k\dm_2^2+|\xi^n_k|^2+\omega^n_k\cdot (I\cdot\omega^n_k))^{\frac{3}{2}}\,.
\end{equation}
Setting $y(\tau):=\dm u^n_{k}(\tau)\dm_{L^2(\Omega_R)}^2 + |\xi^n_{k}(\tau)|^2 + \omega^n_{k}(\tau) \cdot (I \cdot \omega^n_{k} (\tau))$, we obtain
\be \label{eq:phi}
\sfrac{d}{dt}y(t)\le cy^{\frac{3}{2}}(t)\,.
\ee
The above inequality can be integrated in $(0,T'_0)$,  with $T'_0=2(cy^{\frac{1}{2}}(0))^{-1}$, leading to
\be \label{eq: phi1}
y(t)\le \frac{4y(0)}{(2-cty^{\frac{1}{2}}(0))^2}\,.
\ee
Hence, employing \eqref{eq: phi1} for the right-hand side of \eqref{eq:diff in}, and then integrating in $(0,T_0)$, with $T_0:=\frac{T'_0}{2}$,  we arrive at
\begin{equation} \label{eq:est1}
\norm{u^n_{k}(T_0)}_{L^2(\Omega_R)}^2 + \abs{\xi^n_{k}(T_0)}^2 + \omega^n_{k}(T_0) \cdot (I \cdot \omega^n_{k} (T_0))+ 2 \int_{0}^{T_0} \norm{\nabla u^n_{k}(\tau)}_{L^2(\Omega_R)}^2 \, d\tau \le A_0\bigg(1+\frac{3}{c}\bigg)\,.
\end{equation}
We now multiply \eqref{eq:Gal} by $-\lambda_{j}c_{k}^j$ and sum on the index $j$. 
We have
\[
\begin{aligned}
&\sfrac{d}{dt} \norm{\mathbb{D}( u^n_{k})}_{L^2(\Omega_R)}^2+ |\dot \xi^n_k|^2 +\dot\omega^n_k\cdot (I \cdot \dot\omega^n_k)+ \norm{\nabla\cdot \mathbb{T}(u^n_{k},\pi^n_{k})}_{L^2(\Omega_R)}^2=-\omega^n_{k}\times \xi^n_{k} \cdot \dot\xi^n_k + \\ &- \omega^n_{k} \times (I\cdot \omega^n_{k}) \cdot \dot\omega^n_k + \bigl(\mathbb{J}_n(u^n_{k}-\varphi_{\rho}V^n_{k}) \cdot \nabla u^n_{k}+\mathbb J_n(\varphi_\rho)\omega^n_{k} \times u^n_{k}, \nabla\cdot \mathbb{T}(u^n_{k},\pi^n_{k})\bigr)_{L^2(\Omega_R)}\,.
\end{aligned}
\]
We estimate the terms on the right-hand side. 
By H\"older and Young inequalities, we have
\[
\ba {l}
\big|\bigl(\mathbb{J}_n(u^n_{k}-\varphi_{\rho}V^n_{k}) \cdot \nabla u^n_{k}, \nabla\cdot \mathbb{T}(u^n_{k},\pi^n_{k})\bigr)_{L^2(\Omega_R)}\big| \VS \hskip 3cm\le c(n,\rho)(\dm u^n_{k}\dm_{L^2(\Omega_R)}+|\xi^n_{k}|+|\omega^n_{k}|)\dm\n u^n_{k}\dm_{L^2(\Omega_R)}\dm \n\cdot \mathbb{T}(u^n_{k},\pi^n_{k})\dm_{L^2(\Omega_R)}\VS
\hskip 3.2cm\le \vep \dm \n\cdot \mathbb{T}(u^n_{k},\pi^n_{k})\dm_{L^2(\Omega_R)}^2+c(n,\rho,\vep) (\dm u^n_{k}\dm_{L^2(\Omega_R)}+|\xi^n_{k}|+|\omega^n_{k}|)^2\dm\n u^n_{k}\dm_{L^2(\Omega_R)}^2\,,
\ea
\]
\[
\ba {l}
\big|\bigl(\mathbb J_n(\varphi_\rho)\omega^n_{k} \times u^n_{k}, \nabla\cdot \mathbb{T}(u^n_{k},\pi^n_{k})\bigr)_{L^2(\Omega_R)}\big|\le |\omega^n_{k}|\dm u^n_{k}\dm_{L^2(\Omega_R)}\dm \n \cdot \mathbb{T}(u^n_{k},\pi^n_{k})\dm_{L^2(\Omega_R)}\VS \hskip7 cm \le \vep  \dm \n\cdot \mathbb{T}(u^n_{k},\pi^n_{k})\dm_{L^2(\Omega_R)}^2+c(\vep)|\omega^n_{k}|^2\dm u^n_{k}\dm_{L^2(\Omega_R)}^2\,,
\ea
\]
\[
|\omega^n_{k}\times \xi^n_{k} \cdot \dot\xi^n_k|\le \vep |\dot\xi^n_k|^2+c(\vep)|\omega^n_{k}|^2|\xi^n_{k}|^2\,,
\]
\[
|\omega^n_{k} \times (I\cdot \omega^n_{k}) \cdot\dot\omega^n_k | \le \vep \dot\omega^n_k \cdot (I\cdot\dot\omega^n_k) +c(\vep) |\omega^n_{k}|^4\,.
\]
Then, choosing $\vep>0$ sufficiently small, collecting all the estimates and invoking Lemma\,\ref{le:prel-symgrad}, recalling that, via \eqref{eq:est1}, we have
\[
\dm u^n_{k}(t)\dm_{L^2(\Omega_R)}^2+|\xi^n_{k}(t)|^2+|\omega^n_{k}(t)|^2\le A_0(1+3/c)\,,\,\,\text{for all }t\in [0,T_0]\,,
\]
we get
\[
\ba {l}
\sfrac{d}{dt} \dm \mathbb{D}( u^n_{k})\dm_{L^2(\Omega_R)}^2+ \sfrac{1}{2}\bigl(|\dot \xi^n_k|^2 +\dot\omega^n_k\cdot (I \cdot \dot\omega^n_k)+ \dm \nabla\cdot \mathbb{T}(u^n_{k},\pi^n_{k})\dm_{L^2(\Omega_R)}^2\bigr) \VS \hskip 8cm \le c(n,\rho,\vep, A_0)(\dm \mathbb{D}(u^n_{k})\dm_{L^2(\Omega_R)}^2+A_0^2)\,.
\ea
\]
Hence, integrating in time and recalling estimate \eqref{eq:stimederseconde}$_3$, we get
\[
\sup_{t\in(0,T_0)}\dm \mathbb{D}(u^n_{k})(t)\dm_{L^2(\Omega_R)}^2+\int_0^{T_0}\big(\dm D^2 u^n_{k}\dm_{L^2(\Omega_R)}^2+|\dot\xi^n_k|^2+\dot\omega^n_k\cdot I\cdot\dot\omega^n_k\big)\le c(n,\rho,A_0,t)\,.
\]
Finally, we multiply \eqref{eq:Gal} by $\frac{d c_{k}^j(t)}{dt}$ and sum over the index $j=1,\dots,k$. We get
\[
\ba {l}
\sfrac{d}{dt}\dm \mathbb{D}(u^n_{k})\dm_{L^2(\Omega_R)}^2+\Big|\frac{d\xi^n_{k}}{dt}\Big|^2+\frac{d\omega^n_{k}}{dt}\cdot \bigl(I\cdot \frac{d\omega^n_{k}}{dt}\bigr)+\dm \partial_tu^n_{k}\dm_{L^2(\Omega_R)}^2=\omega^n_{k}\times \xi^n_{k}\frac{d\xi^n_{k}}{dt}\VS\hskip 3.5cm+\omega^n_{k}\times (I\cdot \omega^n_{k})\cdot \frac{d\omega^n_{k}}{dt} -\big(\mathbb{J}_n(u^n_{k}-\varphi_\rho V^n_{k})\cdot \n u^n_{k}+\mathbb J_n(\varphi_\rho)\omega^n_{k}\times u^n_{k},\partial_t u^n_{k}\big)\,.
\ea
\]
We estimate the terms on the right-hand side. Recalling \eqref{eq:est1}, we have
\[
\ba {rl}
\bigg|\omega^n_{k}\times \xi^n_{k}\frac{d\xi^n_{k}}{dt}\bigg|   \hskip-0.2cm&\le \vep \bigg|\frac{d\xi^n_{k}}{dt}\bigg|^2+c(\vep,A_0)\,,\VS
\bigg|\omega^n_{k}\times (I\cdot \omega^n_{k})\cdot \frac{d\omega^n_{k}}{dt}\bigg|^2\hskip-0.2cm&\le\vep \frac{d\omega^n_{k}}{dt}\cdot \bigl(I\cdot \frac{d\omega^n_{k}}{dt}\bigl)+c(\vep,A_0)\,,\VS
\big|\big(\mathbb{J}_n(u^n_{k}-\varphi_\rho V^n_{k})\cdot \n u^n_{k}+\mathbb J_n(\varphi_\rho)\omega^n_{k}\times u^n_{k},\partial_t u^n_{k}\big)\big|\hskip-0.2cm &\le \vep \dm \partial_tu^n_{k}\dm_{L^2(\Omega_R)}^2\VS &+c(n,\rho,\vep,A_0)(\dm\mathbb{D}(u^n_{k})\dm_{L^2(\Omega_R)}^2+A_0^2)\,. 
\ea
\]
Hence, choosing $\vep>0$ sufficiently small and collecting the above estimates, via \eqref{eq:est1} we obtain
\[
\ba {l}
\sfrac{d}{dt}\dm \mathbb{D}(u^n_{k})\dm_{L^2(\Omega_R)}^2+\sfrac{1}{2}\Big(\Big|\frac{d\xi^n_{k}}{dt}\Big|^2+\frac{d\omega^n_{k}}{dt}\cdot I\cdot \frac{d\omega^n_{k}}{dt}+\dm \partial_tu^n_{k}\dm_{L^2(\Omega_R)}^2\Big)\VS \hskip 6cm\le c(n,\rho,\vep,A_0)(\dm\mathbb{D}(u^n_{k})\dm_{L^2(\Omega_R)}^2+A_0^2)\,.
\ea
\]
Integrating in time, we get
\[
\sup_{t\in(0,T_0)}\dm \mathbb{D}(u^n_{k})(t)\dm_{L^2(\Omega_R)}^2+\int_0^{T_0}\dm \partial_t u^n_{k}\dm_{L^2(\Omega_R)}^2\, dt\le c(n,\rho,A_0,t)\,.
\]
Since all the estimates deduced do not depend on $R$, we can conclude the proof by following classical arguments, see \cite{Hey, GS}. 
\ep
\begin{rem} \label{rem:MOLL-Exp}
    {\rm { We point out that the presence of the cut-off function $\varphi_\rho$ allows us to state the uniqueness of the solution to \eqref{eq:moll1}, obtained in Theorem\,\ref{thm: Esistenza-moll-1step}, without any relevant effort, as it holds for the Navier-Stokes IBVP.}}
\end{rem}
\begin{rem}
    {\rm  We point out that $(u^n_\rho,\xi^n_\rho,\omega^n_\rho)$ verify, for $\tau\in (0,T_0]$, the following identity
    \begin{equation} \label{eq:RE-rho}
\ba {l} \displ
\sfrac{1}{2} \sfrac{d}{dt}(\norm{u^n_\rho(\tau)}_2^2     \!+ \abs{\xi^n_{\rho}(\tau)}^2 \!+ \omega^n_{\rho}(\tau) \cdot( I \cdot \omega^n_{\rho} (\tau)))+ 2\norm{\mathbb{D} (u^n_{\rho})(\tau)}_2^2\VS\hskip 2.2cm=\!-\sfrac12\!\int_{\partial\Omega} \!\!\!\nu\cdot \mathbb{J}_n(u^n_{\rho}-V^n_{\rho})(\tau) \abs{V^n_{k}}^2 \,d\sigma+\sfrac12\int_{\Omega}\!\!\!\n \cdot \mathbb{J}_n(u^n_\rho-\varphi_\rho V^n_\rho)(\tau)|u^n_\rho(\tau)|^2\,dx\,.
\ea
\end{equation}}
\end{rem}
\subsection{Existence of a regular solution to \eqref{eq:moll}}
{ In this subsection we aim to establish a regular solution to problem \eqref{eq:moll}. We reach our goal by providing \textit{a priori} estimates for the regular solution $(u^n_\rho,\pi^n_\rho,\xi^n_\rho,\omega^n_\rho)$ to system \eqref{eq:moll1}. These estimates will not depend on $\rho$ and, letting $\rho\to\infty$, we will fulfill our purposes.}\par
In order to reach our aims, we need to extend the functions $u^n_\rho$ and $V^n_\rho$. We set
\be \label{PR-rho}
\ba{ll}
\widetilde u^n_\rho(t,x):=&\hskip-0.2cm\begin{cases}
    u^n_\rho(t,x)\,\,\quad \text{if }x\in\Omega\\
    V^n_\rho(t,x)\,\,\,\,\,\,\text{if }x\in\mathcal B\,,
\end{cases}\VS 
\widetilde V^n_\rho(t,x):=&\hskip-0.2cm\xi^n_\rho(t,x)+\omega^n_\rho(t,x)\times x\,,\quad \text{for all }x\in\mathbb R^3\,.
\ea
\ee
Since from now on there will be no possibility of ambiguity, with an abuse of notation we will still use the notation $(u^n_\rho,V^n_\rho)$ instead of $(\widetilde u^n_\rho,\widetilde V^n_\rho)$.\par We start by proving the energy estimate to the solution $(u^n_\rho,\pi^n_\rho,\xi^n_\rho,\omega^n_\rho)$ obtained in Theorem\,\ref{thm: Esistenza-moll-1step}
\begin{lemma} \label{le: RE}
{\sl    Let $(u^n_\rho,\pi^n_\rho,\xi^n_\rho,\omega^n_\rho)$ be the solution to problem \eqref{eq:moll1} obtained in Theorem\,\ref{thm: Esistenza-moll-1step}. Then, for all $t\in [0,T_0]$,
    \be \label{Energy}
\dm u^n_\rho(t)\dm_2^2+|\xi^n_\rho(t)|^2+\omega^n_\rho(t)\cdot( I\cdot \omega^n_\rho(t))+4\int_0^{t}\dm \mathbb D(u^n_\rho)(\tau)\dm_2^2\,d\tau\le A_0\bigg(1+\frac{3}{c}\bigg)\,,
    \ee
    {uniformly in $n\in \mathbb N$, with $c$ independent of $(u^n_\rho,\xi^n_\rho,\omega^n_\rho)$.} }
\end{lemma}
\bp
For the sake of simplicity, in the following we should omit the index $\rho$. We multiply equation \eqref{eq:moll1}$_1$ by $u^n$ and integrate over $\Omega$. We get
\be \label{SE-n}
\frac{1}{2}\frac{d}{dt}(\dm u^n(\tau)\dm_2^2+|\xi^n(\tau)|^2+\omega^n(\tau)\cdot (I\cdot \omega^n(\tau))+2\dm \mathbb D(u^n)\dm_2^2=-\int_{\Omega}\mathbb J_n(u^n-\varphi_\rho V^n)\cdot \n u^n\cdot u^n\,dx\,.
\ee
Integration by parts, assumption \eqref{eq:condizione su rho} and the divergence-free condition on $u^n$ yield to
\[
\int_{\Omega}\mathbb J_n(u^n-\varphi_\rho V^n)\cdot \n u^n\cdot u^n\,dx=\sfrac12\int_{\partial\Omega} \mathbb J_n(u^n-V^n)\cdot \nu |V^n|^2\,dS-\sfrac12\int_{\Omega}\mathbb J_n(\xi^n\cdot\n\varphi_\rho)|u^n|^2\,dx\,.
\]
Employing Lemma\,\ref{le:Stime per rel energia} and Young inequality, we arrive at
\be \label{eq:EE-moll}
\ba{l}\displ
\sfrac{1}{2}\sfrac{d}{d\tau}(\dm u^n(\tau)\dm_2^2+|\xi^n(\tau)|^2+\omega^n(\tau)\cdot (I\cdot \omega^n(\tau))+2\dm \mathbb D(u^n)\dm_2^2\VS\hskip6cm\le c(\dm u^n(\tau)\dm_2^2+|\xi^n(\tau)|^2+\omega^n(\tau)\cdot I\cdot \omega^n(\tau))^{\frac{3}{2}}\,.
\ea
\ee
Setting $y(t):=\dm u^n(t)\dm_2^2+|\xi^n(t)|^2+\omega^n(t)\cdot (I\cdot \omega^n(t))$, we find the differential inequality 
\[
y'(t)\le cy(t)^{\frac{3}{2}}\,,
\]
and, analogously to the analysis developed in Theorem\,\ref{thm: Esistenza-moll-1step}, the previous differential inequality furnishes 
\[
y(t)\le 4y(0)(2-cty(0)^{\frac12})^{-2}\,,\quad \text{for all }t\in[0,T'_0)\,,
\]
independent of $n$ and $\rho$. In particular, for all $t\in [0,T_0]$, we have
\be \label{moll-en}
\dm u^n(t)\dm_2^2+|\xi^n(t)|^2+\omega^n(t)\cdot( I\cdot \omega^n(t))+4\int_0^{t}\dm \mathbb D(u^n)(\tau)\dm_2^2\,d\tau\le A_0\bigg(1+\frac{3}{c}\bigg)\,.
\ee
The lemma is proved. 
\ep
\begin{lemma} \label{le:RG}
  {\sl   Let $(u^n_\rho,\pi^n_\rho,\xi^n_\rho,\omega^n_\rho)$ be the solution to problem \eqref{eq:moll1} obtained in Theorem\,\ref{thm: Esistenza-moll-1step}. Then, for all $t\in [0,T_0]$,
     \be \label{RG}
\dm \mathbb{D}(u^n_\rho(t))\dm_2^2+\int_0^{t}\big[\dm \n\cdot\mathbb T(u^n_\rho,\pi^n_\rho) \dm_2^2+\dm \n \pi^n_\rho\dm_2^2+|\dot\xi^n_\rho|^2+\dot\omega^n_\rho\cdot (I\cdot \dot\omega^n_\rho)\big]\,d\tau\le c(n,A_0,T_0)\,
     \ee
     and, in particular
     \be \label{RG-i}
\dm \mathbb{D}(u^n_\rho(t))\dm_2^2+\int_0^{t}\big[\dm D^2u^n_\rho\dm_2^2+\dm \n \pi^n_\rho\dm_2^2+|\dot\xi^n_\rho|^2+\dot\omega^n_\rho\cdot (I\cdot \dot\omega^n_\rho)\big]\,d\tau\le c(n,A_0,T_0)\,,
     \ee
     and $c$ independent of $(u^n_\rho,\pi^n_\rho,\xi^n_\rho,\omega^n_\rho)$.}
\end{lemma}
\bp
As in the proof of Lemma\,\ref{le: RE}, we omit the index $\rho$ for simplicity. \par
We multiply the equation \eqref{eq:moll1}$_1$ by $-\n \cdot \mathbb T(u^n,\pi^n)$. Recalling the regularity of $u^n_\rho$, expressed in \eqref{eq:SM-i}, we obtain
\[
\begin{aligned}
&\frac{d}{dt} \norm{\mathbb{D}( u^n)}_2^2+ |\dot\xi^n|^2 + \dot\omega^n\cdot (I \cdot \dot\omega^n)+ \norm{\nabla\cdot \mathbb{T}(u^n,\pi^n)}_2^2=-\omega^n\times \xi^n \cdot\dot\xi^n + \\ &- \omega^n \times (I\cdot \omega^n) \cdot \dot\omega^n + \int_\Omega(\mathbb{J}_n(u^n-\varphi_{\rho}V^n) \cdot \nabla u^n+\mathbb J_n(\varphi_\rho)\omega^n \times u^n)\cdot( \nabla\cdot \mathbb{T}(u^n,\pi^n))\,dx\,.
\end{aligned}
\]
We want to get estimates for the terms on the right-hand side that do not depend on $\rho$.
By general properties of mollifiers, H\"{o}lder and Young inequalities, we have
\[
\ba {l}
|(\mathbb J_n(u^n)\cdot \n u^n, \n\cdot \mathbb T(u^n,\pi^n))_{L^2(\Omega)}|\le c(n)\dm u^n\dm_2\dm \n u^n\dm_2\dm \n\cdot\mathbb T(u^n,\pi^n)\dm_2 \VS \hskip7cm \le \vep\dm \n\cdot\mathbb T(u^n,\pi^n)\dm_2^2+c(n,\vep)\dm u^n\dm_2^2\dm \n u^n\dm_2^2\,.
\ea
\]
By analogous arguments, we also have
\[
|(\mathbb J_n(\varphi_\rho\xi^n)\cdot \n u^n, \n\cdot \mathbb T(u^n,\pi^n))_{L^2(\Omega_R)}|\le |\xi^n|\dm\n u^n\dm_2\dm \n \cdot \mathbb T(u^n,\pi^n)\dm_2\le \vep \dm \n \cdot \mathbb T(u^n,\pi^n)\dm_2^2+c(\vep)|\xi^n|^2\dm\n u^n_k\dm_2^2\,.
\]
We now consider 
\[
\int_\Omega (\mathbb J_n(\varphi_\rho V^n_2)\cdot \n u^n-\mathbb J_n(\varphi_\rho)\omega^n\times u^n) \cdot(\n\cdot \mathbb T(u^n,\pi^n))\,dx\,.
\]
Recalling assumption \eqref{eq:condizione su rho}, integration by parts ensures that the previous integral is equal to
\[
\ba {l}\displ
\int_{\partial\Omega} (\mathbb{J}_n(V^n_2)\cdot \n u^n-\mathbb \omega^n\times V^n)\cdot \mathbb T(u^n,\pi^n)\cdot \nu \,dS\VS \hskip4cm -\int_{\Omega}\n(\mathbb{J}_n(\varphi_\rho V^n_2)\cdot \n u^n-\mathbb J_n(\varphi_\rho)\omega^n\times u^n)\cdot \mathbb T(u^n,\pi^n)\,dx=:I_1-I_2\,.
\ea
\]
We can estimate $I_1$ by means of Lemma\,\ref{le: I1}. We have
\[
|I_1|\le \vep\dm \n\cdot \mathbb T\dm_2^2+c(|\omega^n|+|\omega^n|^\frac43+|\omega^n|^4+|\omega^n|^2)\dm \n u^n\dm_2^2\,.
\]
We now estimate $I_2$. Setting $\Phi:=\mathbb{J}_n(\varphi_\rho V^n_2)\cdot \n u^n-\mathbb J_n(\varphi_\rho)\omega^n\times u^n$ and recalling the definition of the stress tensor, we notice that
\[
\int_{\Omega} \n \Phi\cdot \mathbb{T}(u^n,\pi^n)\,dx=\int_{\Omega}\partial_{x_j}\Phi_i\cdot \mathbb T^{i,j}\,dx=\int_{\Omega}\partial_{x_j}\Phi_i( \partial_{x_j}(u^n)_i+\partial_{x_i}(u^n)_j)\,dx-\int_{\Omega}\partial_{x_j}\Phi_i\delta_{i,j}\pi^n\,dx\,.\footnote{We are using the Einstein summation convention and by $(u_n)_i$ and $\Phi_i$ we denote the $i$-th components of the vectors $u^n$ and $\Phi$ respectively.}
\]
Integration by parts and the divergence-free condition on $u^n$ yield to
\[
\int_{\Omega}\partial_{x_j}\Phi_i( \partial_{x_j}(u^n)_i+\partial_{x_i}(u^n)_j)\,dx=\int_{\Omega}\partial_{x_j}\Phi_i \partial_{x_j}(u^n)_i\,dx+\int_{\partial\Omega}\Phi_i\nu_j\partial_{x_i}(u^n)_j\,dS\,.
\]
Hence, we get
\[
\int_{\Omega} \n \Phi\cdot \mathbb{T}(u^n,\pi^n)\,dx=\int_{\Omega} \n \Phi\cdot \n u^n\,dx+\int_{\partial\Omega}\Phi\cdot \n u^n\cdot \nu\,dS-\int_{\Omega}\n\Phi\cdot  \pi^n\mathbb I\,dx=:J_1+J_2-J_3\,.
\]
$J_1$ can be estimated employing Lemma\,\ref{le:J1}. We get
\[
|J_1|\le  \vep\dm\n\cdot\mathbb T(u^n,\pi^n)\dm_2^2+c|\omega^n|\dm u^n\dm_{W^{1,2}(\Omega^\rho)}^2+c(|\omega|+|\omega^n|^2)\dm \n u^n\dm_2^2\,.
\]
{We notice that $J_2$ can be easily estimated via the Gagliardo trace inequality for $\dm \n u^n\dm_{L^2(\partial\Omega)}$ and by estimate \eqref{eq: stima gradiente pressione} employed with $f=\n\cdot\mathbb T(u^n,\pi^n)$, being $\Delta u^n-\n\pi^n=\n\cdot\mathbb T(u^n,\pi^n)$. Hence, we get}
\[
|J_2|\le \vep\dm \n \cdot \mathbb T(u^n,\pi^n)\dm_2^2+c(|\omega^n|+|\omega^n|^\frac43+|\omega^n|^2)\dm \n u^n\dm_2^2\,.
\]
Finally, by virtue of Lemma\,\ref{le:J3}, we get
\[
|J_3|\le \vep\dm \n \cdot\mathbb T(u^n,\pi^n)\dm_2^2+c|\omega^n|^2\dm u^n\dm_{L^2(\Omega^\rho)}^2+c|\omega^n|\dm u^n\dm_{L^2(\Omega^\rho)}\dm \n u^n\dm_2\,.
\]
Eventually, by H\"older and Young inequalities, we have
\[
|\omega^n\times \xi^n \cdot \dot\xi^n|\le \vep |\dot\xi^n|^2+c(\vep)|\omega^n|^2|\xi^n|^2\,,
\]
\[
|\omega^n \times (I\cdot \omega^n) \cdot \dot\omega^n | \le \vep \dot\omega^n \cdot (I\cdot\dot\omega^n )+c(\vep) |\omega^n|^4\,.
\]
Collecting all the estimates and choosing $\vep>0$ sufficiently small, invoking Lemma\,\ref{le:prel-symgrad} and the energy relation \eqref{Energy}, we find that
\be \label{eq:stime gradiente/derivate seconde I}
\ba {l}\displ 
\sfrac{d}{dt} \dm \mathbb{D}( u^n)\dm_2^2+ \sfrac{1}{2}\bigl(|\dot\xi^n|^2 + \dot\omega^n\cdot (I \cdot \dot\omega^n)+ \dm \nabla\cdot \mathbb{T}(u^n,\pi^n)\dm_2^2\bigr)\le c(n,A_0)(1+\dm \mathbb D(u^n)\dm_2^2)\,.
\ea
\ee
Hence, we obtain
\be \label{eq:stime gradiente/derivate seconde II}
\dm \mathbb{D}(u^n(t))\dm_2^2\!+\!\int_0^{t}\!\!\big[\dm \n\cdot\mathbb T( u^n,\pi^n)\dm_2^2\!+\!\dm \n \pi^n\dm_2^2\!+\!|\dot\xi^n|^2\! +\! \dot\omega^n\cdot (I \cdot \dot\omega^n)\big]d\tau\le c(n,A_0,T_0)\,,\,\,\forall t\in[0,T_0]\,,
\ee
and, in particular, {\eqref{RG-i} holds.} 
The lemma is completely proved. 
\ep
We use the notation $\kappa(x)$ for the Stein generalized distance. For the properties of this function, we quote \cite{Stein: HA}. \par {We now state some \textit{a priori} estimates for $\partial_t u^n_\rho(1+|x|)^{-1}$ in $L^2(0,T_0;L^2(\Omega))$ that are independent of $\rho$. These estimates have a key role for proving the uniqueness of a regular solution to \eqref{eq:moll}.}
\begin{lemma} \label{le: DT-peso}
{\sl Let $(u^n_\rho,\pi^n_\rho,\xi^n_\rho,\omega^n_\rho)$ be the solution to problem \eqref{eq:moll1} obtained in Theorem\,\ref{thm: Esistenza-moll-1step}. Then, for all $t\in [0,T_0]$ and for a suitable $K_0>0$,
\be \label{TD}
\int_0^{t} \big(\dm\partial_t u^n_{\rho}(K_0+\kappa(x))^{-1}\dm_2^2+ +|\dot{\xi}^n_\rho|^2+| \dot{\omega}^n_\rho\cdot (I\cdot \dot{\omega}^n_\rho)|\big)\,d\tau\le c(n,A_0,T_0)\,,
\ee
and $c$ independent of $(u^n_\rho,\xi^n_\rho,\omega^n_\rho)$.}
\end{lemma}
\bp
As in the previous lemmas, for simplicity we omit the index $\rho$.\par
We multiply \eqref{eq:moll1}$_1$ by $u^n_t(K_0+\kappa(x))^{-2}$, with $K_0>0$. We set
\be \label{H}
\ba{rl}
H_1&\hskip-0.2cm:=\displ -\int_{\Omega}\mathbb J_n(u^n-\varphi_\rho V^n)\cdot \n u^n\cdot u^n_t(K_0+\kappa)^{-2}\,dx\,,\VS
H_2&\hskip-0.2cm:=-\displ\int_{\Omega}\mathbb J_n(\varphi_\rho)\omega^n\times u^n\cdot u^n_t(K_0+\kappa)^{-2}\,dx\,,\VS
H_3&\hskip-0.2cm:=\displ K_0^{-2}\int_{\partial\Omega}\mathbb T(u^n,\pi^n)\cdot \nu \cdot V^n_t\,dS\,,\VS
H_4&\hskip-0.2cm:=\displ 4\int_\Omega (K_0+\kappa)^{-3}\mathbb D(u^n)\cdot u^n_t\otimes \n \kappa\,dx\,,  \VS
H_5 &\hskip-0.2cm:=\displ -2\int_\Omega (K_0+\kappa)^{-3}(\pi^n-\pi^n_0)\mathbb I\cdot u^n_t\otimes \n \kappa\,dx\,.
\ea
\ee
Recalling the regularity of $(u^n_\rho,\pi^n_\rho,\xi^n_\rho,\omega^n_\rho)$, expressed in \eqref{eq:SM-i}, we get 
\be \label{eq: Stime derivata temporale I}
\ba {l}\displ
\frac{d}{dt}\dm \mathbb D(u^n)(K_0+\kappa)^{-1}\dm_2^2+\dm u^n_t(K_0+\kappa)^{-1}\dm_2^2=\sum_{i=1}^5H_i\,.
\ea
\ee
We estimate the terms on the right-hand side. Remarking that
\[
|\mathbb J_n\ast f|\le \bigg|\mathbb J_n\ast \frac{f}{|x|}\bigg|\bigg(|x|+\frac{1}{n}\bigg)\,,
\]
we have
\[
|H_1|\le \dm \mathbb J_n({\frac{u^n-\varphi_\rho{V^n}}{|x|}})\dm_\infty\int_\Omega \bigg(|x|+\frac{1}{n}\bigg)|\n u^n||u^n_t|(K_0+\kappa(x))^{-2}\,dx\,.
\]
Since $g(x):=\frac{|x|+\frac{1}{n}}{K_0+\kappa(x)}$ is a bounded function and $\dm \mathbb J_n({\frac{u^n-\varphi_\rho{V^n}}{|x|}})\dm_\infty\le c(n,A_0)$, we conclude
\[
|H_1|\le  c(n,A_0)\dm \n u^n\dm_2\dm u^n_t(K_0+\kappa)^{-1}\dm_2\,.
\]
Concerning $H_2$, we easily see that
\[
|H_2|\le  K_0^{-1}|\omega^n|\dm u^n\dm_2\dm u^n_t(K_0+\kappa)^{-1}\dm_2\le K_0^{-1}c(A_0)\dm u^n_t(K_0+\kappa)^{-1}\dm_2\,.
\]
We estimate $H_3$ by taking into account equations \eqref{eq:moll1}$_{5,6}$. We have
\[
H_3=K_0^{-2}\big(-|\dot{\xi}^n|^2-\omega^n\times \xi^n\cdot\dot{\xi}^n-\dot{\omega}^n\cdot (I\cdot \dot{\omega}^n)-\omega^n\times (I\cdot\omega^n)\cdot\dot{\omega}^n\big)\,.
\]
By Young inequality, we have
\[
|\omega^n\times \xi^n\cdot\dot{\xi}^n|\le \vep|\dot{\xi}^n|^2+c(\omega^n\cdot( I\cdot \omega^n))|\xi^n|^2\le \vep|\dot{\xi}^n|^2+c(A_0)
\]
and, analogously
\[
|\omega^n\times (I\cdot\omega^n)\cdot\dot{\omega}^n|\le \vep\dot{\omega}^n\cdot (I\cdot \dot{\omega}^n)+c(A_0)\,.
\]
We now estimate $H_4$. We have, via the Schwarz inequality,
\[
|H_4|\le cK_0^{-2}\dm \mathbb D(u^n)\dm_2\dm u^n_t(K_0+\kappa)^{-1}\dm_2\,.
\]
Finally, being $\Delta u^n-\n\pi^n=\n\cdot\mathbb T(u^n,\pi^n)$, via \eqref{Hardy} and \eqref{eq: stima gradiente pressione}, we estimate $H_5$. We have
\[
\ba{l}
|H_5|\le cK_0^{-1}\dm \frac{\pi^n-\pi^n_0}{K_0+\kappa}\dm_2\dm u^n_t(K_0+\kappa)^{-1}\dm_2\le cK_0^{-1}\dm \n \pi^n\dm_2\dm u^n_t(K_0+\kappa)^{-1}\dm_2\VS\hskip7cm\le cK_0^{-1}(\dm \n \cdot \mathbb T\dm_2+\dm \mathbb D(u^n)\dm_2)\dm u^n_t(K_0+\kappa)^{-1}\dm_2\,.
\ea
\]
{ After increasing $\displ\sum_{i=1}^5 H_i $ on the right-hand side of \eqref{eq: Stime derivata temporale I}, employing Young's inequality, integrating in time and employing the energy estimate \eqref{Energy} for $\displ\int_0^{T_0} \dm \mathbb D(u^n)\dm_2^2\,d\tau$, we get
\be \label{DT-1INT}
\int_0^{T_0}\big(\dm u^n_t(K_0+\kappa(x))^{-1}\dm_2^2+|\dot{\xi}^n|^2+| \dot{\omega}^n\cdot (I\cdot \dot{\omega}^n)|\big)\,dt\le c(n,A_0,T_0)+\vep\int_0^{T_0}\dm \n \cdot\mathbb T(u^n,\pi^n)\dm_2^2\,dt\,,
\ee
with $\vep<1$. Adding \eqref{RG} to \eqref{DT-1INT}, we obtain
\be \label{eq: DT-i}
\int_0^{T_0} \big(\dm u^n_t(K_0+\kappa(x))^{-1}\dm_2^2+|\dot{\xi}^n|^2+| \dot{\omega}^n\cdot (I\cdot \dot{\omega}^n)|\big)\,dt\le c(n,A_0,T_0)
\ee
and this concludes the proof of the lemma.}\ep
The previous estimates lead to the following result.
\begin{tho} \label{thm: esistenza-II}
{\sl     In the assumptions of Theorem\,\ref{thm: Esistenza-moll-1step}, we get a solution $(u^n,\pi^n,\xi^n,\omega^n)$ to \eqref{eq:moll}, a. e. in $(0,T_0)\times\Omega$, such that
    \be \label{SOLM-I-i}
\ba{c}
u^n\in L^\infty(0,T_0;\mathcal V(\Omega))\cap L^2(0,T_0; W^{2,2}(\Omega))\,,\VS
\xi^n,\omega^n\in W^{1,2}(0,T_0)\,,\VS
\n \pi^n\in L^2(0,T_0;L^2(\Omega))\,, \VS
u^n_t(K_0+\kappa(x))^{-1}\in L^2(0,T_0;L^2(\Omega))\,.
\ea
    \ee
Moreover, we deduce
\begin{equation} \label{GSS II-moll}
    \begin{array}{c}
       \xi^n,\omega^n \in C[0,T_0]\,, \,\, \xi^n(0)=\xi_0\,, \,\, \omega^n(0)=\omega_{0}\,, \VS
     u^n\in C([0,T_0]; W^{1,2}(\Omega_R))\,, \,\,\text{for all }R>\text{diam}(\mathcal{B})\,,  \,\, u^n(0,\cdot)=u_0(\cdot)\,.
    \end{array}
\end{equation}}
\end{tho}
\bp
 The proof follows from Lemmas\,\ref{le: RE}-\ref{le: DT-peso}. In fact, since estimates \eqref{Energy}, \eqref{RG} and \eqref{TD}, have been performed on the family of solutions $(u^n_\rho,\pi^n_\rho,\xi^n_\rho,\omega^n_\rho)$ to \eqref{eq:moll1}, we deduce for such solutions the bounds \eqref{SOLM-I-i} and \eqref{GSS II-moll}, uniformly in $\rho$.
Therefore, letting $\rho\to\infty$, we deduce that $(u^n,\pi^n,\xi^n,\omega^n)$ is a solution to \eqref{eq:moll}, a.e. in $[0,T_0]\times\Omega$, satisfying \eqref{SOLM-I-i} and \eqref{GSS II-moll}. The theorem is proved.
\ep
{In accord with what we did for the fields $(u^n_\rho,V^n_\rho)$, setting $V^n(t,x)=\xi^n(t)+\omega^n(t)\times x$, we will assume the fields $(u^n,V^n)$ to be extended to the whole space by setting
\be \label{Prol-n}
u^n|_{\mathcal B}=V^n\,.
\ee
\subsection{Uniqueness of a regular solution to \eqref{eq:moll}}
{ In order to get the uniqueness of the regular solution to problem \eqref{eq:moll} established in Theorem\,\ref{thm: esistenza-II}, the weighted integrability property for the time derivative of the kinetic field \eqref{SOLM-I-i}$_4$ is not sufficient. As it holds for the original problem (see \cite{Mar-Pal}), a further condition is needed, namely 
\be \label{WIP}
\n u^n\in L^2(0,T_0;L^2(\Omega,|x|))\,.
\ee
We obtain the validity of property \eqref{WIP} as soon as we consider an initial datum $(u_0,\xi_0,\omega_0)\in (\mathcal V(\Omega)\cap \widehat W^{1,2}(\Omega))\times\mathbb R^3\times\R^3$.}
\begin{lemma} \label{le: peso-rot}
 {\sl Let $(u^n,\pi^n,\xi^n,\omega^n)$ be the solution to \eqref{eq:moll} deduced in Theorem\,\ref{thm: esistenza-II}. Assume that $\n u_0\in L^2(\Omega,|x|)$. Then, for all $t\in [0,T_0]$, 
    \be \label{int-peso}
     \dm |x|\n u^n(t)\dm_2^2 + \int_0^{t}\dm |x| D^2 u^n(\tau)\dm_2^2\,d\tau \le c(n,A_0,\dm |x|\n u_0\dm_2)\,.
    \ee}
\end{lemma}
\bp
The lemma is proved in a similar way to that employed in \cite[Lemma 2]{Mar-Pal}. We set $b^n(t,x):=\rot u^n(t,x)$.
Given $\text{diam}(\mathcal B)<R_0<R$ and two smooth cut-off functions $\varphi_1$ and $\varphi_2$, with $\varphi_1(x)=1$ if $|x|\le R_0$, $\varphi_1(x)=0$ if $|x|\ge 2R_0$, $\varphi_2(x)=1$ if $|x|\le R$ and $\varphi_2(x)=0$ if $|x|\ge 2R$, we define
\[
\Theta(x):=(1-\varphi_1(x))\varphi_2(x)\,.
\]
We multiply \eqref{eq:moll}$_1$ by $B(\tau,x):=\rot(\Theta^2(x)|x|^2b^n(\tau,x))$ and integrate over $(0,t)\times \Omega$, for all $t\in[0,T_0]$. \par We estimate the terms that involve the mollification:
\[
\ba{l} \displ
\int_0^t\big(\mathbb J_n(u^n-V^n)(\tau)\cdot \n u^n(\tau), B(\tau,x)\big)\,d\tau=\int_0^t\big(\mathbb J_n(u^n)(\tau)\cdot \n u^n(\tau), B(\tau,x)\big)\,d\tau\VS\hskip7.4cm-\int_0^t\big(\mathbb J_n(V^n)(\tau)\cdot \n u^n(\tau), B(\tau,x)\big)\,d\tau\,.
\ea
\]
By virtue of Lemma\,\ref{le: Rotore termine convettivo-u}, we have
\[
\ba{l}\displ
\int_0^t\big(\mathbb J_n(u^n)(\tau)\cdot \n u^n(\tau), B(\tau,x)\big)\,d\tau=\int_0^t\big(\mathbb J_n(u^n)(\tau)\cdot \n b^n(\tau), \Theta^2|x|^2b^n(\tau))\,d\tau \VS \hskip9cm+\int_0^t\big(v(\tau), \Theta^2|x|^2b^n(\tau)\big)\,d\tau\,,
\ea
\]
where the components of the vector field $v(\tau)$ are given by
\[
v^i(\tau)=\vep_{ijk}<(\n u^n(\tau))_k,(\n\mathbb J_n(u^n)(\tau))^j>\,.
\]
Hence, by virtue of H\"{o}lder's inequality, elementary properties of mollifiers and Young's inequality, we arrive at
\[
\Bigg|\int_0^t\big(\mathbb J_n(u^n)(\tau)\cdot \n b^n(\tau), \Theta^2|x|^2b^n(\tau))\,d\tau \Bigg|\le\int_0^t\big( \vep \dm \Theta|x|\n b^n\dm_2^2+c\dm u^n\dm_\infty^2\dm \Theta|x|b^n\dm_2^2\big)\,d\tau\,,
\]
while, invoking Lemma\,\ref{le: peso-prel}, we get
\[
\Bigg|\int_0^t\big(v, \Theta^2|x|^2b^n(\tau)\big)\,d\tau\Bigg|\le \int_0^t c(n)\dm u^n\dm_2\dm \Theta|x|b^n\dm_2^2\,d\tau\,.
\]
Recalling Lemma\,\ref{le: rotore termine convettivo-V} and remarking that $$-\rot (\omega^n\times u^n)=\omega^n\cdot\n u^n\,,\quad \omega^n\times b^n\cdot b^n=0\,,$$ for the mollified term involving the rigid motion $V^n$ we only need to estimate
\[
\int_0^t(\mathbb J_n(V^n)\cdot \n b^n,\Theta^2|x|^2b^n)\,d\tau\,.
\]
Using the divergence theorem, we get
\[
\int_0^t(\mathbb J_n(V^n)\cdot \n b^n,\Theta^2|x|^2b^n)\,d\tau=-\frac12\int_0^t(\mathbb J_n(V^n), |b^n|^2\n (\Theta^2|x|^2))\,.
\]
We recall that
\[
\int_0^t(\mathbb J_n(V^n), |b^n|^2\n (\Theta^2|x|^2))=\int_0^t(\xi^n, |b^n|^2\n (\Theta^2|x|^2))\,d\tau+\int_0^t(\mathbb J_n(V_2^n), |b^n|^2\n (\Theta^2|x|^2))\,d\tau\,.
\]
Concerning the first integral on the right-hand side, analogously to \cite{Mar-Pal}, we get
\[
\Bigg|\int_0^t(\xi^n, |b^n|^2\n (\Theta^2|x|^2))\,d\tau\Bigg|\le \int_0^t \big[c|\xi^n|^2\dm \Theta|x|b^n\dm_2^2+c\dm b^n\dm_2^2\big]\,d\tau\,.
\]
Concerning the last integral, we have
\[
\int_0^t(\mathbb J_n(V_2^n), |b^n|^2\n (\Theta^2|x|^2))\,d\tau=2 \int_0^t\big[(\mathbb J_n(V_2^n), |b^n|^2\Theta\cdot \n \Theta|x|^2)+(\mathbb J_n(V_2^n), |b^n|^2\Theta^2x)\big]\,d\tau\,.
\]
Therefore, recalling that, being $\Theta$ a cut-off function, $\n\Theta$ has compact support, we have
\[
\ba{l}\displ
\Bigg|\int_0^t (\mathbb J_n(V_2^n), |b^n|^2\Theta\cdot \n \Theta|x|^2)\,d\tau\Bigg|\le \int_0^tc|\omega^n|\dm \Theta|x|b^n\dm_2\dm |x|b^n\dm_{L^2(\text{supp}(\n\Theta))}\,d\tau\VS\hskip6.5cm\le \int_0^t\big[ c|\omega^n|^2\dm \Theta|x|b^n\dm_2^2+c\dm b^n\dm_2^2\big]\,d\tau\,.
\ea
\]
Finally, we have
\[
\Bigg|\int_0^t (\mathbb J_n(V_2^n), |b^n|^2\Theta^2x)\,d\tau\le \int_0^t c|\omega^n|\dm \Theta |x|b^n\dm_2^2\,d\tau
\]
We conclude that
\[
\Bigg|\int_0^t(\mathbb J_n(V^n)\cdot \n b^n,\Theta^2|x|^2b^n)\,d\tau\Bigg|\le \int_0^t \big[c(|\xi^n|^2+|\omega^n|+|\omega^n|^2)\dm \Theta|x|b^n\dm_2^2+c\dm b^n\dm_2^2\big]\,d\tau\,.
\]
For the remaining terms, a detailed discussion is provided in \cite[Lemma\,2]{Mar-Pal}. Hence, we arrive at
\[
\dm \Theta|x|b^n(t)\dm_2^2+\int_0^t \dm \Theta|x|\n b^n(\tau)\dm_2^2\,d\tau\le \dm \Theta|x|b^n(0)\dm_2^2+\int_0^t \varphi(\tau)\dm \Theta|x|b^n(\tau)\dm_2^2\,d\tau\,,
\]
with $\varphi\in L^1(0,T_0)$, and we conclude the proof. 
\ep
\begin{lemma} \label{le: SIP}
 {\sl   In the assumptions of Lemma\,\ref{le: peso-rot}, we get
    \be \label{eq: SIP}
    u^n_t\in L^2(0,T_0,L^2(\Omega))\,.
    \ee}
\end{lemma}
\bp
We multiply \eqref{eq:moll}$_1$ by $\varphi_R^2 u^n_t$, where $\varphi_R$ is a smooth cut-off function such that $\varphi_R(x)=1$ if $|x|\le R$ and $\varphi_R(x)=0$ if $|x|\ge 2R$, with $R\ge \text{diam}(\mathcal B)$. Recalling \eqref{Prol-n}, we obtain
\[
\ba {l} \displ
 \sfrac{d}{dt}\dm\varphi_R\mathbb{D}(u^n)\dm_{2}^2+|{\dot{\xi}^n}|^2+( I\cdot \dot{\omega}^n)\cdot\dot{\omega}^n+\norm{\varphi_Ru^n_t}_{2}^2=-\omega^n\times\xi^n\cdot \dot{\xi}^n-\omega^n\times(I\cdot \omega^n)\cdot\dot{\omega}^n\VS\hskip1.4cm-(\mathbb J_n( u^n)\cdot\n u^n, \varphi_R^2 u^n_t) +(\mathbb J_n (V^n)\cdot\n u^n,\varphi_R^2 u^n_t)-(\omega^n\times u^n, \varphi_R^2 u^n_t) - (\mathbb{T}\cdot \varphi_R\n \varphi_R, u^n_t)\,.
\ea
\]
 Concerning the mollified terms, we have
\be \label{SIPDT-i} \ba{l}
|(\mathbb J_n( u^n)\cdot\n u^n, \varphi_R^2 u^n_t)|\le \dm \mathbb J_n( u^n)\dm_\infty\dm \varphi_R \n u^n\dm_2\dm \varphi_R u^n_t\dm_2 \VS\hskip6cm\le c(n)(\dm u^n\dm_2+|\xi^n|+|\omega^n|)\dm \varphi_R \n u^n\dm_2\dm \varphi_R u^n_t\dm_2\,.\ea
\ee
For the term $(\mathbb J_n (V^n)\cdot\n u^n,\varphi_R^2 u^n_t)$, let us point out that
\be  \label{eq: MRT}
\mathbb J_n (V_2^n)=\int_{B(0,\frac{1}{n})}J_{\frac{1}{n}}(z)\omega^n\times (x-z)\,dz=\omega^n\times x-\int_{B(0,\frac{1}{n})}J_{\frac{1}{n}}(z)\omega^n\times z\,dz\,.
\ee
Moreover, we remark that
\be \label{eq: RRT}
\Bigg|\int_{B(0,\frac{1}{n})}J_{\frac{1}{n}}(z)\omega^n\times z\,dz\Bigg|\le c|\omega^n|\,.
\ee
Hence, we deduce
\be \label{SIPDT-ii}
|(\mathbb J_n (V^n)\cdot\n u^n,\varphi_R^2 u^n_t)|\le c(|\xi^n|+|\omega^n|)\dm \varphi_R\n u^n\dm_2\dm \varphi_Ru^n_t\dm_2+|\omega^n|\dm \varphi_R |x|\n u^n\dm_2\dm \varphi_R u^n_t\dm_2\,.
\ee
The terms that do not involve the mollification are actually discussed as made in \cite{Mar-Pal}. Hence, we obtain:
\be \label{SIPDT-iii} \hskip-0.56cm
\begin{array}{rl}
|(\omega^n\times u^n, \varphi_R^2u^n_t)|\hskip-0.2cm&\le c\abs{\omega^n}\norm{\varphi_Ru^n}_2\norm{\varphi_Ru^n_t}_2 \le \vep \norm{\varphi_Ru^n_t}_2^2+ C(\vep)(\abs{\xi_0}^2\!+\!\abs{\omega_0}^2\!+\!\norm{u_0}_2^2)^2\,,\VS
|{\omega^n\times\xi^n\cdot \dot{\xi}^n }|\hskip-0.2cm&\le \vep |\dot{\xi}^n|^2+C(\vep)(\abs{\xi_0}^2+\abs{\omega_0}^2+\norm{u_0}_2^2)^2\,,\VS
\abs{\omega^n\times(I\cdot \omega^n)\cdot\dot{\omega}^n} \hskip-0.2cm&\le \vep \abs{\dot{\omega}^n}^2+C(\vep)(\abs{\xi_0}^2+\abs{\omega_0}^2+\norm{u_0}_2^2)^2\,,\VS
\displ \lim_{R\to\infty}|(\mathbb{T}\cdot \varphi_R\n \varphi_R, u^n_t)|\hskip-0.2cm&=0\,.
\end{array}
\ee
Estimates \eqref{SIPDT-i}, \eqref{SIPDT-ii} and \eqref{SIPDT-iii} allow us to conclude the proof analogously to \cite[Theorem 1]{Mar-Pal}.
\ep
\begin{rem}
    {\rm In order to prove the uniqueness of a regular solution to problem \eqref{eq:moll}, properties \eqref{SOLM-I-i}$_4$ and \eqref{WIP} are sufficient. However, for the sake of the simplicity in the proof of the following theorem, we found it worth to prove also Lemma\,\ref{le: SIP}. Finally, since problem \eqref{eq:moll} is characterized by a mollification of the convective term, we propose the proof of the following uniqueness theorem. The proof is essentially analogous to that of the uniqueness theorem for problem \eqref{eq:model} given in \cite{Mar-Pal}.}
\end{rem}
\begin{tho}\label{thm: uniqueness}
 {\sl   Under the assumptions of Lemma\,\ref{le: peso-rot}, the solution is unique in the class provided by Theorem\,\ref{thm: esistenza-II}.}
\end{tho}
\bp
By virtue of Lemma\,\ref{le: peso-rot}, we get
\be \label{SC-U}
|x|\n u^n\in L^2(0,T_0;L^2(\Omega))\,.
\ee
Let $(u^n,\pi^n,\xi^n,\omega^n)$ and $(\overline u^n,\overline\pi^n,\overline \xi^n,\overline\omega^n)$ be two solutions to \eqref{eq:moll} corresponding to the same initial datum $(u_0,\xi_0,\omega_0)\in(\mathcal V(\Omega)\cap\widehat W^{1,2}(\Omega,|x|))\times \mathbb R^3\times\mathbb R^3$, with $(u^n,\pi^n,\xi^n,\omega^n)$ satisfying \eqref{SOLR-I} and \eqref{SC-U}, while $(\overline u^n,\overline\pi^n,\overline \xi^n,\overline\omega^n)$ satisfies the energy relation. We set
\[
\widehat u^n:=u^n-\overline u^n\,,\quad \widehat \pi^n:=\pi^n-\overline \pi^n\,,\quad \widehat \xi^n:=\xi^n-\overline \xi^n\,,\quad \widehat\omega^n:=\omega^n-\widehat\omega^n\,.
\]
Then, $(\widehat u^n,\widehat \pi^n,\widehat \xi^n,\widehat \omega^n)$ satisfies
\begin{equation} \label{eq:differenza} \small\small
\begin{cases}
    \widehat {u}^n_t=\nabla\cdot\mathbb{T}(\widehat {u}^n,\widehat {\pi}^n)-\mathbb J_n(\widehat {u}^n\!-\!\widehat {V}^n)\cdot\n u^n\!-\!\mathbb J_n(\overline u^n\!-\!\overline V^n)\cdot \n \widehat {u}^n \!-\!\widehat {\omega}^n\!\times u^n\!-\overline{\omega}^n\!\times \widehat {u}^n\,,\\ 
\nabla\cdot \widehat {u}^n=0\,,\quad \forall(t,x)\in (0,T_0]\times\Omega\,,\\
\widehat {u}^n(t,x)=\widehat {\xi}^n(t)+\widehat {\omega}^n(t)\times x\,, \quad \forall(t,x)\in (0,T_0]\times\partial\Omega\,,\\
 \lim_{\abs{x}\to \infty} \widehat {u}^n(t,x)=0 \,,\quad \forall t\in (0,T_0]\,, \\
        \dot{\widehat {\xi}}^n + \widehat {\omega}^n \times \xi^n+\overline{\omega}^n\times\widehat{\xi}^n +  \int_{\partial\Omega} \mathbb{T}(\widehat {u}^n, \widehat {\pi}^n)\cdot \nu=0 \,,\quad \forall t\in (0,T_0]\,,\\
         I \cdot \dot{\widehat {\omega}}^n+ \widehat {\omega}^n \times(I\cdot  \omega^n)+\overline{\omega}^n\times (I\cdot \widehat{\omega}^n)+ \int_{\partial\Omega} x \times \mathbb{T}(\widehat {u}^n,\widehat {\pi}^n) \cdot \nu=0 \,,\quad \forall t\in (0,T_0]\,,\\
         \widehat {\xi}^n(0)= 0\,, \quad \widehat {\omega}^n(0)=0\,, \\
         \widehat {u}^n(0,x)=0\,, \quad \forall x \in \Omega\,.
    \end{cases}
\end{equation}
By virtue of Lemma\,\ref{le: SIP}, we multiply \eqref{eq:differenza}$_1$ by $\widehat {u}^n$ and integrate over $\Omega$. We get
\[
\ba{l}
\sfrac{1}{2}\sfrac{d}{dt}(\norm{\widehat {u}^n}_2^2+|{{\widehat {\xi}^n}}|^2+( I\cdot \widehat {\omega}^n)\cdot\widehat {\omega}^n)+\norm{\mathbb{D}(\widehat {u}^n)}_2^2=\xi^n\times\widehat {\omega}^n\cdot\widehat\xi^n+(I\cdot\widehat{\omega}^n)\times\overline{\omega}^n\cdot\widehat\omega^n\VS \hskip4cm-(\mathbb J_n(\widehat {u}^n-\widehat {V}^n)\cdot\n u^n,\widehat {u}^n)-(\mathbb J_n(\overline u^n-  \overline V^n)\cdot \n \widehat u^n,\widehat u^n) -(\widehat {\omega}^n\times u^n,\widehat {u}^n).
\ea
\]
We estimate the terms that involve the mollification. By H\"{o}lder, Sobolev, interpolation and Young inequalities and employing elementary properties of mollifiers, we arrive at
\be \label{EES-U}
\ba {rl}
|((\mathbb J_n(\widehat {u}^n)\cdot \n u^n, \widehat u^n)|&\hskip-0.2cm\le \vep \dm \n \widehat u^n\dm_2^2+ c\dm \n u^n\dm_2^4\dm \widehat u^n\dm_2^2\,,\VS
|((\mathbb J_n(\widehat {V}^n)\cdot \n u^n, \widehat u^n)|&\hskip-0.2cm\le c(|\widehat\xi|^2+(I\cdot\widehat\omega^n)\cdot\widehat\omega^n)+c(\dm |x|\n u^n\dm_2^2+\dm \n u^n\dm_2^2)\dm \widehat u^n\dm_2^2\,.
\ea
\ee
Finally, integration by parts and Lemma\,\ref{le:tracce-distribuzioni} yield to
\[
|(\mathbb J_n(\overline u^n-  \overline V^n)\cdot \n \widehat u^n,\widehat u^n)|=\Bigg|\frac12\int_{\partial\Omega}\mathbb J_n(\overline u^n-\overline V^n)\cdot \nu |\widehat V^n|^2\,dS\Bigg|\le c(\dm \overline u^n\dm_2+|\overline\xi^n|+|\overline\omega^n|)(|\widehat \xi|^2+(I\cdot\widehat\omega^n)\cdot\widehat\omega^n)\,.
\]
Concerning the terms that do not involve the mollification, we get the same estimates as \cite[Theorem\,3]{Mar-Pal}, we briefly recall them here:
\[
\ba {rl}
|{(\widehat {\omega}^n\times u^n,\widehat {u}^n)}|\hskip-0.2cm &\le c|{\widehat {\omega}^n}|\dm{u^n}\dm_2\dm{\widehat {u}^n}\dm_2\le c\widehat {\omega}^n\cdot(I\cdot\widehat\omega^n)+c\dm{u^n}\dm_2^2\dm{\widehat {u}^n}\dm_2^2\,, \VS
|\xi^n\times\widehat {\omega}^n\cdot\widehat\xi^n|\hskip-0.2cm&\le c(|\widehat\xi^n|^2+\widehat {\omega}^n\cdot(I\cdot\widehat\omega^n))\,, \VS
|(I\cdot\widehat {\omega}^n)\times\overline {\omega}^n\cdot\widehat\omega^n|\hskip-0.2cm&\le c\widehat {\omega}^n\cdot(I\cdot\widehat\omega^n)\,.
 \ea\]
Hence, setting $f(t):=\dm \widehat u^n(t)\dm_2^2+|\widehat\xi^n(t)|^2+(I\cdot\widehat \omega^n(t))\cdot\widehat\omega^n(t)$, we get
\[
f'(t)\le a(t)f(t)\,,
\]
with $a\in L^1(0,T_0)$.
Employing Gronwall Lemma, we conclude the proof.
\ep
We can now prove the following result.
\begin{tho} \label{thm: Esistenza-moll}
{\sl Under the assumptions of Theorem\,\ref{thm: Esistenza-moll-1step} and assuming also $u_0\in \widehat W^{1,2}(\Omega,|x|) $, we get a unique solution $(u^n,\pi^n, \xi^n,\omega^n)$ to \eqref{eq:moll}, a. e. in $(0,T_0)\times\Omega$, such that
\begin{equation} \label{eq:GSS I-moll}
\begin{aligned}
& u^n\in C([0,T_0];\mathcal V(\Omega))\cap  L^2(0,T_0; W^{2,2}(\Omega))\,,\\
& \xi^n,\omega^n \in W^{1,2}(0,T_0)\cap C([0,T_0])\,, \\
& \nabla\pi^n \in L^2(0,T_0; L^2(\Omega))\,, \\
& u^{n}_t \in L^2(0,T_0; L^2(\Omega))\,,\,\, \text{for a suitable }K_0>0\,.
\end{aligned}
\end{equation}
Moreover,
\be \label{SOLM-ii}
|x|\n u^n(t) \in L^2(\Omega)\,,\, \text{for all }t\in[0,T_0]\,,\,\text{and}\,\,|x|D^2u^n\in L^2(0,T_0;L^2(\Omega))\,.
\ee
The solution verifies the following energy equality:
\be\label{ENEQ}
\ba{l}\displ\hskip-0.25cm
\dm u^n(t)\dm_2^2+|\xi^n(t)|^2+\omega^n(t)\cdot (I\cdot\omega^n(t))+2\int_0^t\dm \n u^n(\tau)\dm_2^2\,d\tau=\dm u_0\dm_2^2+|\xi_0|^2+\omega_0\cdot(I\cdot\omega_0)\VS\hskip3cm-\int_0^t\int_{\partial\Omega} \nu\cdot \mathbb J_n(u^n(\tau,\sigma)-V^n(\tau,\sigma))|V^n(\tau,\sigma)|\,dS\,d\tau,\,\text{for all }t\in(0,T_0]\,.
\ea
\ee}
\end{tho}
\bp
{ Existence and uniqueness of the solution $(u^n,\pi^n,\xi^n,\omega^n)$, as long as properties \eqref{eq:GSS I-moll} and \eqref{SOLM-ii} follow immediately from Theorem\,\ref{thm: esistenza-II}, Lemmas\,\ref{le: peso-rot}-\ref{le: SIP} and Theorem\,\ref{thm: uniqueness}. \par Concerning the energy equality, it can be easily obtained by multiplying equation \eqref{eq:moll}$_1$ by $u^n$ and integrating on $(0,t)\times\Omega$. The Theorem is proved. }
\ep 
{ Without losing the generality,  we are going to define the following extensions of the fields $(u^n,V^n)$:
\[
\widetilde u^n(t,x):=\begin{cases}
    u^n(t,x)\,,\quad \text{if }x\in\overline\Omega\,,\\
    V^n(t,x)\,, \quad \text{if }x\in\mathcal B\,,
\end{cases}
\]
\[
\widetilde V^n(t,x):= V^n(t,x)\,,\quad \text{for all }x\in \mathbb R^3\,.
\]
We state the following result.
\begin{lemma} \label{le: conv-ext}
   {\sl For all $\widetilde\psi\in H(\R^3)$ the sequence $\{(\widetilde u^n(t),\widetilde\psi)_1\}_n$ is uniformly bounded in $n\in\N$ and equicontinuous in $t\in [0,T_0]$.}
\end{lemma}}
\bp
{ We verify the claim for all $\widetilde\psi\in\mathcal K_0(\R^3)$ and then conclude by a density argument. We recall that a function $\widetilde \psi \in \mathcal K_0(\R^3)$ is such that
\[
\widetilde \psi(x)= \overline\psi_1+\overline\psi_2\times x\,,\,\,\text{in a neighborhood of }\mathcal B\,.
\]
Let $0\le s\le t\le T_0$. We have
\[
\ba{l}\displ
(\widetilde u^n(t),\widetilde \psi)_1\!-\!\!(\widetilde u^n(s),\widetilde \psi)_1\!=\!\int_\Omega\! (u^n(t)\!-\!u^n(s))\cdot \widetilde\psi\,dx\!+\!(\xi^n(t)\!-\!\xi^n(s))\!\cdot\!\overline \psi_1\!+ (I\!\cdot\! (\omega^n(t)\!-\!\omega^n(s)))\!\cdot\overline\psi_2\VS\hskip0.7cm=-2\int_s^t\int_\Omega \mathbb D(u^n)\cdot\mathbb D(\widetilde \psi)\,dx\,d\tau-\int_s^t\int_\Omega \mathbb J_n(\widetilde u^n-\widetilde V^n)\cdot \n u^n\cdot\widetilde\psi\,dx\,d\tau-\int_s^t\int_\Omega\omega^n\times  u^n\cdot \widetilde\psi\,dx\,d\tau\VS \hskip9cm -\int_s^t(\overline\psi_1\cdot \xi^n\times \omega^n+\overline\psi_2\cdot (I\cdot\omega^n)\times\omega^n)\,d\tau\,.
\ea
\]
Hence, we get
\be \label{eq: Stima AA}
\ba{l}\displ
|(\widetilde u^n(t),\widetilde \psi)_1\!-\!(\widetilde u^n(s),\widetilde \psi)_1|\le 2\dm \mathbb D(\widetilde\psi)\dm_2\!\int_s^t\!\dm \mathbb D(u^n)\dm_2\,d\tau \VS\hskip2cm+\dm\widetilde \psi\dm_\infty\!\int_s^t\!\dm \widetilde u^n\!-\!\widetilde V^n\dm_{\null_{L^2(\text{supp}\,\widetilde\psi)}}\dm \n u^n\dm_{\null_{L^2(\text{supp}\,\widetilde\psi)}}+\dm \widetilde\psi\dm_2\int_s^t|\omega^n|\dm u^n\dm_2\,d\tau\VS\hskip7cm+c(|\overline \psi_1|+|\overline\psi_2|) \int_s^t|\omega^n|(|\omega^n|+|\xi^n|)\,.
\ea
\ee
Since $\widetilde \psi$ has compact support and $\dm \widetilde V^n(\tau)\dm_{ L^2(K)}\le c(K)(|\xi^n(\tau)|+|\omega^n(\tau)|)$, for all compact $K\subset\mathbb R^3$, then we have
\be \label{eq: Stima AA convettivo}
\dm\widetilde u^n-\widetilde V^n\dm_{\null_{L^2(\text{supp}\,\widetilde\psi)}}\le (\dm u^n\dm_2+\dm \widetilde V^n\dm_{\null_{L^2(\text{supp}\,\widetilde\psi)}})\le c(\text{supp}\,\widetilde \psi)(\dm u^n\dm_2+|\xi^n|+|\omega^n|)\,.
\ee
Moreover, multiplying equation \eqref{eq:moll}$_1$ by $u^n$ and integrating over $\Omega$, we get 
\[
\frac{1}{2}\frac{d}{dt}(\dm u^n(\tau)\dm_2^2+|\xi^n(\tau)|^2+\omega^n(\tau)\cdot (I\cdot \omega^n(\tau)))+2\dm \mathbb D(u^n)\dm_2^2=-\frac12 \int_{\partial\Omega} \mathbb J_n(u^n-V^n)\cdot \nu |V^n|^2\,dS\,.
\]
Hence, for all $t\in[0,T_0]$, we get the following bound:
\be \label{EBound}
\dm u^n(t)\dm_2^2+|\xi^n(t)|^2+\omega^n(t)\cdot( I\cdot \omega^n(t)))+4\int_0^{t}\dm \mathbb D(u^n)(\tau)\dm_2^2\,d\tau\le A_0\bigg(1+\frac{3}{c}\bigg)
\ee
Therefore, using Schwarz inequality, \eqref{eq: Stima AA convettivo} and the uniform bound \eqref{EBound}, we obtain
\[
\ba{l}
|(\widetilde u^n(t),\widetilde \psi)_1\!-\!(\widetilde u^n(s),\widetilde \psi)_1|\le c(A_0+A_0^{\frac12})(\dm\mathbb D(\widetilde\psi)\dm_2(t-s)^{\frac{1}{2}}+\dm \widetilde\psi\dm_\infty(t-s)^{\frac{1}{2}}\VS \hskip7cm+\dm \widetilde\psi\dm_2(t-s)+(|\overline\psi_1|+|\overline\psi_2|)(t-s))\,.
\ea
\]
Hence, the sequence $\{(\widetilde u^n(t),\widetilde \psi)_1\}_{n\in \N}$ is equicontinuous in $t\in[0,T_0]$, for all $\widetilde\psi \in \mathcal K_0(\R^3)$. Using a density argument, we can extend the equicontinuity for all $\widetilde \psi\in H(\R^3)$. Concerning the uniform boundedness of $\{(\widetilde u^n(t),\widetilde \psi)_1\}$, it follows from the uniform bound \eqref{EBound}. The lemma is completely proved.}
\ep
\begin{rem} \label{rem:cv-H}
    {\rm  The uniform boundedness and equicontinuity of the sequence \( \{(\widetilde u^n(t),\widetilde \psi)_1\}_{n\in\N}\), with $t\in [0,T_0]$ and $\widetilde \psi\in H(\R^3)$, ensures, by virtue of the Ascoli-Arzel\`{a} Theorem, the existence of a subsequence, extracted from $\{(\widetilde u^n(t),\widetilde \psi)_1\}$ and still denoted by $\{(\widetilde u^n(t),\widetilde \psi)_1\}$, that converges uniformly in $[0,T_0]$ to a function $F(t,\widetilde\psi)$, that is continuous in $t\in[0,T_0]$. Moreover, since $\{\widetilde u^n(t)\}$ is a bounded sequence of $H(\R^3)$ for all $t\in [0,T_0]$, there exists a subsequence, extracted from the Ascoli-Arzel\`{a} subsequence, weakly converging in $H(\R^3)$ to a function $\widetilde u(t)$. Therefore, we get $F(t,\widetilde \psi)=(\widetilde u(t),\widetilde\psi)_1$, with $\widetilde u(t)\in H(\R^3)$, for all $t\in [0,T_0]$. In the following, we will always refer to the subsequence $\{\widetilde u^n\}$ instead of the original sequence. }
\end{rem}
{In order to prove the strong convergence for the sequence $\{\widetilde u^n\}$ in $L^2(0,T_0;L(B_R))$, by employing Corollary\,\ref{GEN-CF}, we are going to prove the convergence property 
\[
(\widetilde u^n(t),\widetilde\varphi)_1\to (\widetilde u(t),\widetilde\varphi)_1\,, \,\,\text{for all }\widetilde\varphi\in L(\R^3)\,.
\]
In fact, the following result holds.
\begin{lemma}\label{le: TF-ext}
{\sl For all $\widetilde\varphi\in L(\R^3)$, the following convergence property holds:
\[
(\widetilde u^n(t),\widetilde\varphi)_1\to (\widetilde u(t),\widetilde \varphi)_1\,,\,\,\text{for all }t\in[0,T_0]\,.
\]}
\end{lemma}
\bp
By virtue of Lemma\,\ref{le: conv-ext} and the related Remark\,\ref{rem:cv-H} the convergence holds for all test functions $\widetilde\psi\in H(\R^3)$, we show that we can extend the set of test functions. Following Lemma\,\ref{le:HD}, we have
\[
\widetilde \varphi=\widetilde\psi\oplus\n p\,,
\]
with $\widetilde\psi\in H(\R^3)$ and $\n p\in G(\R^3)$. We have
\[
(\widetilde u^n(t),\widetilde\varphi)_1=(\widetilde u^n(t),\widetilde \psi)_1+(\widetilde u^n(t),\n p)_1\,.
\]
The first term of the right-hand side converges to $(\widetilde u(t),\widetilde\psi)_1$, while, integrating by parts and using the divergence-free condition for $\widetilde u^n(t)$, we get
\[
\ba{l}\displ
(\widetilde u^n(t),\n p)_1=\int_\Omega \widetilde u^n(t)\cdot \n p\,dx-\int_{\partial\Omega}p\nu \,dS\,\cdot \xi^n(t)-\int_{\partial\Omega}px\times \nu \,dS\,\cdot \omega^n(t)\VS\hskip0.8cm= \int_{\partial\Omega} p\nu\,dS\cdot  \xi^n(t)+\int_{\partial\Omega} px\times\nu \,dS\cdot\omega^n(t)-\int_{\partial\Omega}p\nu \,dS\,\cdot \xi^n(t)-\int_{\partial\Omega}px\times \nu \,dS\,\cdot \omega^n(t)=0\,.
\ea
\]
Hence, we get
\[
(\widetilde u^n(t),\widetilde\varphi)_1=(\widetilde u^n(t),\widetilde \psi)_1\to (\widetilde u(t),\widetilde\psi)_1\,.
\]
Since $\widetilde u(t)$ is weakly divergence-free, we get
\[
(\widetilde u(t),\widetilde\psi)_1=(\widetilde u(t),\widetilde\varphi)_1\,,
\]
and finally
\[
(\widetilde u^n(t),\widetilde\varphi)_1\to (\widetilde u(t),\widetilde\varphi)_1 \,,\quad \forall \widetilde\varphi\in L(\R^3)\,.
\]
The lemma is proved.
\ep
We now prove a strong convergence property for the sequence $\{\widetilde u^n\}_n$
\begin{lemma} \label{le: SC-BR}
   {\sl $\{\widetilde u^n\}_n$ converges to $\widetilde u$ strongly in $L^2(0,T_0; L(B_R))$ for all $R>\text{diam}(\mathcal B)$.}
\end{lemma}
\bp
We know that $\{\widetilde u^n\}$ is a bounded sequence in $L^2(0,T;W(B_R))$.
Moreover, the energy relation ensures that $\{\widetilde u^n(t)\}$, for all $t$, is a bounded sequence in $L(B_R)$. We are going to prove that  $(\widetilde u^n(t),\varphi)_{1,R}\to (\widetilde u(t),\varphi)_{1,R}$, for all $\varphi\in L(B_R)$. In fact, denoting by $\mathbf{1}_{\Omega_R}$ the characteristic function of $\Omega_R$, we have
\[
\ba{l}\displ
(\widetilde u^n(t),\varphi)_{1,R}=\int_{\Omega_R }\widetilde u^n(t)\cdot \varphi\,dx+\xi^n(t)\cdot \overline\varphi_1+\omega^n(t)\cdot (I\cdot \overline\varphi_2)\VS \hskip3cm=\int_{\Omega}\widetilde u^n(t,x)\cdot \mathbf{1}_{\Omega_R}(x)\varphi(x)\,dx+\xi^n(t)\cdot \overline\varphi_1+\omega^n(t)\cdot (I\cdot \overline\varphi_2)\,.
\ea
\] 
 The function $\widetilde\varphi=\mathbf{1}_{\Omega_R}\varphi$ belongs to $L(\R^3)$ and has characteristic vectors $(\overline\varphi_1,\overline\varphi_2)$, so $\widetilde\varphi|_{\mathcal B}=\overline\varphi_1+\overline\varphi_2\times x$. Since, by virtue of Lemma\,\ref{le: TF-ext} we have
\[
(\widetilde u^n(t),\widetilde\varphi)_1\to (\widetilde u(t),\widetilde \varphi)_1\,,
\]
we get our claim. \par
By virtue of Corollary\,\ref{GEN-CF}, we conclude that $\{\widetilde u^n\}$ converges strongly to $\widetilde u$ in $L^2(0,T;L(B_R))$.
\ep
\begin{rem} \label{rem: utilde}
   {\rm The function $\widetilde u$ is a function of the kind
    \[
    \widetilde u(t,x)=\begin{cases}
        u(t,x)\,, \,\,\text{   if }x\in\Omega\,,\\
        \xi(t)+\omega(t)\times x\,,\,\,\text{if }x\in\mathcal B\,,
    \end{cases}
    \]
    with $u(t)\in \mathcal H(\Omega)$. Lemma\,\ref{le: SC-BR} ensures that
    \[
    u^n\to u\,\,\text{strongly in }L^2(0,T_0;L^2(\Omega_R))
    \]
    and
    \[
    (\xi^n,\omega^n)\to (\xi,\omega)\,\,\text{strongly in }L^2(0,T_0)\,.
    \]
    In particular, we have
    \[
    \widetilde V^n(t,x)\to \xi(t)+\omega(t)\times x\,\,\text{strongly in }L^2(0,T_0;L^2(B_R))\,,
    \]
    and
    \[
     (\xi^n(t),\omega^n(t))\to (\xi(t),\omega(t))\,\,\text{ for a. e. }t\in (0,T_0)\,.
    \]}
\end{rem}
}
\subsection{Extension of the local solution for large times}
Now, our task is to prove that for all $T>0$ there exists $n(T)\in\mathbb N$ such that for all $n\ge n(T)$ the solution $(u^n,\pi^n,\xi^n,\omega^n)$ exists for all $t\in [0,T]$. \par We start by proving the following lemma.
\begin{lemma} \label{le:convergenza integrale di superficie}
{\sl    Let $\{(u^n,\pi^n,\xi^n,\omega^n)\}$ be the sequence ensured by Theorem\,\ref{thm: Esistenza-moll}.
    Then,
    \begin{equation}\label{eq:limpro}
        \lim_{n\to \infty} \int_{0}^{T_0} \int_{\partial\Omega} \nu\cdot \mathbb{J}_n(u^n(t,\sigma)-V^n(t,\sigma)) \abs{V^n(t,\sigma)}^2 \, dS\, dt=0\,.
    \end{equation}}
\end{lemma}
\bp
Firstly, we remark that, for all compact $K\subset \mathbb{R}^3$,
\begin{equation} \label{eq:bound MR}
\sup_{(0,T_0)} \norm{\widetilde V^n(t)}_{W^{1,\infty}(K)} < Z,
\end{equation}
for some positive constant $Z=Z(K,A_0)$, independent of $\widetilde V^n$. \par We denote by $\overline B$ a closed ball such that $\mathcal B\cup \text{supp}\,\mathbb J_n\subset \overline B$, for all \par
We point out that \eqref{eq:limpro} can be rewritten as
\[
\lim_{n\to \infty} \int_{0}^{T_0} \int_{\partial\Omega} \nu\cdot \mathbb{J}_n(\widetilde u^n(t,\sigma)-\widetilde V^n(t,\sigma)) |\widetilde V^n(t,\sigma)|^2 \, dS\, dt=0\,.
\]
We define 
\[
I_1:=\int_{0}^{T_0} \int_{\partial\Omega} \nu\cdot\int_{B(0,\frac{1}{n})}J_{\frac{1}{n}}(z)(\widetilde u^n(t,\sigma-z)-\widetilde u^n(t,\sigma))\, dz |\widetilde V^n(t,\sigma)|^2\, dS\, dt\,,
\]
\[
I_2:=\int_{0}^{T_0} \int_{\partial\Omega} \nu\cdot(\widetilde u^n(t,\sigma)-\widetilde u(t,\sigma))|\widetilde V^n(t,\sigma)|^2\, dS\, dt\,,
\]
\[
I_3:=\int_{0}^{T_0} \int_{\partial\Omega} \nu\cdot(\widetilde V(t,\sigma)-\widetilde V^n(t,\sigma)) |\widetilde V^n(t,\sigma)|^2\, dS\, dt\,,
\]
\[
I_4:=\int_{0}^{T_0} \int_{\partial\Omega} \nu\cdot\int_{B(0,\frac{1}{n})}J_{\frac{1}{n}}(z)(\widetilde V^n(t,\sigma)-\widetilde V^n(t,\sigma-z))\, dz |\widetilde V^n(t,\sigma)|^2\, dS\, dt\,.
\]
In order to prove the limit property \eqref{eq:limpro}, we perform the following identities
\[
\begin{aligned}
&\int_{0}^{T_0} \int_{\partial\Omega} \nu\cdot \mathbb{J}_n(\widetilde u^n(t,\sigma)-\widetilde V^n(t,\sigma)) |\widetilde V^n(t,\sigma)|^2 \, dS\, dt=\\ &=\int_{0}^{T_0} \int_{\partial\Omega} \nu\cdot\int_{B(\sigma,\frac{1}{n})}J_{\frac{1}{n}}(\sigma-y)(\widetilde u^n(t,y)-\widetilde V^n(t,y))\, dy |\widetilde V^n(t,\sigma)|^2\, dS\, dt= \\
&=\int_{0}^{T_0} \int_{\partial\Omega} \nu\cdot\int_{B(0,\frac{1}{n})}J_{\frac{1}{n}}(z)(\widetilde u^n(t,\sigma-z)-\widetilde V^n(t,\sigma-z))\, dz |\widetilde V^n(t,\sigma)|^2\, dS\, dt=\sum_{i=1}^4 I_i\,.
\end{aligned} 
\]
Now, our task is to prove that each integral term $I_i$ converges to $0$ as $n\to\infty$. \par
Employing Lemma\,\ref{le:tracce-distribuzioni} and recalling estimate \eqref{eq:bound MR}, the following estimates hold for $I_2$ and $I_3$:
\[
\ba {rl}
\abs{I_2}&\hskip-0.2cm\displ\le Z\int_{0}^{T_0} \norm{u^n(t)-u(t)}_{L^2(\Omega\cap \overline B)} \, dt\,,\VS
\abs{I_3}&\hskip-0.2cm\displ\le Z\int_{0}^{T_0} \norm{\widetilde V(t)-\widetilde V^n(t)}_{L^2(\overline B)} \, dt\,.
\ea
\]
Then, by virtue of Lemma\,\ref{le: SC-BR}, we easily conclude their convergence to zero. \par
We estimate $I_1$. Employing the Fubini Theorem, estimate \eqref{eq:bound MR} and the trace inequality in Lemma\,\ref{le:tracce-distribuzioni}, we get 
\[
\ba{l}\displ
\abs{I_1}=\Bigg|\int_{0}^{T_0} \int_{B(0,\frac{1}{n})} J_{\frac{1}{n}}(z)\int_{\partial\Omega}\nu\cdot (\widetilde u^n(t,\sigma-z)-\widetilde u^n(t,\sigma))\,  |\widetilde V^n(t,\sigma)|^2\, dS\,dz\, dt\Bigg|\VS\hskip5cm\le Z\int_{0}^{T_0} \int_{B(0,\frac{1}{n})} J_{\frac{1}{n}}(z) \norm{\widetilde u^n(t,\cdot-z)-\widetilde u^n(t)}_{L^2(\Omega\cap \overline B)} \, dz \, dt\,.
\ea
\]
Setting
\[
I_{1,1}:=\int_{0}^{T_0} \int_{B(0,\frac{1}{n})} J_{\frac{1}{n}}(z) \norm{\widetilde u^n(t,\cdot-z)-\widetilde u(t,\cdot-z)}_{L^2(\Omega\cap\overline B)} \, dz \, dt\,,
\]
\[
I_{1,2}:=\int_{0}^{T_0} \int_{B(0,\frac{1}{n})} J_{\frac{1}{n}}(z) \norm{\widetilde u(t,\cdot-z)-\widetilde u(t)}_{L^2(\Omega\cap\overline B)}\, dz \, dt\,,
\]
\[
I_{1,3}:=\int_{0}^{T_0} \norm{\widetilde u(t)-\widetilde u^n(t)}_{L^2(\Omega\cap\overline B)} \, dt\,,
\]
we have
\[
\abs{I_1}\le cZ(I_{1,1}+I_{1,2}+I_{1,3})\,.
\]
It is readily seen that $\displ\lim_{n\to +\infty} I_{1,3}=0$. Concerning $I_{1,1}$, we have that
\[
\ba {rl}\displ
I_{1,1}&\hskip-0.2cm \le \displ\int_0^{T_0}\sup_{|z|<\frac{1}{n}}\biggl(\int_{\Omega\cap\overline B} \abs{\widetilde u^n(t,x-z)-\widetilde u(t,x-z)}^2\,dx\biggr)^{\frac{1}{2}} \, dt \VS &\hskip-0.2cm \le\displ \int_0^{T_0}\norm{\widetilde u^n(t)-\widetilde u(t)}_{L^2(\overline B\setminus\overline{\overline B})} \,dt,
\ea
\]
being $\overline{\overline B}\subset \mathcal B\setminus \text{supp}\,\mathbb J_n$, for all $n\in\mathbb N$. By virtue of Lemma\,\ref{le: SC-BR}, the last integral converges to $0$ as $n\to\infty$. \par Moreover, concerning $I_{1,2}$, we get
\[
I_{1,2}\le \int_0^{T_0} \sup_{\abs{z}<\frac{1}{n}} \norm{\widetilde u(t,\cdot-z)-\widetilde u(t)}_{L^2(\Omega\cap\overline B)} \,dt
\]
 Lemma\,\ref{le: proprieta spazi lebesgue} ensures that $\displ\lim_{n} \sup_{\abs{z}<\frac{1}{n}} \norm{\widetilde u(t,\cdot-z)-\widetilde u(t)}_{L^2(\Omega\cap\overline B)}=0$ and, moreover, being the $L^2$-norm of $\widetilde u$ bounded, we conclude the convergence to zero of $I_{1,2}$ by the Lebesgue Theorem of dominated convergence.
Finally, let us consider $I_4$. 
\[
\abs{I_4} \le C \int_{0}^{T_0} \sup_{\abs{z}< \frac{1}{n}}\norm{\widetilde V^n(t)-\widetilde V^n(t,\cdot-z)}_{L^2(\Omega\cap\overline B)} \, dt \le cZ(J_{1,1}+J_{1,2}+J_{1,3}), 
\]
where
\[
J_{1,1}:= \int_0^{T_0} \norm{\widetilde V^n(t)-\widetilde V(t)}_{L^2(\Omega\cap\overline B)} \, dt,
\]
\[
J_{1,2}:= \int_0^{T_0} \sup_{\abs{z}< \frac{1}{n}}\norm{\widetilde V(t)-\widetilde V(t,\cdot-z)}_{L^2(\Omega\cap\overline B)} \, dt,
\]
\[
J_{1,3}:= \int_0^{T_0} \sup_{\abs{z}< \frac{1}{n}}\norm{\widetilde V(t,\cdot-z)-\widetilde V^n(t,\cdot-z)}_{L^2(\Omega\cap\overline B)} \, dt.
\]
By the same arguments used for $I_{1,1}$, $I_{1,2}$, $I_{1,3}$, we will get that $J_{1,1}$, $J_{1,2}$, $J_{1,3}$ admit $0$ as limit letting $n\to\infty$. This completes the proof.
\ep
Now, we are in a position to prove the main result of this subsection, that is an extension theorem for the local solution $(u^n,\pi^n,\xi^n,\omega^n)$, ensured by Theorem\,\ref{thm: Esistenza-moll}, if $n$ is sufficiently large. 
{ \begin{tho} \label{thm: Prolungamento}
  {\sl  We state the existence of a divergent sequence $\{T_h\}_{h\in\mathbb N}$ and of a subsequence $\{(u^i,\pi^i,\xi^i,\omega^i)\}_{i\ge n(T_h)}$ of the sequence $\{(u^n,\pi^n,\xi^n,\omega^n)\}_{n\in\mathbb N}$, ensured by Theorem\,\ref{thm: Esistenza-moll}, whose elements are defined on $[0,T_h]$. }
\end{tho}}
\bp 
We multiply equation \eqref{eq:moll}$_1$ by $u^n$ and integrate over $\Omega$. We get the following energy equality
\be \label{eq: uguaglianza energia}
\sfrac{d}{dt}(\dm u^n(\tau)\dm_2^2\!+\!|\xi^n(\tau)|^2\!+\!(I\!\cdot\!\omega^n(\tau))\!\cdot\!\omega^n(\tau))\!+\!2\dm\n u^n(\tau)\dm_2^2=\!-\!\!\int_{\partial\Omega} \mathbb{J}_n(u^n\!-\! V^n)(\tau)\cdot \nu |V^n(\tau)|^2\, dS\,.
\ee
Then, employing Lemma\,\ref{le:Stime per rel energia} to estimate  the right-hand side of \eqref{eq: uguaglianza energia} and setting $y(t)=\dm u^n(t)\dm_2^2+|\xi^n(t)|^2+\omega^n(t)\cdot I\cdot \omega^n(t)$, we obtain the following differential inequality:
\begin{equation} \label{eq: diseq diff energia}
    y'(t)\le c y^{\frac{3}{2}}(t)\,.
\end{equation}
  Setting $A_0=y(0)$, we point out that the above inequality can be integrated on the time interval $[0,T'_0)$, where
\[
T'_0=2(cA_0^{\frac{1}{2}})^{-1}\,.
\]
In particular, setting $T_0:=\frac{1}{2}T'_0$, it can be integrated on $[0,T_0]$.\footnote{Here, the symbol $T_0$ actually denotes a quantity that is { greater} than the previous one given in Theorem\,\ref{thm: Esistenza-moll-1step}. This is due to the fact that on the right-hand side of \eqref{eq: uguaglianza energia} there is not the second integral that appears on the right-hand side of \eqref{eq:RE-rho}.} We integrate in $[0,T_0]$ the energy equality \eqref{eq: uguaglianza energia}
and we get
\be \label{eq:prima iterazione}
\ba {l} \displ
y(T_0)+2\int_0^{T_0} \dm \n u^n(\tau)\dm \,d\tau = A_0-\int_0^{T_0}\int_{\partial\Omega} \mathbb{J}_n(u^n- V^n)(\tau)\cdot \nu |V^n(\tau)|^2\, dS\,d\tau\,.
\ea
\ee
 Lemma\,\ref{le:convergenza integrale di superficie} ensures that there exists $n_0\in\mathbb{N}$ such that, for all $n\ge n_0$, 
\be \label{eq:supIiterazione}
\Bigg|\int_0^{T_0}\int_{\partial\Omega} \mathbb{J}_n(u^n- V^n)(\tau)\cdot \nu |V^n(\tau)|^2\, dS\,d\tau\Bigg|\le \frac{1}{2}A_0\,.
\ee
Set $A_1:=y(T_0)$. We deduce that
\be \label{eq: prima maggiorazione}
A_1+2\int_0^{T_0} \dm \n u^n(\tau)\dm_2^2\,d\tau \le A_0\biggl(1+\frac{1}{2}\biggr)\,,
\ee
for all $n\ge n_0$. \par Moreover, the following bounds hold:
\[
\dm \n u^n(T_0)\dm_2^2\le c(n,A_0,\dm\n u_0\dm_2)\,,\quad
\dm|x| \n u^n(T_0))\dm_2^2\le c(n,A_0,\dm |x|\n u_0\dm_2)\,,
\]
so that $(u^n(T_0),\xi^n(T_0),\omega^n(T_0))\in(\mathcal V(\Omega)\cap \widehat W^{1,2}(\Omega,|x|))\times\mathbb R^3\times\mathbb R^3$. Now, by assuming $(u^n(T_0),\xi^n(T_0),\omega^n(T_0))$ as the initial datum for problem \eqref{eq:moll}, by virtue of Theorem\,\ref{thm: Esistenza-moll}, we get a unique solution in the interval $[T_0,T'_1)\equiv [T_0,T_0+2(cA_1^{\frac{1}{2}})^{-1})$. In particular, the solution exists in $[T_0,T_1]$, where $T_1=T_0+(cA_1^{\frac{1}{2}})^{-1}$. Integrating the energy equality \eqref{eq: uguaglianza energia} on the interval $[T_0,T_1]$, we get
\[
\ba {l} \displ
y(T_1)+2\int_{T_0}^{T_1} \dm \n u^n(\tau)\dm \,d\tau = A_1-\int_{T_0}^{T_1}\int_{\partial\Omega} \mathbb{J}_n(u^n- V^n)(\tau)\cdot \nu |V^n(\tau)|^2\, dS\,d\tau\,.
\ea
\]
Lemma\,\ref{le:convergenza integrale di superficie} ensures the existence of $n_1\ge n_0$ such that, for all $n\ge n_1$
\[
\Bigg|\int_{T_0}^{T_1}\int_{\partial\Omega} \mathbb{J}_n(u^n- V^n)(\tau)\cdot \nu |V^n(\tau)|^2\, dS\,d\tau\Bigg|\le \frac{1}{4}A_0\,.
\]
By \eqref{eq: prima maggiorazione} we can infer that, for all $n\ge n_1$,
\[
A_1\le A_0\biggl(1+\frac{1}{2}\biggr)-2\int_0^{T_0}\dm \n u^n(\tau)\dm_2^2\,d\tau\,.
\]
Setting $A_2:=y(T_1)$, we conclude that
\[
\ba{l}\displ
A_2+2\int_{0}^{T_1}\dm \n u^n(\tau)\dm_2^2\,d\tau\le A_0\biggl(1+\frac{1}{2}+\frac{1}{4}\biggr)\,,\VS\dm \n u^n(T_1)\dm_2^2\le c(n,A_1,\dm\n u(T_0)\dm_2)\,,\,\,\,
\dm|x| \n u^n(T_1))\dm_2^2\le c(n,A_1,\dm |x|\n u(T_0)\dm_2)\,, \,\, \,.
\ea
\]
for all $n\ge n_1$.\par
By iterating the previous process, we construct three sequences, one is a sequence $\{n_h\}_{h\in\mathbb{N}}$ of positive integers, which is increasing, then there is
\[
\{A_h\}_{h\in\mathbb N}:=\{y(T_{h-1})\}_{h\in\mathbb N}\,,
\]
bounded, being 
\be \label{eq:iterazione h}
A_h+2\int_0^{T_{h-1}} \dm \n u^n(\tau)\dm_2^2 \, d\tau \le A_0\sum_{j=0}^h 2^{-j}\le 2A_0\,,
\ee
instead,
\be \label{eq: Th}
\{T_h\}_{h\in\mathbb N}:=\{T_{h-1}+(cA_h^{\frac{1}{2}})^{-1}\}_{h\in\mathbb N} 
\ee
divergent holds. We justify the divergence of \eqref{eq: Th}. By definition, for any $h\in\mathbb N$, and by virtue of the iterative procedure,
\[
T_h=T_0+ \sum_{j=1}^h (cA_j^{\frac{1}{2}})^{-1}\,.
\]
By virtue of \eqref{eq:iterazione h}, we get
\[
T_h \ge T_0+c^{-1}A_0^{-\frac{1}{2}}\sum_{j=1}^h \biggl(\sum_{l=0}^j2^{-l}\biggr)^{-\frac{1}{2}}\,.
\]
Since
\[
\sum_{l=0}^j 2^{-l}=\frac{1-(1/2)^{j+1}}{1-(1/2)}=2\biggl(1-\frac{1}{2^{j+1}}\biggr)\,,
\]
we get
\[
T_h \ge T_0+ c\sum_{j=1}^h \biggl(1-\frac{1}{2^{j+1}}\biggr)^{-\frac{1}{2}}\,.
\]
By the divergence of the series $\displ \sum_{j=1}^\infty \biggl(1-\frac{1}{2^{j+1}}\biggr)^{-\frac{1}{2}}$, we deduce the divergence of $\{T_h\}_{h\in\mathbb N}$ and we conclude the proof.
\ep
\subsection{Vorticity estimates for the solution to \eqref{eq:moll}}
Let $\{(u^i,\pi^i,\xi^i,\omega^i)\}_{i\ge n(T_h)}$ be the subsequence ensured by Theorem\,\ref{thm: Prolungamento} defined on the interval $[0,T_h]$. 
In this subsection, we show that the integrability assumption $\rot u_0\in L^1(\Omega)$ ensures that $\{\rot u^i(t,x)\}_{i\ge n(T_h)}\subset L^1(|x|\ge R_0+t^{\frac{1}{2}})$, with $R_0>\text{diam}(\mathcal B)$, for all $t \in [0,T_h]$. 
\par
From now on, we set $\mathfrak A:=c(\dm \rot u_0\dm_1+A_0^{\frac12}+A_0)$ and $\mathfrak B:=c\mathfrak A^{-1}\,.$\par
Our task is to prove the following theorem.
\begin{tho} \label{thm:rotoreL1}
{\sl Let $R_0$ be greater than $\text{diam}(\mathcal B)$. Assume that $\rot  u_0\in L^1(\Omega)$. Then, for all $t\in (0,T_h)$,
\be \label{eq:stima rotore}
\dm \rot u^i(t)\dm_{L^1(|x|\ge R_0+t^{\frac{1}{2}})} \le  \mathfrak A(1+t^{\frac{1}{4}})\,.
\ee}
\end{tho}
\bp
Set $\overline R:=\text{diam}(\mathcal B)$ and
let $\overline{R}<\overline R_0<2\overline R_0<R_0<R$. For all $t\in (0,T_h)$, we define
\[
\zeta(t,x)=(1-\zeta_0^2(t,x))^2\zeta_1^2(t,x)\,,
\]
where the cut-off functions $\zeta_0$ and $\zeta_1$ are defined by
\[
\zeta_0(t,x):=\begin{cases}
    1\,, &\text{if }|x|\le R_0+t^{\frac{1}{2}}\,,\\
    \frac{2(R_0+t^{\frac{1}{2}})-|x|}{R_0+t^{\frac{1}{2}}}\,, &\text{if }R_0+t^{\frac{1}{2}}\le |x|\le 2(R_0+t^{\frac{1}{2}})\,,\\
    0\,, &\text{if } |x|\ge 2(R_0+t^{\frac{1}{2}})\,,
\end{cases}
\]
and
\[
\zeta_1(t,x):=\begin{cases}
    1\,, &\text{if }|x|\le R+t^{\frac{1}{2}}\,,\\
    \frac{2(R+t^{\frac{1}{2}})-|x|}{R+t^{\frac{1}{2}}}\,, &\text{if }R+t^{\frac{1}{2}}\le |x|\le 2(R+t^{\frac{1}{2}})\,,\\
    0\,, &\text{if } |x|\ge 2(R+t^{\frac{1}{2}})\,.
\end{cases}
\]
We also consider the function $h(x):=1-\varphi_{\overline R_0}(x)$, with $\varphi_{\overline R_0}$ smooth cut-off function of radius $\overline R_0$, hence $\varphi_{\overline R_0}(x)=1$ if $|x|\le \overline R_0$ and $\varphi_{\overline R_0}(x)=0$ if $|x|\ge 2\overline R_0$. \par
Moreover, we notice that $\text{supp }\n\zeta\cap\text{supp }\n h=\emptyset$ and $\text{supp }\zeta\subset \text{supp }h$.\par
Set $b^i(\tau,x):=\rot u^i(\tau,x)$. 
We multiply \eqref{eq:moll}$_1$ by $A(\tau,x):=\rot\Big(\zeta(t,x) \frac{b^i(\tau,x)}{(\vep+|b^i(\tau,x)|^2)^{\frac{1}{2}}}\Big)$ and, for any $t\in (0,T_h)$, we integrate over $(0,t)\times\mathbb R^3$. Denoting by $J_\gamma$ a Friedrichs space mollifier, we get
\[
\ba {rl} \displ
\int_0^t\int_{\mathbb{R}^3} u^i_\tau\cdot \rot\Big(\zeta \frac{b^i}{(\vep+|b^i|^2)^{\frac{1}{2}}}\Big) \, dx\, d\tau \hskip-0.2cm&=\dm \zeta(\vep+|b^i|^2)^{\frac{1}{2}}(t)\dm_1-\dm \zeta(\vep+|b^i|^2)^{\frac{1}{2}}(0)\dm_1\,,\VS
\int_0^t\int_{\mathbb{R}^3}-\Delta u^i\cdot \rot\Big(\zeta \frac{b^i}{(\vep+|b^i|^2)^{\frac{1}{2}}}\Big)\,dx\,d\tau\hskip-0.2cm&=\displ \int_0^t\int_{\mathbb{R}^3}-\Delta (hu^i)\cdot \rot\Big(\zeta \frac{b^i}{(\vep+|b^i|^2)^{\frac{1}{2}}}\Big)\,dx\,d\tau\VS\hskip-0.2cm&=\displ\int_0^t\lim_{\gamma\to 0 }\int_{\mathbb{R}^3}-\Delta J_\gamma(hu^i) \cdot \rot\Big(\zeta \frac{b^i}{(\vep+|b^i|^2)^{\frac{1}{2}}}\Big)\,dx\,d\tau\VS \hskip-0.2cm&=\displ \int_0^t\lim_{\gamma\to 0 }\int_{\mathbb{R}^3}-\Delta J_\gamma(hb^i+\n h\times b^i) \cdot\zeta \frac{b^i}{(\vep+|b^i|^2)^{\frac{1}{2}}}\,dx\,d\tau\VS \hskip-0.2cm &=\displ \int_0^t\lim_{\gamma\to 0 }\int_{\mathbb{R}^3}\n J_\gamma(hb^i+\n h\times b^i) \cdot \n\Big(\zeta \frac{b^i}{(\vep+|b^i|^2)^{\frac{1}{2}}}\Big)\,dx\,d\tau\VS \hskip-0.2cm &=\displ \int_0^t\int_{\mathbb{R}^3}\n b^i \cdot \n\Big(\zeta \frac{b^i}{(\vep+|b^i|^2)^{\frac{1}{2}}}\Big)\,dx\,d\tau=\sum_{i=1}^3 I_i\,,
\ea 
\]
where, after straightforward computations, we set
\[
\ba {rl}\displ
I_1(t,\vep)\hskip-0.2cm&:=\displ\int_0^t \int_{\mathbb{R}^3} \n b^i\cdot \n b^i \frac{\zeta}{(\vep+|b^i|^2)^{\frac{1}{2}}}\,dx\,d\tau\,,\VS
I_2(t,\vep)\hskip-0.2cm&:=-\displ\int_0^t \int_{\mathbb{R}^3} (\n b^i\cdot  b^i)^2 \frac{\zeta}{(\vep+|b^i|^2)^{\frac{3}{2}}}\,dx\,d\tau\,,\VS
I_3(t,\vep)\hskip-0.2cm&:=\displ\int_0^t\int_{\mathbb{R}^3} \n b^i\cdot \n \zeta\otimes \frac{b^i}{(\vep+|b^i|^2)^{\frac{1}{2}}}\,dx\,d\tau\,.
\ea
\]
We point out that in the previous computations, being the function $g(\tau,x):=h(x)u^i(\tau,x)$ defined for all $x\in\mathbb R^3$, the introduction of the cut-off function $h$ is necessary by virtue of mollification properties in Sobolev spaces. Actually, the previous identities hold true by recalling that the choice of the cut-off functions $h$ and $\zeta$ is performed in a way that 
\[
\ba{rl}
\text{supp }\zeta&\hskip-0.2cm\subset \text{supp }h\,,\VS
\text{supp }\n^i\zeta&\hskip-0.2cm\subset \text{supp }h\,,\quad i=1,2\,,\VS
\text{supp }\n^i\zeta&\hskip-0.2cm\cap\, \text{supp }\n^jh=\emptyset\,,\quad i,j=1,2\,.
\ea
\]
We remark that $I_1(t,\vep)+I_2(t,\vep)\ge0$ for all $\vep>0$, while integration by parts yields to
\[
I_3(t,\vep)=-\int_0^t\int_{\mathbb{R}^3}\Delta\zeta(\vep+|b^n|^2)^{\frac{1}{2}}\,dx\,d\tau\,.
\]
We now focus on the mollified terms of the right-hand side of \eqref{eq:moll}$_1$. We have
\[
-\int_0^t\Big(\mathbb{J}_i(u^i)\cdot \n u^i,\rot\Big(\zeta\frac{b^i}{(\vep+|b^i|^2)^{\frac{1}{2}}}\Big)\Big)\,d\tau=-\int_0^t\Big(\rot(\mathbb{J}_i(u^i)\cdot \n u^i),\zeta\frac{b^i}{(\vep+|b^i|^2)^{\frac{1}{2}}}\Big)\,d\tau\,.
\]
By Lemma\,\ref{le: Rotore termine convettivo-u}, we have
\[
\ba {l}\displ
-\int_0^t\Big(\rot(\mathbb{J}_i(u^i)\cdot \n u^i),\zeta\frac{b^i}{(\vep+|b^i|^2)^{\frac{1}{2}}}\Big)\,d\tau=-\int_0^t\Big(\mathbb{J}_i(u^i)\cdot \n b^i,\zeta\frac{b^i}{(\vep+|b^i|^2)^{\frac{1}{2}}}\Big)\,d\tau\VS\hskip6.6cm-\int_0^t\Big(v,\zeta\frac{b^i}{(\vep+|b^i|^2)^{\frac{1}{2}}}\Big)\,d\tau=:-(I_4(t,\vep)+I_5(t,\vep))\,,
\ea
\]
where, by virtue of Lemma\,\ref{le: Rotore termine convettivo-u}, we recall that
\[
v^j(\tau)=\vep_{jkl}<(\n u^i(\tau))_l,(\n\mathbb J_n(u^i)(\tau))^k>\,.
\]
Considering the term $\mathbb J_i(V^i)\cdot \n u^i$, we have that
\[
\int_0^t\Big(\mathbb{J}_i(V^i)\cdot \n u^i,\rot\Big(\zeta\frac{b^i}{(\vep+|b^n|^2)^{\frac{1}{2}}}\Big)\Big)\,d\tau=\int_0^t\Big(\rot(\mathbb{J}_i(V^i)\cdot \n u^i),\zeta\frac{b^i}{(\vep+|b^i|^2)^{\frac{1}{2}}}\Big)\,d\tau\,.
\]
By Lemma\,\ref{le: rotore termine convettivo-V}, we find that
\[
\ba {l}\displ
\int_0^t\Big(\rot(\mathbb{J}_i(V^i)\cdot \n u^i),\zeta\frac{b^i}{(\vep+|b^i|^2)^{\frac{1}{2}}}\Big)\,d\tau=\int_0^t\Big(\mathbb{J}_i(V^i)\cdot \n b^i,\zeta\frac{b^i}{(\vep+|b^i|^2)^{\frac{1}{2}}}\Big)\,d\tau\VS \hskip 4cm-\int_0^t\Big(\omega^i\times b^i,\zeta\frac{b^i}{(\vep+|b^i|^2)^{\frac{1}{2}}}\Big)\,d\tau-\int_0^t\Big(\omega^i\cdot \n u^i,\zeta\frac{b^i}{(\vep+|b^i|^2)^{\frac{1}{2}}}\Big)\,d\tau\,.
\ea
\]
By elementary properties of scalar triple product, we have
\[
\int_0^t\Big(\omega^i\times b^i,\zeta\frac{b^i}{(\vep+|b^i|^2)^{\frac{1}{2}}}\Big)\,d\tau=0\,.
\]
Finally, we consider the term $\omega^i\times u^i$. We can easily say that
\[
-\int_0^t\Big(\omega^i\times u^i,\rot\Big(\zeta\frac{b^i}{(\vep+|b^i|^2)^{\frac{1}{2}}}\Big)\Big)\,d\tau=\int_0^t\Big(\omega^i\cdot \n u^i,\zeta\frac{b^i}{(\vep+|b^i|^2)^{\frac{1}{2}}}\Big)\,d\tau\,.
\]
Hence, we arrive at
\[
\ba{l}\displ
\int_0^t\Big(\mathbb{J}_i(V^i)\cdot \n u^i,\rot\Big(\zeta\frac{b^i}{(\vep+|b^i|^2)^{\frac{1}{2}}}\Big)\Big)\,d\tau-\int_0^t\Big(\omega^i\times u^i,\rot\Big(\zeta\frac{b^i}{(\vep+|b^i|^2)^{\frac{1}{2}}}\Big)\Big)\,d\tau\VS\hskip7cm=\int_0^t\Big(\mathbb{J}_i(V^i)\cdot \n b^i,\zeta\frac{b^i}{(\vep+|b^i|^2)^{\frac{1}{2}}}\Big)\,d\tau=:I_6(t,\vep)\,.
\ea
\]
Concerning the gradient of the pressure field, since $A(\tau,x)\in J^2(\Omega_R)$, we have
\[
-\int_0^t(\n \pi^i,A(\tau))\,d\tau=0\,.
\]
Summarizing, by virtue of the Beppo Levi Theorem, letting $\vep \to 0$ and setting $I_i(t)=I_i(t,0)$, we have
\[
\dm \zeta b^i(t)\dm_1\le\dm \zeta b^i(0)\dm_1-I_3(t)-I_4(t)-I_5(t)+I_6(t)\,.
\]
We should estimate $I_i$, $i=3,\dots, 6$. \par  Firstly, let us remark that 
\[
\n\zeta=-4\zeta_1^2(1-\zeta_0^2)\zeta_0\n \zeta_0+2(1-\zeta_0^2)^2\zeta_1\n \zeta_1
\]
and
\[
\ba{l}
\Delta \zeta = 
- 4\zeta_0 \zeta_1^2 (1 - \zeta_0^2) \Delta \zeta_0 
+ 2\zeta_1 (1 - \zeta_0^2)^2 \Delta \zeta_1 + \left( 12\zeta_0^2\zeta_1^2 - 4\zeta_1^2 \right) |\nabla \zeta_0|^2 \VS\hskip4cm+ 2(1 - \zeta_0^2)^2 |\nabla \zeta_1|^2- 16\zeta_0 \zeta_1 (1 - \zeta_0^2) (\nabla \zeta_0 \cdot \nabla \zeta_1)\,.
\ea
\]
Hence, we can say that
\[
|\n \zeta|\le c(|\n \zeta_0|+|\n \zeta_1|)
\]
and
\[
|\Delta\zeta|\le c(|\Delta\zeta_0|+|\Delta\zeta_1|+|\n \zeta_0||\n\zeta_1|+|\n\zeta_0|^2+|\n\zeta_1|^2)\,.
\]
Hence, setting
\[
\ba {rl}
I_{3,1}&\hskip-0.2cm:=\displ\int_0^t\int_{\text{supp}(\n^i\zeta)}(|\Delta\zeta_0|+|\n\zeta_0|^2)|b^i|\,dx\,d\tau\,, \VS
I_{3,2}&\hskip-0.2cm:=\displ \int_0^t\int_{\text{supp}(\n^i\zeta)}(|\Delta\zeta_1|+|\n\zeta_1|^2)|b^i|\,dx\,d\tau\,,\VS
I_{3,3}&\hskip-0.2cm:=\displ\int_0^t\int_{\text{supp}(\n^i\zeta)}|\n\zeta_0|\cdot|\n\zeta_1||b^i|\,dx\,d\tau\,,
\ea
\]
we get
\[
\lim_{\vep\to 0} |I_3| \le c(I_{3,1}+I_{3,2}+I_{3,3})\,.
\]
Concerning $I_4$ and $I_6$, knowing that $\n\cdot u^i=\n\cdot V^i=0$, integration by parts yields 
\[
I_4=-\int_0^t\int_{\mathbb R^3} \n \zeta\cdot \mathbb J_i(u^i) |b^i|\,dx\,d\tau
\]
and
\[
I_6=-\int_0^t\int_{\mathbb R^3} \n \zeta\cdot \mathbb J_i(V^i) |b^i|\,dx\,d\tau\,.
\]
Since $V^i(\tau,x)=\xi^i(\tau)+\omega^i(\tau)\times x$ and $\n \zeta\cdot \omega^i(\tau)\times x=0$ for all $\tau\in [0,T_0]$, we find that
\[
I_6=-\int_0^t\int_{\mathbb R^3} \n \zeta\cdot  \xi^i |b^i|\,dx\,d\tau+\int_0^t\int_{\mathbb R^3} \n \zeta\cdot \int_{B(0,\frac{1}{i})} J_{\frac{1}{i}}(z)\omega^i\times z\,dz\, |b^i|\,dx\,d\tau\,.
\]
Hence, setting 
\[
\ba {rl} 
I_{4,1}&\hskip-0.2cm:=\displ \int_0^t\int_{\text{supp}(\n\zeta)} |\n \zeta_0||\mathbb J_i(u^i) ||b^i|\,dx\,d\tau\,,\VS
I_{4,2}&\hskip-0.2cm:=\displ \int_0^t\int_{\text{supp}(\n\zeta)} |\n \zeta_1|| \mathbb J_i(u^i) ||b^i|\,dx\,d\tau\,,
\ea
\]
we have
\[
|I_4|\le c(I_{4,1}+I_{4,2})\,,
\]
and, setting 
\[
\ba {rl} 
I_{6,1}&\hskip-0.2cm:=\displ \int_0^t\int_{\text{supp}(\n\zeta)}|\n \zeta_0| \Bigg|\xi^i+ \int_{B(0,\frac{1}{i})} J_{\frac{1}{i}}(z)\omega^i\times z\,dz\Bigg||b^i|\,dx\,d\tau\,,\VS
I_{6,2}&\hskip-0.2cm:=\displ \int_0^t\int_{\text{supp}(\n\zeta)} |\n \zeta_1|\Bigg|\xi^i+ \int_{B(0,\frac{1}{i})} J_{\frac{1}{i}}(z)\omega^i\times z\,dz\Bigg||b^i|\,dx\,d\tau\,,
\ea
\]
we have
\[
|I_6|\le c(I_{6,1}+I_{6,2})\,,
\]
Setting
\[
\ba {rl}
d_0&\hskip-0.2cm:= R_0+t^{\frac{1}{2}}\,,\VS
d&\hskip-0.2cm:=R+t^{\frac{1}{2}}\,,
\ea
\]
we find that
\[
\ba {rl} \displ
 I_{3,1}&\hskip-0.2cm\le\displ c\int_0^t d_0^{-\frac{1}{2}}\dm b^i\dm_2\,d\tau\le c\int_0^t d_0^{-\frac{1}{2}}\dm \n u^i\dm_2\,d\tau=:H_1\,,\VS
I_{3,2}&\hskip-0.2cm    \le c\displ\int_0^t d^{-1}\dm \n u^i\dm_2\,d\tau=:H_2\,,\VS
I_{3,3}&\hskip-0.2cm\le \displ c\int_0^t d_0^{-\frac{1}{2}}d^{-1}\dm b^i\dm_2\,d\tau\le c\int_0^t d_0^{-\frac{1}{2}}d^{-1}\dm \n u^i\dm_2\,d\tau=:H_3\,, \VS
I_{4,1}&\hskip-0.2cm\le \displ c\int_0^t d_0^{-1}\dm u^i\dm_2\dm b^i\dm_2\,d\tau\le c\int_0^t d_0^{-1}\dm u^i\dm_2\dm \n u^i\dm_2\,d\tau=:H_4\,, \VS
I_{4,2}&\hskip-0.2cm\le\displ c\int_0^t d^{-1}\dm u^i\dm_2\dm b^i\dm_2\,d\tau\le c\int_0^t d^{-1}\dm u^i\dm_2\dm \n u^i\dm_2\,d\tau=:H_5\,, \VS
|I_5|&\hskip-0.2cm\le\displ c\int_0^t \dm \n u^i\dm_2^2 \,d\tau=:H_6\,, \VS
I_{6,1}+I_{6,2}&\hskip-0.2cm\le\displ c\int_0^t(d_0^{-1}+d^{-1})( |\xi^i| +|\omega^i|)\dm\n u^i\dm_2\,d\tau=:H_7\,,
\ea
\]
Collecting all the estimates, we arrive at
\be \label{eq:RotL1-prel}
\dm \zeta b^i(t)\dm_1\le\dm \zeta \rot u_0\dm_1+\sum_{j=1}^7 H_j\,.
\ee
Letting $R\to +\infty$, via the Beppo Levi Theorem, we deduce that
\be \label{rotL1-prel2}
\ba{l}
\dm (1-\zeta_0)b^i(t)\dm_1\le\dm (1-\zeta_0) \rot u_0\dm_1\VS \hskip2cm+\int_0^t (cd_0^{-\frac{1}{2}}\dm \n u^i\dm_2+d_0^{-1} (\dm u^i\dm_2+| \xi^i|+|\omega^i|)\dm \n u^i \dm_2 + \dm \n u^i \dm_2^2) \, d\tau \,.
\ea
\ee
Hence, using the Schwarz inequality in \eqref{rotL1-prel2}, we find
\[
\dm (1-\zeta_0)b^i(t)\dm_1  \le \dm \rot u_0\dm_1+c\frac{t^{\frac{1}{2}}}{(R_0+t^{\frac{1}{2}})^{\frac{1}{2}}}(\dm u_0\dm_2+|\xi_0|+|\omega_0|)+\bigg[\frac{t^{\frac{1}{2}}}{(R_0+t^{\frac{1}{2}})}+1\bigg](\dm u_0\dm_2+|\xi_0|+|\omega_0|)^2\,.
\]
Finally, we conclude that 
\[
\dm b^i(t)\dm_{L^1(|x|  \ge R_0+t^{\frac{1}{2}})}\le \mathfrak A(1+t^{\frac{1}{4}})\,.
\]
The theorem is proved.
\ep
The previous theorem, suitably combined with some results on the Riesz singular kernels, leads to an estimate of $\norm{u^i(t)}_{L^{\frac{3}{2}}_w}$. In fact, we have
\begin{tho} \label{thm: 3/2weak}
  {\sl  Let $\{(u^i,\pi^i,\xi^i,\omega^i)\}$ the subsequence stated in Theorem\,\ref{thm: Prolungamento} defined on the interval $[0,T_h]$. Assume that $\rot u_0\in  L^1(\Omega)$. Then, for all $t\in (0,T_h)$,
    \begin{equation}
        \norm{u^i(t)}_{L^{\frac{3}{2}}_w} \le \mathfrak A (1+t^{\frac{1}{4}})\,.
    \end{equation}}
    \end{tho}
    \bp
    Let $t\in (0,T_h)$. Set $\widehat{\zeta}:=1-\zeta_0^2$, where $$\zeta_0(t,x):=\begin{cases}
    1\,, &\text{if }|x|\le R_0+t^{\frac{1}{2}}\,,\\
    \frac{2(R_0+t^{\frac{1}{2}})-|x|}{R_0+t^{\frac{1}{2}}}\,, &\text{if }R_0+t^{\frac{1}{2}}\le |x|\le 2(R_0+t^{\frac{1}{2}})\,,\\
    0\,, &\text{if } |x|\ge 2(R_0+t^{\frac{1}{2}})\,.
\end{cases}$$ 
Let us recall the following well-known identity
\[
\Delta(\widehat\zeta u^i)=-\rot \rot (\widehat\zeta u^i)+\nabla(\nabla \cdot (\widehat\zeta u^i))\,.
\]
The last identity yields to
\[
\widehat\zeta u^i= \frac{1}{4\pi} \int_{\mathbb{R}^3} \frac{1}{\abs{x-y}}(-\rot \rot (\widehat\zeta u^i)+\nabla(\nabla \cdot (\widehat\zeta u^i)))\, dy\,.
\]
An integration by parts yields to
\[
\widehat\zeta u^i= \frac{1}{4\pi} \int_{\mathbb{R}^3} \biggl[\nabla (\frac{1}{\abs{x-y}}) \times\rot (\widehat\zeta u^i)- \nabla (\frac{1}{\abs{x-y}}) \nabla \cdot (\widehat\zeta u^i)\biggr]\, dy.
\]
We remark that, by the incompressibility condition, we have
\[
\n\cdot (\widehat\zeta u^i)=\n\widehat\zeta\cdot u^i\,,
\]
and $\n\widehat\zeta$ has compact support. \par The above considerations and Theorem\,\ref{thm:rotoreL1} allow us to say that both $\rot (\widehat\zeta u^i)$ and $\nabla \cdot (\widehat\zeta u^i)$ are in $L^1(\mathbb{R}^3)$ and $\nabla (\frac{1}{\abs{x-y}})$ is a Riesz potential. Hence, via the Sobolev Theorem \cite{Sob, Mir}, we find that 
\[
T_1(\rot (\widehat\zeta u^i)):=\int_{\mathbb{R}^3} \nabla (\frac{1}{\abs{x-y}}) \times\rot (\widehat\zeta u^i)\, dy
\]
and
\[
T_2(\nabla\cdot (\widehat\zeta u^i)):=-\int_{\mathbb{R}^3}  \nabla (\frac{1}{\abs{x-y}}) \nabla \cdot (\widehat\zeta u^i)\, dy
\]
are linear and continuous maps between $L^1(\mathbb{R}^3)$ and $L^{\frac{3}{2}}_{w}(\mathbb{R}^3)$. Then, we get that
\[
\ba {l}
\norm{\widehat\zeta u^i}_{\frac{3}{2}w} = \norm{T_1(\rot (\widehat\zeta u^i ))+T_2(\n\cdot(\widehat\zeta u^i))}_{\frac{3}{2}w}\le \dm T_1(\rot (\widehat\zeta u^i ))\dm_{\frac{3}{2}w}+\dm  T_2(\n\cdot(\widehat\zeta u^i))\dm_{\frac{3}{2}w}\VS \hskip10.5cm \le c(\dm \rot(\widehat\zeta u^i)\dm_1+\dm \n\cdot (\widehat\zeta u^i)\dm_1)\,.
\ea
\]
By Theorem\,\ref{thm:rotoreL1}, we can say that
\[
\dm \rot(\widehat\zeta u^i)\dm_1\le \mathfrak A(1+t^{\frac{1}{4}})\,.
\]
Recalling that
\[
\text{supp }\nabla\widehat\zeta\subset[R_0+t^{\frac{1}{2}},2(R_0+t^{\frac{1}{2}})], \quad |\nabla \widehat\zeta|\le (R_0+t^{\frac{1}{2}})^{-1}
\]
and the well-known identity
\[
\nabla\cdot (\widehat\zeta u^i)=\widehat\zeta\nabla \cdot u^i+ u^i\cdot \nabla\widehat\zeta=u^i\cdot \nabla\widehat\zeta\,,
\]
according to incompressibility condition, we have that
\[
\ba {l}\displ
\norm{\nabla\cdot (\widehat\zeta u^i(t))}_1 =\norm{u^i(t)\cdot\nabla \widehat\zeta}_1 \le \frac{c}{d_0(t)}\int_{d_0(t)\le|x|\le 2d_0(t)} |u^i(t)|\,dx,
\ea
\]
with $$ d_0(t)=R_0+t^{\frac{1}{2}}\,.$$
Considering that $\dm u^i(t)\dm_2\le cA_0^{\frac{1}{2}}$ and that, via the Schwarz inequality we have
$$
\int_{d_0(t)\le|x|\le 2d_0(t)} |u^i(t)|\,dx\le cd^{\frac{3}{2}}_0(t)\bigg(\int_{d_0(t)\le|x|\le 2d_0(t)} |u^i(t)|^2\,dx\bigg)^{\frac{1}{2}}\,,
$$
we find 
\[
\norm{\nabla\cdot (\widehat\zeta u^i(t))}_1\le cA_0^{\frac{1}{2}}(1+t^{\frac{1}{4}})\,.
\]
Hence, we conclude that
\[
\norm{u^i(t)}_{L^{\frac{3}{2}}_w(\abs{x}\ge R_0+t^{\frac{1}{2}})} \le \mathfrak A (1+t^{\frac{1}{4}})\,.
\]
 By virtue of the Schwarz inequality, it is also true that 
        \[ \small
        \norm{u^i(t)}_{L^{\frac{3}{2}}_w(\{\abs{x}\le R_0+t^{\frac{1}{2}}\}\cap\Omega)} \le c(R_0+t^{\frac{1}{2}})^{\frac{1}{2}}\norm{u^i(t)}_2\le cA_0^{\frac{1}{2}}(R_0+t^{\frac{1}{2}})^{\frac{1}{2}}\le cA_0^{\frac{1}{2}}(1+t^{\frac{1}{4}})\,.
        \]
The proof is complete.
    \ep
For the subsequence stated in Theorem\,\ref{thm: Prolungamento} we obtained the energy equality 
\begin{equation*} \label{eq:enes}
\begin{array}{l}\displ
\norm{u^i(t)}_2^2 +\abs{\xi^i(t)}^2+ \omega^i (t)\cdot (I \cdot \omega^i(t)) + 2\int_0^t \norm{\nabla u^i(\tau)}_2^2 \, d\tau \VS\hskip2cm= \norm{u_0}_2^2+|\xi_0|^2+\omega_0\cdot (I\cdot\omega_0)-\int_0^t\int_{\partial\Omega} \nu\cdot \mathbb J_{i}(u^i(\tau)-V^i(\tau)) \abs{V^i(\tau)}^2 \, d\sigma\,d\tau\,.
\end{array}
\end{equation*}
Moreover, for all $t\in [0,T_h]$, we have
\be \label{eq: stima-h}
\norm{u^i(t)}_2^2 +\abs{\xi^i(t)}^2+ \omega^i (t)\cdot (I \cdot \omega^i(t)) + 2\int_0^t \norm{\nabla u^i(\tau)}_2^2 \, d\tau\le 2A_0\,.
\ee
Estimate \eqref{eq: stima-h} is the starting point to prove the next lemma.
\begin{lemma} \label{le: new energy estimate}
  {\sl  For all $t\in[0,T_h]$
    \begin{equation} \label{eq:enes2}
\begin{array}{l}\displ
\norm{u^i(t)}_2^2 +\abs{\xi^i(t)}^2+ \omega^i (t)\cdot( I \cdot \omega^i(t) )+ \int_0^t \bigg[\mathfrak B\frac{\norm{u^i(\tau)}_2^6}{(1+\tau^{\frac{1}{4}})^4}+\dm \n u^i(\tau)\dm_2^2\bigg] \, d\tau\le 2A_0\,.
\end{array}
\end{equation}}
\end{lemma}
\bp
By Lemma\,\ref{le: interpolazione weak}, we can say that
\[
\norm{u^i(t)}_2 \le c\norm{u^i(t)}_6^a \norm{u^i(t)}_{\frac{3}{2}w}^{1-a}\,,
\]
where $a=\frac{1}{3}$ is obtained via the relation
\[
\frac{1}{2}=(1-a)\frac{2}{3}+a\frac{1}{6}\,.
\]
Using Sobolev embedding theorem, we get
\[
\norm{u^i(t)}_2 \le c\norm{\nabla u^i(t)}_2^{\frac{1}{3}} \norm{u^i(t)}_{\frac{3}{2}w}^{\frac{2}{3}}\,.
\]
From the latter and by virtue of Theorem\,\ref{thm: 3/2weak}, we deduce
\be \label{eq: Dis interpolazione}
\norm{\nabla u^i(t)}_2^2 \ge \frac{1}{c}\frac{\norm{u^i(t)}_2^6}{\norm{u^i(t)}^{4}_{\frac{3}{2}w}} \ge \mathfrak B\frac{\norm{u^i(t)}_2^6}{(1+t^{\frac{1}{4}})^4}\,.
\ee
Recalling inequality \eqref{eq: stima-h}, we get \eqref{eq:enes2}.
\ep
\section{Proof of the main result: Theorem\,\ref{thm:MR}} \label{sec: proof}
In this final section, we prove the main result of our paper.\par  We divide the proof into two steps. We first prove the existence of a weak solution and then deduce its partial regularity.\par
Before proving our main theorem, we need the following two lemmas.
\begin{lemma} \label{le: Preliminari-struttura} {\sl Let $\{T_h\}_{h\in\N_0}$ be the sequence stated in Theorem\,\ref{thm: Prolungamento}. Let $\chi$ be a positive real number and let $T^*=\displ\min_{h\in\N} \{T_h\,:\,T_h\geq \exp[\sfrac{2A_0}\chi]-1\}$. Let $\{(u^i,\pi^i,\xi^i,\omega^i)\}_{i\ge n(T^*)}$ be the subsequence defined on $[0,T^*]$. Then, there exists a $\theta_i\in (0,\exp[\frac{2A_0}\chi]-1)\equiv (0,\theta)$ such that \be\label{UN-BOUD}\mathfrak B\dm u^i(\theta_i)\dm_2^6+\dm\n u^i(\theta_i)\dm_2^2\leq \chi\,.\ee   } \end{lemma}
\bp
Assume that the statement is false. Then, by virtue of Lemma\,\ref{le: new energy estimate}, we get
$$\ba{ll}\displ2A_0\hskip-0.2cm&\displ=\chi\sfrac{2A_0}\chi=\chi\hskip-0.3cm\intll0{\exp[\frac{2A_0} \chi]-1}\hskip-0.3cm(t+1)^{-1}dt< \hskip-0.3cm\intll0{\exp[\frac{2A_0} \chi]-1}\hskip-0.3cm(t+1)^{-1}\big[\mathfrak B\dm u^i(t)\dm_2^6+\dm\n u^i(t)\dm_2^2\big]dt\VSE<\hskip-0.3cm\intll0{\exp[\frac{2A_0} \chi]-1}\hskip-0.3cm\big[\mathfrak B(t+1)^{-1}\dm u^i(t)\dm_2^6+\dm\n u^i(t)\dm_2^2\big]dt\leq 2A_0\,,\ea$$
which is a contradiction.
\ep
\begin{lemma}\label{le: globale mollificato}
  {\sl  Let $(u_0,\xi_0,\omega_0)\in(\mathcal V(\Omega)\cap\widehat W^{1,2}(\Omega,|x|))\times\mathbb R^3\times\mathbb R^3$, with $\gamma(u_0-\xi_0-\omega_0\times x)=0$. Let $\theta_i\le \exp[\sfrac{2A_0}\chi]-1$ established in Lemma\,\ref{le: Preliminari-struttura}.
    Then, for all $i\ge n(T^*)$, $(u^i,\pi^i,\xi^i,\omega^i)$  admits a unique extension in $(\theta_i,\infty)$. The extension enjoys the regularity properties detected by the metrics in \rf{SOLR-I} and the properties hold uniformly in $i$. Moreover, we get  $\displ[\theta,\infty)\subseteq {\underset{i}\cap (\theta_i,\infty)}$, where $\theta$ is furnished in Lemma\,\ref{le: Preliminari-struttura}.}
\end{lemma}
\bp
First of all, let us point out that for all $\delta>0$ there exists $\chi>0$ such that
\be \label{eq: delta}
\delta=\chi^\frac12(1+c)+(\mathfrak A\chi)^\frac16\,.
\ee
Lemma\,\ref{le: Preliminari-struttura} ensures that there exists $\theta_i \in (0,T^*)$ such that
\[
\dm u^i(\theta_i)\dm_{1,2}+|\xi^i(\theta_i)|+((I\cdot \omega^i(\theta_i))\cdot\omega^i(\theta_i))^{\frac{1}{2}}\leq \delta\,,
\]
for all $i\ge n(T^*)$.\par
 Hence, assuming $(u^i(\theta_i),\xi^i(\theta_i),\omega^i(\theta_i))\in (\mathcal V(\Omega)\cap \widehat W^{1,2}(\Omega,|x|))\times\mathbb R^3\times\mathbb R^3$ as the initial datum for problem \eqref{eq:moll}, Theorem\,\ref{thm: Esistenza-moll} ensures that there exists a unique solution defined on a maximal time interval $(\theta_i,\theta_i+T(\theta_i))$. \par Our aim is to get $T(\theta_i)=\infty$. \par Recalling the statement of Lemma\,\ref{le: Gronwall generalizzato}, setting $g(t):=\dm \mathbb D(u^i)(t)\dm_2^2$, we need to show that\footnote{We are going to employ the proof obtained by Galdi in \cite{Ga:23}, suitably modified due to the presence of the mollifier.}
 \begin{description}
     \item[i)] $\displ\lim_{t\to \theta_i +T(\theta_i)} (\dm u^i(t)\dm_2^2+|\xi^i(t)|^2+(I\cdot\omega^i(t))\cdot\omega^i(t))\le \delta^2$\,,
     \item[ii)]$\displ \lim_{t\to \theta_i+T(\theta_i)}\int_{\theta_i}^t \dm \n u^i(\tau)\dm_2^2\,d\tau\le\delta^2$\,,
     \item[iii)]$ \displ \lim_{t\to \theta_i+T(\theta_i)}\dm  g(t)\dm_2\le M\delta$\,.
 \end{description}
 In order to get the validity of the first item, we invoke the energy equality:
 \begin{equation} \label{eq: EE-glob}
\begin{array}{l}\displ
\norm{u^i(t)}_2^2 +\abs{\xi^i(t)}^2+ \omega^i (t)\cdot (I \cdot \omega^i(t)) + 2\int_{\theta_i}^t \norm{\nabla u^i(\tau)}_2^2 \, d\tau \VS\hskip5cm= \norm{u^i(\theta_i)}_2^2+|\xi^i(\theta_i)|^2+\omega^i(\theta_i)\cdot (I\cdot\omega^i(\theta_i))\VS\hskip5.4cm-\int_{\theta_i}^t\int_{\partial\Omega} \nu\cdot \mathbb J_i(u^i(\tau)-V^i(\tau)) \abs{V^i(\tau)}^2 \, dS\,d\tau\,,
\end{array}
\end{equation}
for all $t\in (\theta_i,\theta_i+T(\theta_i))$. Employing Lemma\,\ref{le:Stime per rel energia} and estimate \eqref{eq: moto rigido-gradiente}, we get
\[
\Bigg|\int_{\theta_i}^t\int_{\partial\Omega} \nu\cdot \mathbb J_i(u^i(\tau)-V^i(\tau)) \abs{V^i(\tau)}^2\,d\sigma\,d\tau\Bigg| \le c_0\int_{\theta_i}^t (\dm u^i(\tau)\dm_2+ |\xi^i(\tau)|+((I\cdot\omega^i(\tau))\cdot\omega^i(\tau))^{\frac12})\dm \n u^i(\tau)\dm_2^2\,d\tau\,.
\]
Since there are no restrictions on $\delta$, we may take it such that $\delta<(2c_0)^{-1}$. We want to prove that for all $t\in [\theta_i,\theta_i+T(\theta_i))$, $$\dm u^i(t)\dm_2+ |\xi^i(t)|+((I\cdot\omega^i(t))\cdot\omega^i(t))^{\frac12}<(\sqrt{2}c_0)^{-1}\,.$$ Recalling that $(u^i,\pi^i,\xi^i,\omega^i)$ enjoys properties \eqref{eq:GSS I-moll}$_{1,2}$, by continuity, if this is not the case, there exists $\overline t\in (\theta_i,\theta_i+T(\theta_i))$ such that $\dm u^i(\overline t)\dm_2+ |\xi^i(\overline t)|+((I\cdot\omega^i(\overline t))\cdot\omega^i(\overline t))^{\frac12}=(\sqrt{2}c_0)^{-1}$. Hence, taking $t=\overline t$ in \eqref{eq: EE-glob} and remarking that
\[
\ba{l} \displ
\Bigg|\int_{\theta_i}^{\overline t}\int_{\partial\Omega} \nu\cdot \mathbb J_i(u^i(\tau)-V^i(\tau)) \abs{V^i(\tau)}^2\,d\sigma\,d\tau\Bigg| \VS\hskip0.5cm\le c_0\int_{\theta_i}^{\overline t} \big[\dm u^i(\tau)\dm_2+ |\xi^i(\tau)|+((I\cdot\omega^i(\tau))\cdot\omega^i(\tau))^{\frac12}\big]\dm \n u^i(\tau)\dm_2^2\,d\tau\le\frac{1}{\sqrt{2}} \int_{\theta_i}^{\overline t} \dm \n u^i(\tau)\dm_2^2\,d\tau\,,
\ea
\]
we arrive at
\[
 \ba {l} \displ
\norm{u^i(\overline t)}_2^2 +\abs{\xi^i(\overline t)}^2+ \omega^i (\overline t)\cdot( I \cdot \omega^i(\overline t)) + \int_{\theta_i}^{\overline t} \norm{\nabla u^i(\tau)}_2^2 \, d\tau\VS\hskip4cm\le \norm{u^i(\theta_i)}_2^2+|\xi^i(\theta_i)|^2+\omega^i(\theta_i)\cdot (I\cdot\omega^i(\theta_i)) <(2c_0)^{-2}
\ea
\]
Therefore, we get
\[
\dm u^i(\overline t)\dm_2+ |\xi^i(\overline t)|+((I\cdot\omega^i(\overline t))\cdot\omega^i(\overline t))^{\frac12}<(\sqrt{2}c_0)^{-1}\,,
\]
which is a contradiction. We conclude that
\[
\lim_{t\to \theta_i+T(\theta_i)}\dm u^i(t)\dm_2+ |\xi^i(t)|+((I\cdot\omega^i(t))\cdot\omega^i(t))^{\frac12}\le(\sqrt{2}c_0)^{-1}\,.
\]
Concerning item {\bf ii)}, taking into account item {\bf i)}, we easily arrive at
\[
\norm{u^i(t)}_2^2 +\abs{\xi^i(t)}^2+ \omega^i (t)\cdot (I \cdot \omega^i(t)) +\int_{\theta_i}^t \norm{\nabla u^i(\tau)}_2^2\,d\tau\le \norm{u^i(\theta_i)}_2^2+|\xi^i(\theta_i)|^2+\omega^i(\theta_i)\cdot (I\cdot\omega^i(\theta_i)) \,,
\]
for all $t\in (\theta_i,\theta_i+T(\theta_i))$. Therefore, the claim is easily obtained.\par
We finally discuss item {\bf iii)}.\par
We multiply equation \eqref{eq:moll}$_1$ by $-\n\cdot\mathbb T(u^i,\pi^i)$. We get
\[
\ba{l} \displ
\frac{d}{dt}\dm \mathbb D(u^i)\dm_2^2+\dm \n\cdot\mathbb T\dm_2^2=\int_{\partial\Omega} V^i_t\cdot \mathbb T\cdot\nu\,dS+\int_\Omega \mathbb J_i(u^i)\cdot\n u^i\cdot (\n\cdot\mathbb T)\,dx\VS\hskip6cm-\int_\Omega(\mathbb J_i(V^i)\cdot\n u^i-\omega^i\times u^i)\cdot(\n\cdot\mathbb T)\,dx\,.
\ea
\]
We estimate the terms on the right-hand side. Recalling equations \eqref{eq:moll}$_{5,6}$, we have
\[
\int_{\partial\Omega} V^i_t\cdot \mathbb T\cdot\nu\,dS=-|\dot{\xi}^i|^2-\dot\omega^i\cdot(I\cdot\dot\omega^i)+\xi^i\times\omega^i\cdot\dot\xi^i+(I\cdot\omega^i)\times\omega^i\cdot\dot\omega^i\,.
\]
Hence, invoking Lemma\,\ref{le:prel-symgrad}, we get
\[
|\xi^i\times\omega^i\cdot\dot\xi^i+(I\cdot\omega^i)\times\omega^i\cdot\dot\omega^i|\le \vep(|\dot\xi^i|^2+(I\cdot\dot\omega^i)\cdot\dot\omega^i)+ c\dm\n u^i\dm_2^4
\]
Concerning the second integral, by H\"{o}lder, Sobolev and Young inequalities and a fundamental property of mollifiers, we have
\[
\ba{l}\displ
\Bigg|\int_\Omega \mathbb J_i(u^i)\cdot\n u^i\cdot (\n\cdot\mathbb T)\,dx\Bigg|\leq \dm \n\cdot\mathbb T\dm_2\dm \n u^i\dm_3\dm \mathbb J_i(u^i)\dm_6\leq \vep \dm  \n\cdot\mathbb T\dm_2^2+c\dm \n u^i\dm_3^2\dm \mathbb D(u^i)\dm_2^2\,.
\ea
\]
Moreover, by an interpolation inequality, estimate \eqref{eq: stima gradiente pressione} and again Young's inequality, we have
\[
\dm \n u^i\dm_3^2\dm \mathbb D(u^i)\dm_2^2\le \vep \dm \n \cdot\mathbb T\dm_2^2 +c (\dm\mathbb D(u^i)\dm_2^4+\dm\mathbb D(u^i)\dm_2^6)\,.
\]
We now consider the integral
\[
\ba{l}\displ
\int_\Omega(\mathbb J_i(V^i)\cdot\n u^i-\omega^i\times u^i)\cdot(\n\cdot\mathbb T)\,dx=\int_\Omega(\mathbb J_i(V^i)\cdot\n u^i-\omega^i\times u^i)\cdot\Delta u^i\,dx\VS\hskip5cm-\int_\Omega(\mathbb J_i(V^i)\cdot\n u^i-\omega^i\times u^i)\cdot\n\pi^i\,dx=:I-J\,.
\ea
\]
Remarking that
\be \label{TPROD-sol}
\n \big[ \mathbb J_i(V^i)\cdot \n u^i- \omega^i\times u^i\big]= <\n u^i,\mathbb J_i(\n V^i)> +\mathbb J_i(V^i)\cdot\n\n u^i-\n (\omega^i\times u^i)\,,\footnote{In fact, we have
\[
\ba{l}
\n \big[ \mathbb J_i(V^i)\cdot \n u^i- \omega^i\times u^i\big]_{h,k}=\partial_{x_k}\big[\partial_{x_l}(u^i)_h\mathbb J_i(V^i)_l-(\omega^i\times u^i)_h\big]\VS\hskip4cm=\partial_{x_k}\partial_{x_l}(u^i)_h\mathbb J_i(V^i)_l+\partial_{x_l}(u^i)_h\partial_{x_k}\mathbb J_i(V^i)_l-\n(\omega^i\times u^i)_{h,k}\,,
\ea
\]
and we easily get the above identity.}
\ee
integration by parts yields to
\[
\ba{l}\displ
I=-\int_\Omega\n\big[\mathbb J_i(V^i)\cdot\n u^i-\omega^i\times u^i\big]\cdot\n u^i\,dx+\int_{\partial\Omega}\big[\mathbb J_i(V^i)\cdot\n u^i-\omega^i\times u^i\big]\cdot\n u^i\cdot\nu\,dS\VS\hskip2cm=-\int_\Omega\big[<\n u^i,\mathbb J_i(\n V^i)> -\n (\omega^i\times u^i)\big]\cdot\n u^i\,dx-\int_\Omega\mathbb J_i(V^i)\cdot\n\n u^i\cdot\n u^i\,dx\VS\hskip7cm+\int_{\partial\Omega}\big[\mathbb J_i(V^i)\cdot\n u^i-\omega^i\times u^i\big]\cdot\n u^i\cdot\nu\,dS\,,
\ea
\]
while
\[
\ba{l}\displ
-J=\int_\Omega\n\big[\mathbb J_i(V^i)\cdot\n u^i-\omega^i\times u^i\big]\cdot(\pi^i-\pi^i_0)\mathbb I\,dx-\int_{\partial\Omega} \big[\mathbb J_i(V^i)\cdot\n u^i-\omega^i\times u^i\big]\cdot (\pi^i-\pi^i_0)\mathbb I\cdot\nu\,dS\VS\hskip2cm=\int_\Omega\big[<\n u^i,\mathbb J_i(\n V^i)>-\n (\omega^i\times u^i)\big]\cdot(\pi^i-\pi^i_0)\mathbb I\,dx+ \int_\Omega\mathbb J_i(V^i)\cdot\n\n u^i\cdot(\pi^i-\pi^i_0)\mathbb I\,dx\VS\hskip7.5cm-\int_{\partial\Omega} \big[\mathbb J_i(V^i)\cdot\n u^i-\omega^i\times u^i\big]\cdot (\pi^i-\pi^i_0)\mathbb I\cdot\nu\,dS\,.
\ea
\]
Remarking that
\[
\ba{rl}
<\n u^i,\mathbb J_i(\n V^i)>\cdot\n u^i&\hskip-0.2cm=\displ\sum_{k=1}^3 \omega^i\times (\n u^i)_k\cdot(\n u^i)_k=0\,,\VS
\big[<\n u^i,\mathbb J_i(\n V^i)>-\n (\omega^i\times u^i)\big]\cdot \mathbb I&\hskip-0.2cm=0\,, \ea\]
and, via the divergence-free condition,
\[
 \mathbb J_i(V^i)\cdot\n\n u^i\cdot(\pi^i-\pi^i_0)\mathbb I=0\,,
\]
we get
\[
\ba{l}\displ
\int_\Omega(\mathbb J_i(V^i)\cdot\n u^i-\omega^i\times u^i)\cdot(\n\cdot\mathbb T)\,dx=-\frac12\int_\Omega\mathbb J_i(V^i)\cdot\n|\n u^i|^2\,dx\VS\hskip1.7cm-\int_\Omega \n(\omega^i\times u^i)\cdot \n u^i\,dx+\int_{\partial\Omega} (\mathbb J_i(V^i)\cdot\n u^i-\omega^i\times V^i)\cdot (\n u^i-(\pi^i-\pi^i_0)\mathbb I)\cdot\nu\,dS\,.
\ea
\]
Since $\n\cdot V^i=0$, we have
\[
\int_\Omega\mathbb J_i(V^i)\cdot\n|\n u^i|^2\,dx=\int_{\partial\Omega} \mathbb J_i(V^i)\cdot \nu |\n u^i|^2\,dS\,.
\]
Recalling estimate \eqref{eq: stima gradiente pressione}, where we mean $f=\n\cdot\mathbb T(u^i,\pi^i)$, employing Lemma\,\ref{le:prel-symgrad}, the Gagliardo trace inequality and Young's inequality, we arrive at
\[
\Bigg|\int_{\partial\Omega} \mathbb J_i(V^i)\cdot \nu |\n u^i|^2\,dS\Bigg|\le c(|\xi^i|+|\omega^i|)\dm \n u^i\dm_{L^2(\partial\Omega)}^2\le \vep\dm \n\cdot\mathbb T\dm_2^2+c(\dm \mathbb D(u^i)\dm_2^3+\dm \mathbb D(u^i)\dm_2^4)\,.
\]
Moreover, by analogous arguments we have
\[
\Bigg|\int_{\partial\Omega} (\mathbb J_i(V^i)\cdot\n u^i-\omega^i\times V^i)\cdot \n u^i\cdot\nu\,dS\Bigg|\le \vep \dm \n\cdot\mathbb T\dm_2^2+c(\dm \mathbb D(u^i)\dm_2^3+\dm \mathbb D(u^i)\dm_2^4)\,,
\]
and
\[
\ba{l}\displ
\Bigg|\int_{\partial\Omega} (\mathbb J_i(V^i)\cdot\n u^i-\omega^i\times V^i)\cdot(\pi^i-\pi^i_0)\mathbb I\cdot\nu\,dS\Bigg|\VS\hskip3cm\le c(|\xi^i|+|\omega^i|)(|\xi^i|+|\omega^i|+\dm \n u^i \dm_{L^2(\partial\Omega)})\dm \pi^i-\pi^i_0\dm_{L^2(\partial\Omega)}\,.
\ea
\]
By virtue of Lemma\,\ref{le:prel-symgrad}, the Gagliardo trace inequality for $\dm\pi^i-\pi^i_0\dm_{L^2(\partial\Omega)}$ and inequality \eqref{Hardy}, remarking that $|\omega^i|^\frac43\dm \n u^i\dm_2^2\le c(\dm \n u^i\dm_2^3+\dm \n u^i\dm_2^4)$, we arrive at
\[
\Bigg|\int_{\partial\Omega} (\mathbb J_i(V^i)\cdot\n u^i-\omega^i\times V^i)\cdot(\pi^i-\pi^i_0)\mathbb I\cdot\nu\,dS\Bigg|\le \vep \dm \n \cdot \mathbb T\dm_2^2+c(\dm \n u^i\dm_2^3+\dm \n u^i\dm_2^4+\dm \n u^i\dm_2^6)\,.
\]
Finally, employing again Lemma\,\ref{le:prel-symgrad}, we have
\[
\Bigg|\int_\Omega \n(\omega^i\times u^i)\cdot\n u^i\,dx\Bigg|\le c\dm \n u^i\dm_2^3
\]
We now multiply \eqref{eq:moll}$_1$ by $u^i_t(K_0+\kappa(x))^{-2}$. Setting
\[
\ba{rl}
\widetilde H_1&\hskip-0.2cm:=\displ -\int_{\Omega}\mathbb J_i(u^i-V^i)\cdot \n u^i\cdot u^i_t(K_0+\kappa)^{-2}\,dx\,,\VS
\widetilde H_2&\hskip-0.2cm:=-\displ\int_{\Omega}\omega^i\times u^i\cdot u^i_t(K_0+\kappa)^{-2}\,dx\,,\VS
\widetilde H_3&\hskip-0.2cm:=\displ K_0^{-2}\int_{\partial\Omega}\mathbb T(u^i,\pi^i)\cdot \nu \cdot V^i_t\,dS\,,\VS
\widetilde H_4&\hskip-0.2cm:=\displ 4\int_\Omega (K_0+\kappa)^{-3}\mathbb D(u^i)\cdot u^i_t\otimes \n \kappa\,dx\,,  \VS
\widetilde H_5 &\hskip-0.2cm:=\displ -2\int_\Omega (K_0+\kappa)^{-3}(\pi^i-\pi^i_0)\mathbb I\cdot u^i_t\otimes \n \kappa\,dx\,.
\ea
\]
we get
\[
\frac{d}{dt}\dm \mathbb D(u^i)(K_0+\kappa)^{-1}\dm_2^2+\dm u^i_t(K_0+\kappa)^{-1}\dm_2^2=\sum_{k=1}^5 \widetilde H_k\,,
\]
We estimate $\widetilde H_1$. We have
\[
|\widetilde H_1|\le \dm \mathbb J_i({\frac{u^i-V^i}{|x|}})\dm_\infty\int_\Omega \bigg(|x|+\frac{1}{n}\bigg)|\n u^i||u^i_t|(K_0+\kappa(x))^{-2}\,dx\,.
\]
Using Lemma\,\ref{CM-Lemma}, we get
\[
|\widetilde H_1|\le  c(\dm \n u^i\dm_2^{\frac12}\dm D^2u^i\dm_2^{\frac{1}{2}}+|\xi^i|+|\omega^i|)\dm \n u^i\dm_2\dm u^i_t(K_0+\kappa)^{-1}\dm_2\,.
\]
Hence, employing Lemma\,\ref{le:prel-symgrad} and estimate \eqref{eq: stima gradiente pressione} with $f=\n\cdot\mathbb T$, we arrive at
\[
|\widetilde H_1|\le c(\dm \n u^i\dm_2^4+\dm \n u^i\dm_2^6)+\vep(\dm \n\cdot\mathbb T\dm_2^2+\dm u^i_t(K_0+\kappa)^{-1}\dm_2^2)\,.
\]
The terms $\widetilde H_k$, $k=2,\dots 5$, can be estimated analogously to the integrals $H_k$, already estimated in Lemma\,\ref{le: DT-peso}.\par
Choosing $\vep>0$ sufficiently small, we conclude that
\[
\ba{l}
\frac{d}{dt}(\dm \mathbb D(u^i)\dm_2^2+\dm \mathbb D(u^i)(K_0+\kappa)^{-1}\dm_2^2)+\sfrac12(\dm \n\cdot\mathbb T\dm_2^2+\dm u^i_t(K_0+\kappa)^{-1}\dm_2^2+|\dot\xi^i|^2+(I\cdot\dot\omega^i)\cdot\dot\omega^i)\VS\hskip5cm\leq c(\dm\mathbb D(u^i)\dm_2^2+ \dm \mathbb D(u^i)\dm_2^3 +\dm \mathbb D(u^i)\dm_2^4+\dm \mathbb D(u^i)\dm_2^6)\,.
\ea
\]
Hence by using iteratively Schwarz inequality, we get
\be \label{Le18-dis finale} 
\ba{l}\displ
\frac{d}{dt}(\dm \mathbb D(u^i)\dm_2^2+\dm \mathbb D(u^i)(K_0+\kappa)^{-1}\dm_2^2)+\sfrac12(\dm \n\cdot\mathbb T\dm_2^2+\dm u^i_t(K_0+\kappa)^{-1}\dm_2^2+|\dot\xi^i|^2+(I\cdot\dot\omega^i)\cdot\dot\omega^i)\VS\hskip10cm\leq c_1\dm\mathbb D(u^i)\dm_2^2+c_2\dm \mathbb D(u^i)\dm_2^6\,,
\ea
\ee
for some positive constants $c_1$ and $c_2$ independent of $u^i$. Set $M:=2\max\{1,c_1,c_2\}$, let $\eta\in (0,M^{-\frac{1}{\alpha-1}})$. If 
\[
\dm u^i(\theta_i)\dm_{1,2}+|\xi^i(\theta_i)|+((I\cdot \omega^i(\theta_i))\cdot\omega^i(\theta_i))^{\frac{1}{2}}\le \eta^\frac12\,,
\]
by virtue of Lemma\,\ref{le: Gronwall generalizzato} we arrive at
\[
\lim_{t\to \theta_i+T(\theta_i)}\dm \n u^i(t)\dm_2^2\le M\eta<\infty\,.
\]
Since there are no restrictions on $\delta$, we can choose it such that $\delta<\min\{\eta^\frac12,(2c_0)^{-1}\}$. Hence, if $\dm u^i(\theta_i)\dm_{1,2}+|\xi^i(\theta_i)|+((I\cdot \omega^i(\theta_i))\cdot\omega^i(\theta_i))^{\frac{1}{2}}\le \delta$, the items ${\bf i)-iii)}$ are satisfied.
\par
Therefore, we get, for all $T>\theta_i$,
\be\label{SOLREG-APPR}\ba{c}\xi^i(t)\,,\;\omega^i(t)\in C(\theta_i,T)\mbox{\; and\; }\dot\xi^i(t)\,,\;\dot\omega^i(t)\in L^2(\theta_i,T)\,,\VS u^i\!\in\! L^\infty\!(\theta_i,T;\mathcal V(\OO))\!\cap\! L^2(\theta_i,T;W^{2,2}(\OO)),u^i\in C(\theta_i,T; W^{1,2}(\Omega_R))\,\,\text{for all R}>\text{diam}(\mathcal B)\,,\VS\,u^i_t(1\!+\!|x|)^{-1}\!\!\in \!L^2(\theta_i,T;L^2(\OO)),\VS \lim_{t\to\theta_i}\xi^i(t)=\xi^i(\theta_i)\,,\quad\lim_{t\to\theta_i}\omega^i(t)=\omega^i(\theta_i)\,,\quad\lim_{t\to\theta_i}\dm u^i(t)-u^i(\theta_i)\dm_{2}=0\,.\ea\ee Finally, since Lemma \,\ref{le: Preliminari-struttura} ensures that $\theta_i\le \theta$, we get $\displ[\theta,\infty)\subseteq {\underset{i\ge n(T^*)}\cap (\theta_i,\infty)}$ and the lemma is completely proved.
\ep
We are now in a position to prove our main result.
\subsection{Existence of a weak solution}
 \par Let $\{T_h\}_{h\in\N_0}$ be the sequence stated in Theorem\,\ref{thm: Prolungamento} and $\{(u^i,\pi^i,\xi^i,\omega^i)\}_{i\ge n(T^*)}$ be the subsequence, ensured by Theorem\,\ref{thm: Prolungamento}, defined on the interval $[0,T^*]$. By virtue of Theorem\,\ref{thm: Prolungamento},  for all $t\in(0,T^*]$, we have
\be \label{eq: Energy-Limit1}
\dm u^i(t)\dm_2^2+|\xi^i(t)|^2+\omega^i(t)\cdot (I\cdot\omega^i(t))+2\int_0^{t}\dm \n u^i\dm_2^2\,d\tau\le 2A_0\,,
\ee
while, remarking that $\theta_i\le \theta$, by virtue of Lemma\,\ref{le: globale mollificato} and Lemma\,\ref{le:prel-symgrad}, we have
\begin{equation} \label{eq: stime metriche forti}
    \ba {c}
    \displ \int_\theta^{\infty} \!\!(|\xi^i(t)|^2\!+\!\omega^i(t)\cdot\!(I\!\cdot\omega^i(t))\!+\!\dm \n u^i(t)\dm_2^2)\,dt \leq c\delta^2\,,\VS\hskip-0.3cm
    \int_\theta^{\infty}(|\dot{\xi}^i(t)|^2\!+\!\dot\omega^i(t)\!\cdot\!(I\!\cdot\!\dot\omega^i(t))\!+\!\dm D^2u^i(t)\dm_2^2\!+\!\dm \n\pi^i(t)\dm_2^2\!+\!\dm u^i_t(K_0+\kappa)^{-1}\dm_2^2)dt\!\leq \!c(M\delta\!+\!M^3\delta^3)\,,
    \ea
\end{equation}
where $M$ and $\delta$ are deduced in Lemma\,\ref{le: globale mollificato}, in accord with Lemma\,\ref{le: Gronwall generalizzato} and Lemma\,\ref{le: Preliminari-struttura}. \par
{Let us consider the following extensions of the fields $(u^i,V^i)$:
\[
\widetilde u^i(t,x):=\begin{cases}
    u^i(t,x)\,,\quad \text{if }x\in\overline\Omega\,,\\
    V^i(t,x)\,, \quad \text{if }x\in\mathcal B\,,
\end{cases}
\]
\[
\widetilde V^i(t,x):= V^i(t,x)\,,\quad \text{for all }x\in \mathbb R^3\,.
\]
By virtue of Lemma\,\ref{le: conv-ext}, Remark\,\ref{rem:cv-H} and Lemma\,\ref{le: TF-ext}, we state the existence of a subsequence of $\{(\widetilde u^i,\widetilde V^i)\}_{i\ge n(T^*)}$, that for simplicity we still denote by $\{(\widetilde u^i,\widetilde V^i)\}$, such that, for all $\widetilde\varphi\in L(\R^3)$,
\[
(\widetilde u^i(t),\widetilde\varphi)_1\to (\widetilde u(t),\widetilde\varphi)_1\,, \,\,\text{for all }t\in[0, T]\,.
\]
The function $\widetilde u$ is such that
\[
\widetilde u(t,x):=\begin{cases}
    u(t,x)\,,\quad \text{if }x\in\overline\Omega\,,\\
    V(t,x)=\xi(t)+\omega(t)\times x\,, \quad \text{if }x\in\mathcal B\,,
\end{cases}
\]
and $u(t)\in \mathcal H(\Omega)$, with $\phi_\nu [u(t)]=\phi_\nu [V(t)]$, for all $t\in [0,T]$. Moreover, we denote by $\widetilde V$ the extension of $V$ to the whole space.\par}
{Going back to \eqref{eq: Energy-Limit1} and \eqref{eq: stime metriche forti}, from the subsequence $\{(\widetilde u^i,\widetilde V^i)\}$ we can extract a subsequence, that for simplicity we still denote by $\{(\widetilde u^i,\widetilde V^i)\}$, weakly converging to a function $\widetilde{\widetilde  u}$ in $L^2(0,T; W(\R^3))$. In particular, the convergence is weak in $L^2(0,T;L(\R^3))$. So, we deduce that $\widetilde{\widetilde u}= \widetilde u$ a. e. in $[0,T]$ and $\gamma(\widetilde u(t))=V(t,x)=\xi(t)+\omega(t)\times x$ for almost all $t\in[0,T]$. The previous identity holds in the trace sense in $H^\frac12(\partial\Omega)$.} \par {Moreover, by virtue of Lemma\,\ref{le: SC-BR}, we also have, for all $R>\text{diam}(\mathcal B)$,
\[
\int_0^T \big[\dm u^i(t)-u(t)\dm_{L^2(\Omega_R)}^2+|\xi^i(t)-\xi(t)|^2+|\omega^i(t)-\omega(t)|^2\big]\,dt\to 0\,.
\]
Hence, we deduce that
\be \label{SCONV}
\ba{c}
u^i\to u\,\,\text{strongly in }L^2(0,T;L^2(\Omega_R))\,,\VS 
\widetilde V^i\to \widetilde V\,\,\text{strongly in }L^2(0,T;L^2(B_R))\,,\VS \text{and, in particular, }
\xi^i(t)\to \xi(t)\,,\,\,\omega^i(t)\to\omega(t)\,,\,\,\, \text{a. e. in } [0,T]\,.
\ea
\ee} Finally, we prove that $(u,\xi,\omega)$ is a weak solution to problem \eqref{eq:model} in the sense of Definition\,\ref{def:WS}. \par
 We multiply equation \eqref{eq:moll}$_1$ by a test function $\varphi\in\mathscr C(\mathbb R^3_{T})$ and we integrate over $(0,T)\times \Omega$. We get
\[
\begin{aligned}
                &\int_0^{T}\! \!\Big[\int_{\Omega} \big[u^i\!\cdot \varphi_t \!-2\mathbb D( u^i)\!\cdot \!\mathbb D( \varphi) -(\mathbb J_i(u^i\!-\!V^i)\!\cdot \!\nabla u^i) \cdot \varphi - \omega^i\!\times u^i\!\cdot \varphi\big]\,dx+\frac{d\overline{\varphi}_1}{dt}\cdot \xi^i\! +\frac{d\overline{\varphi}_1}{dt}\cdot\! (I\!\cdot \omega^i)\Big]dt\\
                & \hskip1cm= -\int_0^{T} [\overline{\varphi}_1\cdot\xi^i\times\omega^i + \overline{\varphi}_2\cdot(I\cdot\omega^i)\times\omega^i]\, dt - \int_{\Omega}\varphi(0)\cdot u_0\, dx -\overline{\varphi}_1(0)\cdot\xi_0 -\overline{\varphi}_2(0)\!\cdot\!( I\!\cdot \!\omega_0)\,.
            \end{aligned}
\]
For the linear terms, {via the weak and strong convergence properties previously established,} the convergence is immediate. We discuss the convergence of 
\[
\int_0^{T} \int_{\Omega}(\mathbb J_i(u^i-V^i)\cdot \nabla u^i) \cdot \varphi\,dx\,dt=\int_0^{T} \int_{\Omega}(\mathbb J_i(\widetilde u^i-\widetilde V^i)\cdot \nabla \widetilde u^i) \cdot \varphi\,dx\,dt=\int_0^{T}\int_\Omega \mathbb (\varphi\otimes \mathbb J_i(\widetilde u^i-\widetilde V^i))\cdot \nabla \widetilde u^i \,.
\]
We have
\[
\ba{l}\displ
\int_0^{T}\int_\Omega  (\varphi\otimes \mathbb J_i(\widetilde u^i))\cdot \nabla\widetilde u^i \,dx\,dt=\int_0^{T}\int_\Omega  (\varphi\otimes\mathbb J_i(\widetilde u^i-\widetilde u))\cdot \nabla\widetilde u^i\,dx\,dt \VS\hskip7cm+\int_0^{T}\int_\Omega  (\varphi\otimes\mathbb J_i(\widetilde u))\cdot \nabla\widetilde u^i \,dx\,dt=:L_1+L_2\,.
\ea
\]
We notice that
\[
L_2=\int_0^{T}\int_\Omega  (\varphi\otimes\mathbb J_i(\widetilde u)-\widetilde u)\cdot \nabla\widetilde u^i \,dx\,dt+\int_0^{T}\int_\Omega  (\varphi \otimes\widetilde u)\cdot \nabla\widetilde u^i \,dx\,dt=:L_{2,1}+L_{2,2}\,.
\]
Since $\varphi$ has compact support, properties \eqref{SCONV} and the weak convergence of $\{\widetilde u^i\}$ in $L^2(0,T;W(\R^3))$ allow us to say that $L_{2,2}$ converges to
\[
\int_0^{T}\int_\Omega  (\varphi\otimes\widetilde u)\cdot \nabla\widetilde u\,dx\,dt=\int_0^{T}\int_\Omega\widetilde u\cdot \nabla\widetilde u\cdot\varphi \,dx\,dt=\int_0^{T}\int_\Omega u\cdot \nabla u\cdot\varphi \,dx\,dt\,.
\]
Concerning $L_{2,1}$, we have
\[
|L_{2,1}|\le \int_0^{T}\dm \mathbb J_i(\widetilde u)-\widetilde u\dm_2\dm \varphi\dm_{\infty}\dm \n\widetilde u^i\dm_2\to 0\,,\quad \text{if }n\to+\infty\,.
\]
Finally, considering $L_1$, we have
\[
|L_1|\le \int_0^{T}\dm \varphi\dm_{\infty}\dm \mathbb J_i(\widetilde u^i-\widetilde u)\dm_{\null_{L^2(\text{supp}\,\varphi)}}\dm \n\widetilde u^i\dm_2\le \int_0^{T}\dm \varphi\dm_{\infty}\dm \widetilde u^i-\widetilde u\dm_{\null_{L^2(\text{supp}\,\varphi)}}\dm \n\widetilde u^i\dm_2
\]
Letting $i\to+\infty$, by virtue of \eqref{SCONV} the last integral converges to $0$. \par
We now study the convergence of 
\[
\int_0^{T}\int_\Omega\mathbb J_i(\widetilde V^i)\cdot \n\widetilde u^i\cdot \varphi\,dx\,dt=\int_0^{T}\int_\Omega\mathbb J_i(\widetilde V^i-\widetilde V)\cdot \n\widetilde u^i\cdot \varphi\,dx\,dt+\int_0^{T}\int_\Omega\mathbb J_i(\widetilde V)\cdot \n\widetilde u^i\cdot \varphi\,dx\,dt=:L_3+L_4\,.
\]
We have
\[
|L_3|\le \int_0^{T}\dm \mathbb J_i(\widetilde V^i-\widetilde V)\dm{\null_{L^2(\text{supp}\,\varphi)}}\dm \varphi\dm_\infty\dm \n\widetilde u^i\dm_2\le c(\text{supp}\,\varphi)\int_0^{T}(|\xi^i-\xi|+|\omega^i-\omega|)\dm \varphi\dm_\infty\dm \n\widetilde u^i\dm_2\,.
\]
 Letting $i\to+\infty$, by virtue of \eqref{SCONV} the last integral converges to $0$. 
We conclude by discussing the contribution of $L_4$. We have
\[
L_4=\int_0^{T}\int_\Omega(\mathbb J_i(\widetilde V)-\widetilde V)\cdot \n\widetilde u^i\cdot \varphi\,dx\,dt+\int_0^{T}\int_\Omega\widetilde V\cdot \n\widetilde u^i\cdot \varphi\,dx\,dt=:L_{4,1}+L_{4,2}\,.
\]
Concerning $L_{4,1}$, we get
\[
|L_{4,1}|\le \int_0^{T}\dm \mathbb J_i(\widetilde V)-\widetilde V\dm{\null_{L^2(\text{supp}\,\varphi)}}\dm \varphi\dm_\infty\dm \n\widetilde u^i\dm_2\to 0\,,\quad \text{if }i\to+\infty\,.
\]
Finally, we consider the integral $L_{4,2}$. We recall that $\widetilde u=\widetilde{\widetilde u}$ a. e. in $t\in [0,T]$, with $\widetilde{\widetilde u}$ weak limit of $\{\widetilde u^i\}$ in $L^2(0,T;W(\R^3))$. Recalling also that $\varphi$ has compact support and properties \eqref{SCONV}, we get the convergence 
\[
L_{4,2}\to \int_0^{T}\int_\Omega\widetilde V\cdot \nabla\widetilde u \cdot\varphi\,dx\,dt=\int_0^{T}\int_\Omega V\cdot \nabla u \cdot\varphi\,dx\,dt \,.
\]
We now discuss the convergence of the quadratic term
\[
\int_0^T \int_\Omega \omega^i\times u^i\cdot \varphi\,dx\,dt\,.
\]
Since $\varphi$ has compact support, the convergence properties \eqref{SCONV}$_{1,3}$ allow us to conclude that 
\[
\int_0^T \int_\Omega \omega^i\times u^i\cdot \varphi\,dx\,dt\to \int_0^T \int_\Omega \omega\times u\cdot \varphi\,dx\,dt\,.
\]
 \par{Concerning the convergence property \eqref{CDI}, we notice that the regularity of the initial datum ensures that the solution is regular and unique on an interval of time $[0,\theta_0)$, $\theta_0>0$. Therefore, the claimed convergence holds. We deduce that the three functions $(u,\xi,\omega)$ are a weak solution in the sense of Definition\,\ref{def:WS}.\par Finally, in order to get \eqref{REG}, we consider the energy equality \eqref{ENEQ} that is valid, in particular, for the subsequence $\{(u^i,\xi^i,\omega^i)\}$. Passing to the minimum limit as $i\to \infty$ and invoking Lemma\,\ref{le:convergenza integrale di superficie}, we obtain \eqref{REG}.}
\subsection{Partial regularity of the weak solution}
We now prove that the weak solution constructed in the previous section enjoys a partial regularity property. Via estimates \eqref{eq: stime metriche forti}, we deduce the existence of a subsequence of $\{u^i\}$ (still denoted by $\{u^i\}$), extracted from the subsequence converging to the weak solution $(u,\xi,\omega)$ obtained in the previous subsection, weakly converging to a function $\widehat u$ in $L^2(\theta,T;W^{2,2}(\Omega))$, {with $\widehat u_t(K_0+\kappa)^{-1}\in L^2(\theta,T;L^2(\Omega))$}, and a subsequence of $\{(\xi^i,\omega^i)\}$ weakly converging to $(\widehat\xi,\widehat\omega)$ in $W^{1,2}(\theta,T)$, for all $T>\theta$. Then, by virtue of Corollary\,\ref{GEN-CF}, we get $u^i\to \widehat u$ strongly in $ L^2(\theta,T;W^{1,2}(\Omega_R))$ and $(\xi^i,\omega^i)\to (\widehat\xi,\widehat\omega)$ a.e. in $(\theta,T)$. Therefore, by virtue of the uniqueness of the weak limit in $L^2(\theta,T;\mathcal V(\Omega))$, we deduce that $\widehat u=u$ and $(\widehat \xi,\widehat \omega)=(\xi,\omega)$ a.e. in $t\in (\theta,T)$. Hence, we obtain that the weak solution $(u,\xi,\omega)$, constructed in the previous subsection, is such that $u$ belongs to $L^2(\theta,\infty;W^{2,2}(\Omega))\cap L^{\infty}(\theta,\infty;\mathcal V(\Omega))$, and  {we also get $u_t(K_0+\kappa)^{-1}\in L^2(\theta,\infty;L^2(\Omega))$. Moreover, we get $\xi,\omega\in W^{1,2}(\theta,T)\cap C([\theta,T])$, for all $T>\theta$. We conclude that the solution is regular in $(\theta,\infty)$.}
The proof of our main result is concluded.
\subsection{Proof of Corollary\,\ref{ASBEH}}
By virtue of Theorem\,\ref{thm: 3/2weak} and Lemma\,\ref{le: globale mollificato}, for all $i\ge n(T^*)$ and for all $q\in (\frac32,2)$ the sequence $\{u^i\}_i \subset L^\infty(\theta,\theta+T; L^2(\Omega))\cap L^q(\theta,\theta+T; L^q(\Omega)) $, for all $T>0$. Moreover, the bound of $u^i$ in $L^q(\theta,\theta+T; L^q(\Omega))\cong L^q((\theta,\theta+T)\times \Omega)$ is uniform with respect to $i$. Then, there exists a weak limit $\widehat{\widehat u}$ of a subsequence extracted from $\{u^i\}_i$ in $L^q(\theta,\theta+T; L^q(\Omega))$. Since $\{u^i\}\to u$ strongly in $L^2(\theta,\theta+T; L^2(\Omega_R))$ and $L^2(\theta,\theta+T; L^2(\Omega_R))$ is continuously embedded in $ L^q(\theta,\theta+T; L^q(\Omega_R))$, we get $u=\widehat{\widehat u}$ a. e. in $(\theta,\theta+T)$. In particular, we deduce that there exists $t_0\in (\theta,\theta+T)$ such that $\dm u(t_0)\dm_q$ is finite. Therefore, employing the arguments in \cite{Gal-Mar}, we conclude the proof.

\end{document}